\documentclass[12pt,letterpaper,a4paper]{article}
\usepackage[includeheadfoot,
marginratio={1:1,2:3},
width=525pt,
height=786pt,]{geometry}

\usepackage{amsmath}
\usepackage{amsfonts}
\usepackage{amssymb}
\usepackage{graphicx}
\usepackage{cancel}
\usepackage{empheq}
\usepackage{color}
\usepackage{hyperref}

\numberwithin{equation}{section}
\usepackage{float}
\restylefloat{table}
\restylefloat{figure}
\usepackage[utf8]{inputenc}
\usepackage{amsmath, amsthm, amssymb, amsfonts}
\usepackage[font=small, labelfont=bf]{caption}
\usepackage{amsmath, amsthm, amssymb, amsfonts}
\usepackage{multirow}
\usepackage{graphicx}
\usepackage{float}
\usepackage[numbers,sort&compress]{natbib}
\usepackage{enumerate}
\usepackage{amsmath}
\usepackage{amsfonts}
\usepackage{amssymb}
\usepackage{graphicx}
\usepackage{cancel}
\usepackage{empheq}
\usepackage{color}
\usepackage{hyperref}
\usepackage{float}
\restylefloat{table}
\restylefloat{figure}
\usepackage{longtable}
\usepackage{lipsum} 

\usepackage{amsfonts}
\usepackage{amssymb}
\usepackage{comment}
\usepackage{graphicx}
\usepackage{amsmath}
\usepackage{tabu}
\usepackage{dsfont}
\usepackage{float}
\usepackage{amsthm}
\usepackage{slashed}
\usepackage{setspace}
\usepackage[labelformat=simple]{subcaption}

\usepackage{pgfplots}
\usepackage{tikz}
\usepackage{tikz-cd}
\usetikzlibrary{shapes.misc,shadows,fit,decorations.pathmorphing,positioning,trees,decorations.markings,decorations.pathreplacing,calc,shapes,patterns,arrows,positioning,arrows.meta}
\usepackage{xcolor}
\usepackage{pgfplots}
\usepackage{pgfplotstable}
\usepackage{xcolor}
\usepackage{makecell}
\usepackage{diagbox}
\usepackage{environ}
\usepackage[utf8]{inputenc}
\usepackage{amsmath, amsthm, amssymb, amsfonts}
\usepackage{amsmath, amsthm, amssymb, amsfonts}
\usepackage{multirow}
\usepackage{graphicx}
\usepackage{float}
\usepackage[numbers,sort&compress]{natbib}
\usepackage{enumerate}
\usepackage{enumitem}
\usepackage[capitalize,sort]{cleveref}
\usepackage{hhline}
\usepackage{mathtools}
\usepackage{longtable}
\usepackage{makecell}
\usepackage{tabularx}
\usepackage{cellspace}
\usepackage{alphalph}
\usepackage{lscape}
\usepackage{slashed}
\usepackage{setspace}
\usepackage{pifont}

\crefname{figure}{Figure}{Figures}
\crefname{table}{Table}{Tables}

\def\be{\begin{equation}}
\def\ee{\end{equation}}
\def\bea{\begin{eqnarray}}
\def\eea{\end{eqnarray}}
\def\bes{\begin{subequations}}
	\def\ees{\end{subequations}}

\newcommand{\bmat}{\left(\begin{array}}
	\newcommand{\emat}{\end{array}\right)}

\def\ov{\overline}

\def\ov{\overline}
\def\1{{\bf 1}}
\def\2{{\bf 2}}
\def\3{{\bf 3}}
\def\4{{\bf 4}}
\def\6{{\bf 6}}

\newcommand{\beq}{\begin{equation}}
\newcommand{\eeq}{\end{equation}}

\def\ov{\overline}

\allowdisplaybreaks[2]
\numberwithin{equation}{section}

\def\rmt{\rm t}
\def\rmz{\rm z}

\def\be{\begin{equation}}
\def\ee{\end{equation}}
\def\bea{\begin{eqnarray}}
\def\eea{\end{eqnarray}}
\def\bes{\begin{subequations}}
	\def\ees{\end{subequations}}

\usepackage{multicol}
\usepackage{float}
\usepackage{caption}
\allowdisplaybreaks[2]
\numberwithin{equation}{section}

\begin{document}
	{\hfill
		%
		\hfill
		}

	\vspace{1.0cm}
	\begin{center}
		{\Large Systematic exploration of the non-geometric flux landscape
   }
		\vspace{0.4cm}
	\end{center}

	\vspace{0.35cm}
	\begin{center}
		Shehu AbdusSalam$^\diamond$, Xin Gao$^\dagger$, George K. Leontaris$^{\ddagger}$, and Pramod Shukla$^\star$ \footnote{Email: abdussalam@sbu.ac.ir, xingao@scu.edu.cn, leonta@uoi.gr, pshukla@jcbose.ac.in}
	\end{center}

\vspace{0.1cm}
\begin{center}
{$^\diamond$ Department of Physics, Shahid Beheshti University, Tehran, Iran.}\\
 \vskip0.5cm
 {$^\dagger$ College of Physics, Sichuan University, Chengdu 610065, China.}\\
  \vskip0.5cm
{$^\ddagger$ Physics Department, University of Ioannina,
University Campus, Ioannina 45110, Greece.}
 \vskip0.5cm
{$^\star$ Department of Physical Sciences, Bose Institute,\\
Unified Academic Campus, EN 80, Sector V, Bidhannagar, Kolkata 700091, India.}
\end{center}
\vspace{1cm}

\abstract{
Given the huge size of the generic four-dimensional scalar potentials arising from the type II supergravities based on toroidal orientifolds, it is even hard to analytically solve the  extremization conditions, and therefore the previous studies have been mainly focused on taking some numerical approaches. In this work, using the so-called {\it axionic flux polynomials} we demonstrate that the scalar potential and the extremization conditions can be simplified to a great extent, leading to the possibility of performing an analytic exploration of the flux landscape. In this regard, we consider the isotropic case of a type IIB model based on the standard ${\mathbb T}^6/({\mathbb Z}_2 \times {\mathbb Z}_2)$ orientifold having the three-form fluxes $F_3/H_3$ and the non-geometric $Q$-flux. This model results in around 300 terms in the scalar potential which depend on 6 moduli/axionic fields and 14 flux parameters. Considering that the axionic flux polynomials can take either zero or non-zero values results in the need of analyzing $2^{14}=16384$ candidate configurations, and we find that more than 16200 of those result in No-Go scenarios for Minkowskian/de-Sitter vacua. Based on our systematic exploration of non-tachyonic flux vacua, we present a detailed classification of such No-Go scenarios as well as the leftover ``undecided" configurations for which we could not conclude about the presence/absence of the stable Minkowskian/de-Sitter vacua.
}

\clearpage

\tableofcontents


\section{Introduction}
\label{sec_intro}

Non-geometric fluxes in type IIB superstring compactifications constitute a fascinating  feature which might provide
new possibilities in the search of de-Sitter (dS) vacua. They play a decisive role in shaping the scalar potential of the effective field theories, leading to novel physical implications and offering new avenues for moduli stabilisation in string theory.
Due to their remarkable role for possible extensions of  the string landscape,  they have attracted a lot of  attention  of the community working in string theory model building~
\cite{Aldazabal:2006up, deCarlos:2009qm,Danielsson:2012by, Blaback:2013ht, Damian:2013dq, Damian:2013dwa, Hassler:2014mla,Blumenhagen:2015qda, Blumenhagen:2015kja,Blumenhagen:2015jva, Blumenhagen:2015xpa,  Li:2015taa,Plauschinn:2020ram,Shukla:2022srx,Damian:2023ote}; see \cite{Plauschinn:2018wbo} for a review. Non-geometric fluxes emerge from applying T-duality transformation   that links  different string theories.
 For the case of type IIB and type IIA string theories in particular, under T-duality a geometric flux corresponds to a non-geometric one  that  cannot be captured purely through Riemannian geometry.
 One of the main motivations for introducing these new kind of fluxes is to restore T-duality which is lost between type IIA and type IIB superstring theories
in the presence of standard NS-NS and RR $p$-form fluxes only. More specifically, in this procedure,  three new kinds of fluxes  are created when a chain of successive T-duality transformations is applied to the three-form NS-NS flux $H_{abc}$.
 The first T-duality in this chain generates an internal spin connection with components denoted as $\omega_{bc}^a$,
 which has a geometric interpretation of the twisted torus.
 When a second T-duality is applied, the so called non-geometric $Q$-flux is incorporated with components $Q_{c}^{ab}$. This flux  is associated with topological properties of the internal six dimensional space in a way that does not have a global geometric interpretation. The complete T-duality between type {IIA} and {IIB} is restored with a third  T-dual transformation which gives rise to the non-geometric $R$-flux\footnote{An explicit dictionary between the type IIA and type IIB fluxes under the T-duality can be found in \cite{Shukla:2019wfo}.}. Schematically,
 the above  chain with geometric and non-geometric fluxes
 is summarized below
 \begin{equation}
 H_{abc}\;\xrightarrow[]{{T_a}}\; \omega^a_{bc}\;\xrightarrow[]{{T_b}}\;Q^{ab}_c\;\xrightarrow[]{{T_c}}\;R^{abc}~.
 \label{Tdualnongeo}
 \end{equation}
In parallel to the above T-duality operations, S-duality transformations must be implemented in order to achieve modular completion of type IIB superstring compactifications. For example, considering a type IIB orientifold setup with $h^{1,1}_- = 0 = h^{2,1}_+$, four S-dual pairs of fluxes are incorporated to have a U-dual completion of the flux superpotential, and these are denoted as \cite{Aldazabal:2006up}
 \begin{equation}
 (F, H), \quad (Q, P), \quad (P^\prime, Q^\prime), \quad (H^\prime, F^\prime).
 \label{Sdualfluxes}
 \end{equation}
The implications  of these generalised fluxes in the scalar potential of  the effective theory derived form Type II superstring compactifications have attracted a lot of attention during the last two decades~\cite{Derendinger:2004jn,Shelton:2005cf,Grana:2012rr,Dibitetto:2012rk, Danielsson:2012by, Blaback:2013ht, Damian:2013dq, Damian:2013dwa, Hassler:2014mla, Ihl:2007ah, deCarlos:2009qm, Danielsson:2009ff, Blaback:2015zra, Dibitetto:2011qs, Plauschinn:2018wbo,Damian:2023ote}. Consequently,  an alternative model building phenomenology has been initiated under this novel flux  background~\cite{Hassler:2014mla, Blumenhagen:2015qda, Blumenhagen:2015kja,Blumenhagen:2015jva,  Li:2015taa, Blumenhagen:2015xpa}. In this regard, toroidal orientifolds, despite being simple, have served as a fruitful and promising ground for model building studies such as moduli stabilization and the search of physical (dS) vacua \cite{Kachru:2003aw, Balasubramanian:2005zx, Angelantonj:2002ct, Sagnotti:1987tw, Grana:2005jc, Blumenhagen:2006ci, Douglas:2006es, Denef:2005mm, Blumenhagen:2007sm}. In the initial stages, model building efforts using non-geometric fluxes have been made mostly via considering the four-dimensional (4D) effective scalar potentials arising from the K\"ahler potentials and the flux superpotentials \cite{Danielsson:2012by, Blaback:2013ht, Damian:2013dq, Damian:2013dwa, Blumenhagen:2013hva, Villadoro:2005cu, Robbins:2007yv, Ihl:2007ah, Gao:2015nra}. However,  a proper understanding of the higher-dimensional origin of such 4D non-geometric scalar potentials was lacking until recent developments took place~\cite{Villadoro:2005cu, Blumenhagen:2013hva, Gao:2015nra,Shukla:2015rua, Shukla:2015bca,Gao:2017gxk,Shukla:2019wfo,Leontaris:2023lfc,Leontaris:2023mmm}.

Given the existence of some close connections between the 4D effective potentials of type II supergravities and the symplectic geometries, the requirement of CY metric can be bypassed by using (a combination of) period vectors and moduli space metrics~\cite{Ceresole:1995ca, Taylor:1999ii, D'Auria:2007ay}.
This strategy of circumventing the need of the  CY metric was subsequently adopted for a number of type IIA/IIB models with various generalised fluxes, e.g.
see~\cite{Shukla:2015hpa,Blumenhagen:2015lta,Shukla:2016hyy,Shukla:2019wfo,Shukla:2019dqd,Shukla:2019akv,Leontaris:2023lfc,Leontaris:2023mmm,Biswas:2024ewk} for type IIB based models, \cite{Gao:2017gxk,Shukla:2019wfo,Marchesano:2020uqz,Prieto:2024shz} for type IIA based models, and \cite{Marchesano:2021gyv} for F-theory based models. For example, the generic 4D scalar potential in type IIB model induced by the standard RR and NS-NS flux pair $(F, H)$ can be equivalently computed by considering the flux superpotential  \cite{Gukov:1999ya,Dasgupta:1999ss} or by the dimensional reduction of the 10D kinetic pieces by using the period matrices \cite{Taylor:1999ii, Blumenhagen:2003vr}. One can generalize this  background by  consistently including the four S-dual pairs of fluxes given in~(\ref{Sdualfluxes})
 in the generalized superpotential \cite{Shelton:2005cf, Aldazabal:2006up,Aldazabal:2008zza,Font:2008vd,Guarino:2008ik, Hull:2004in, Kumar:1996zx, Hull:2003kr,Aldazabal:2010ef, Shukla:2015rua, Lombardo:2016swq, Lombardo:2017yme}. Such fluxes act as some parameters in the 4D effective supergravity dynamics and can induce a diverse set of superpotential couplings in the model building context \cite{Derendinger:2004jn,Grana:2012rr,Dibitetto:2012rk, Danielsson:2012by, Blaback:2013ht, Damian:2013dq, Damian:2013dwa, Hassler:2014mla, Ihl:2007ah, deCarlos:2009qm, Danielsson:2009ff, Blaback:2015zra, Dibitetto:2011qs}.

In the context of model building, the major challenges faced in any non-geometric setup can be encoded by posing the following two questions:
\begin{enumerate}

\item
In a given explicit orientifold construction, how many and which types of fluxes can be consistently turned-on ? For example, implementing the successive chain of T- and S-dualities for a toroidal type IIB $\mathbb T^6/(\mathbb Z_2 \times \mathbb Z_2)$ orientifold model leads to a U-dual completed holomorphic flux superpotential with 128 flux parameters \cite{Aldazabal:2006up, Aldazabal:2008zza, Aldazabal:2010ef, Lombardo:2016swq, Lombardo:2017yme}.

This challenge of ``how many consistent fluxes are allowed ?" is usually encoded in satisfying the tadpole cancellation conditions and the Bianchi identities \cite{Robbins:2007yv, Ihl:2007ah,Shukla:2016xdy,Plauschinn:2021hkp}. In this regard it is worth mentioning that there are two inequivalent formulations of the Bianchi identities (known as cohomology formulation and standard formulation \cite{Robbins:2007yv, Ihl:2007ah,Shukla:2016xdy}) and addressing this for the beyond-toroidal cases with prime-fluxes in (\ref{Sdualfluxes}) is still an open question.

\item

How to handle the enormously huge size of the scalar potential induced through the generalized flux superpotential in typical models ? For example, it has recently been found that the scalar potential of the moduli fields $S, T_\alpha, U^i$ induced in a toroidal type IIB $\mathbb T^6/(\mathbb Z_2 \times \mathbb Z_2)$ orientifold model results in 76276 terms \cite{Leontaris:2023lfc,Leontaris:2023mmm} making it impractically large even for solving the extremization conditions !

For addressing this issue, a relatively compact formulation of the scalar potential has been presented in \cite{Biswas:2024ewk} using a set of {\it novel  axionic fluxes}, which are some combinations of axions and standard fluxes. It has been found that the complicated scalar potential with 76276 terms can be equivalently expressed in terms of only 2816 terms \cite{Biswas:2024ewk} ! This creates a hope to perform a systematic scan of physical vacua using an analytic approach.

\end{enumerate}

\noindent
These effective potential studies and their compact formulations in terms of presenting the master formulae can be directly relevant for making an analytic exploration of physical vacua in non-geometric models, e.g. on the lines of studies presented in \cite{Shukla:2016xdy, Shukla:2019dqd, Shukla:2019akv, Marchesano:2020uqz,Shukla:2022srx,Prieto:2024shz}.

In the context of realizing de-Sitter solutions and possible obstructions on the way of doing it in the string theory (inspired) setups, the following three points should be taken into account: 
\begin{itemize}
\item{Existence: Having a simple model building approach with the minimal ingredients at hand, several de-Sitter no-go scenarios have been found from time to time, e.g. see \cite{Maldacena:2000mw, Hertzberg:2007wc, Hertzberg:2007ke, Haque:2008jz, Flauger:2008ad, Caviezel:2008tf,  Covi:2008ea, deCarlos:2009fq, Caviezel:2009tu, Danielsson:2009ff, Danielsson:2010bc, Wrase:2010ew,  Shiu:2011zt, McOrist:2012yc, Dasgupta:2014pma, Gautason:2015tig, Junghans:2016uvg, Andriot:2016xvq, Andriot:2017jhf}. Such no-go results have played a key role in the revival of the swampland conjectures \cite{Ooguri:2006in, Obied:2018sgi, Danielsson:2018ztv,Shukla:2019dqd,Shukla:2019akv,Marchesano:2020uqz}. However, in contrary to the (minimal) de-Sitter no-go scenarios, in the meantime there have been several proposals for realizing (stable) de-Sitter vacua \cite{Kachru:2003aw, Burgess:2003ic, Achucarro:2006zf, Westphal:2006tn, Silverstein:2007ac, Rummel:2011cd, Cicoli:2012fh, Louis:2012nb, Cicoli:2013cha, Cicoli:2015ylx, Cicoli:2017shd, Akrami:2018ylq, Antoniadis:2018hqy,Antoniadis:2018ngr, Antoniadis:2019rkh, Basiouris:2020jgp, Antoniadis:2020stf, Cicoli:2018kdo, Crino:2020qwk, Cicoli:2021dhg, Andriot:2022way,Heckman:2019dsj, Heckman:2018mxl}.}

\item{Stability: Checking the stability a de-Sitter solution is a quite crucial task as the simple attempts of evading the no-go results may turn out to be resulting in some tachyonic solutions \cite{Flauger:2008ad,Hertzberg:2007ke, Haque:2008jz, Danielsson:2009ff, Danielsson:2011au, Chen:2011ac, Danielsson:2012et,Junghans:2016uvg}. Such a possibility of having the tachyonic de-Sitter solutions has led to the refinements in the original swampland conjectures \cite{Ooguri:2006in, Obied:2018sgi}, further leading to numerous amount of follow-up studies \cite{Garg:2018reu, Agrawal:2018own, Andriot:2018wzk, Andriot:2018ept, Denef:2018etk, Conlon:2018eyr, Roupec:2018mbn, Murayama:2018lie, Choi:2018rze, Hamaguchi:2018vtv, Olguin-Tejo:2018pfq, Blanco-Pillado:2018xyn,Lin:2018kjm, Han:2018yrk, Raveri:2018ddi, Dasgupta:2018rtp, Danielsson:2018qpa, Andriolo:2018yrz, Dasgupta:2019gcd, Andriot:2019wrs,Palti:2019pca}.}

\item{Viability: Given that very little is known about the whole series of corrections which may contribute to the scalar potential, it is important to ask questions regarding the overall consistency of the setup, say in terms of challenges in scale separation and field excursions in moduli space \cite{Blumenhagen:2017cxt, Blumenhagen:2018nts, Blumenhagen:2018hsh, Palti:2017elp, Conlon:2016aea, Hebecker:2017lxm, Klaewer:2016kiy, Baume:2016psm, Landete:2018kqf, Cicoli:2018tcq, Font:2019cxq, Grimm:2018cpv, Hebecker:2018fln, Banlaki:2018ayh, Junghans:2018gdb,Junghans:2020acz,Apers:2022zjx}, tadpole conjecture \cite{Bena:2020xrh,Plauschinn:2021hkp,Plauschinn:2020ram}, and ways to avoid it \cite{Marchesano:2021gyv}.}

\end{itemize}

\noindent
In continuation to previous studies \cite{Shukla:2016xdy, Shukla:2019dqd, Shukla:2019akv, Marchesano:2020uqz,Shukla:2022srx,Prieto:2024shz}, regarding the utility of the axionic flux approach of the scalar potential, we plan to present a systematic exploration of the stable flux vacua in the light of finding No-Go scenarios for realizing Minkowskian and de-Sitter solutions. We demonstrate this idea in a concrete toroidal type IIB model based on the standard $\mathbb T^6/(\mathbb Z_2 \times \mathbb Z_2)$ orientifold having the 3-form fluxes $(F_3, H_3)$ and the non-geometric $Q$-flux. It turn out that this model generically induces an effective scalar potential with 2422 terms expressed in terms of 14 axions/moduli fields and 40 flux parameters \cite{Blumenhagen:2013hva}. After imposing the isotropy conditions, one gets a smaller scalar potential with 309 terms expressed in terms of 6 axions/moduli fields and 14 flux parameters. However, it has been observed recently that using the so-called axionic flux polynomials this isotropic scalar potential can be compactly expressed in only 40 terms \cite{Biswas:2024ewk}. Moreover, we find that the set of Bianchi identities continue to hold after standard fluxes being promoted to the axionic fluxes. This subsequently helps one to rewrite the complicated derivatives of the scalar potential as well as the Hessian components in a rather simpler and compact form. In this work we aim to analyze this compact scalar potential in an analytic approach and attempt to find the Minkowskian/de-Sitter solutions which are non-tachyonic. Surprisingly, we find that all of the candidate configurations (which corresponds to checking around 16200 out of 16384 configurations) result in No-Go scenarios for realizing stable Minkowskian/de-Sitter vacua. We classify such No-Go results in a systematic way and present the details of the `Reduced flux landscape' which remain undecided in our analysis.

The article is organized as follows: In Section \ref{sec_Pre} we review the necessary ingredients of type II orientifold models, and set  our notations for the various fluxes, and the moduli fields. In section~\ref{sec_analytic} we develop an analytic approach with a concrete methodology of applying a step-by-step consistency check  to exclude candidate configurations which prevent Minkowskian/de-Sitter vacua. Demonstrate the use of axionic flux polynomials, by the end of this step we show that the possible candidate configurations reduce by 98\%. Subsequently we adopt a numerical approach to further explore the reduced flux landscape in Section \ref{sec_numeric} which results in some additional No-Go cases. Finally we conclude in Section \ref{sec_conclusions} with a summary of the results about classifying the various No-Go scenarios. In addition, we provide three appendices; the first appendix \ref{sec_appendix1} collects the relevant information about the analogous type IIA setup which is T-dual to the type IIB model we studied, while the second appendix \ref{sec_appendix2} collects all the Hessian components needed for the checking the stability of the candidate flux vacua. The third appendix \ref{sec_appendix3} consists of the 161 (out of a total of 16384) configurations for which de-Sitter No-Go could not be confirmed.


\section{Preliminaries}
\label{sec_Pre}
The F-term scalar potential governing the dynamics of the ${\cal N}=1$ low energy effective supergravity can be computed from the K\"ahler potential ($K$) and the flux induced holomorphic superpotential $(W)$ by considering the following well known formula,
\bea
\label{eq:Vtot}
& & \hskip-1cm V=e^{K}\Big(K^{{A} \ov{B}} \, (D_{A} W) \, (\ov D_{\ov {B}} \ov{W}) -3 |W|^2 \Big)  \,,\,
\eea
where the covariant derivatives for a chiral coordinate $A$ is defined as $D_A W = \partial_A W + (\partial_T K) \, W$. This general expression (\ref{eq:Vtot}) has resulted in a series of the so-called ``master-formulae" for the scalar potential for a given set of K\"ahler- and the super-potentials; e.g.~see \cite{Cicoli:2007xp,Shukla:2015hpa,Shukla:2016hyy,Cicoli:2017shd,Gao:2017gxk,Shukla:2019wfo,AbdusSalam:2020ywo,Cicoli:2021dhg,Cicoli:2021tzt,Marchesano:2020uqz,Leontaris:2022rzj,Leontaris:2023lfc, Leontaris:2023mmm, Biswas:2024ewk}

\subsection{Non-geometric fluxes in Type II orientifold setups}

In this subsection we briefly summarize the relevant ingredients of the type II orientifold models, and refer \cite{Shukla:2019wfo} to the readers interested in more details, including the $T$-dual completion of the flux superpotential. Here basically we focus on collecting the relevant information about the K\"ahler potential and the flux superpotential.

\subsubsection{Type IIA model}
Following the notations of \cite{Shukla:2019wfo}, the set of complexified ${\cal N}=1$ chiral coordinates $\left\{{\rm T}^a, \, {\rm N}^0, \, {\rm N}^k, {\rm U}_\lambda \right\}$ necessary for describing the moduli dynamics in the 4D type IIA effective supergravity can be defined as below,
\bea
\label{eq:N=1_coordsIIA}
& & {\rm T}^a = \, {\rm b}^a - i\, \, \rmt^a, \qquad {\rm N}^0 = \, \xi^0 + \, i \, ({\rmz}^0)^{-1}, \qquad {\rm N}^{k} =\, \xi^{k} + \, i \, ({\rmz}^0)^{-1} \, {\rmz}^k , \\
& & {\rm U}_\lambda = \xi_\lambda -\, i \, ({\rmz}^0)^{-1} \left(\frac{1}{2} k_{\lambda\rho\kappa} \, {\rmz}^\rho {\rmz}^\kappa - \frac{1}{2}\, \hat{k}_{\lambda k m} {\rmz}^k {\rmz}^m - \tilde{\rm p}_{k\lambda} \, {\rm z}^k - \tilde{\rm p}_\lambda \right) \, . \nonumber
\eea
Here ${\rm t}^a$ denotes the two-cycle volume moduli, while ${\rm b}^a$ denotes the NS-NS 2-form $B_2$ axions. Further, $\{{\rm z}^0, {\rm z}^k, {\rm z}_\lambda\}$ correspond to dilaton and complex structure moduli while $\{\xi^0, \xi^k, \xi_\lambda\}$ are the components of the RR $3$-form ${\rm C}_3$ following from the orientifold projection via an anti-holomorphic involution. In addition, $\kappa_{abc}$ is the triple intersection number of the CY threefold while $\{k_{\lambda\rho\kappa}, \hat{k}_{\lambda k m}\}$ are the components of the triple intersection number corresponding to the mirror CY consistent with the orientifold construction.

Subsequently, the K\"ahler potential and the generalized flux superpotential can be given as,
\bea
\label{eq:KIIA}
& & \hskip-1cm K_{\rm IIA} = -\ln\left(\frac{4}{3} \, \kappa_{abc}\, \rmt^a \,\rmt^b \, \rmt^c + 2\, {\rm p}_0 \right) - 4 \, \ln ({\rmz}^0)^{-1} - 2 \ln\left(\frac{1}{6} \, k_{\lambda\rho\kappa} \, {\rmz}^\lambda {\rmz}^\rho {\rmz}^\kappa + \frac{\tilde{\rm p}_0}{4}\right),
\eea
and
\bea
\label{eq:WgenIIA}
& & \hskip-1cm \sqrt{2}\, W_{\rm IIA} = \biggl[\ov{e}_0 + {\rm T}^a \, \ov{e}_a + \frac{1}{2} \kappa_{abc} {\rm T}^a {\rm T}^b m^c  + \frac{1}{6}\, \kappa_{abc}\, {\rm T}^a \, {\rm T}^b\, {\rm T}^c\, \, m^0 - i \, {\rm p}_0 \, m^0 \biggr]\, \\
& & \hskip0.5cm - \, {\rm N}^0 \biggl[\ov{\rm H}_0 +  {\rm T}^a \, \ov{w}_{a0}\, + \frac{1}{2} \kappa_{abc} {\rm T}^b {\rm T}^c \, {\rm Q}^a{}_0 + \frac{1}{6}\, \kappa_{abc} {\rm T}^a {\rm T}^b {\rm T}^c \, {\rm R}_0 - i \, {\rm p}_0 \, {\rm R}_0 \biggr] \,\nonumber\\
& & \hskip0.5cm - \, {\rm N}^k \, \biggl[\ov{\rm H}_k + {\rm T}^a \, \ov{w}_{ak}\, + \frac{1}{2} \kappa_{abc} {\rm T}^b {\rm T}^c \, {\rm Q}^a{}_k + \frac{1}{6}\, \kappa_{abc} {\rm T}^a {\rm T}^b {\rm T}^c \, {\rm R}_k - i \, {\rm p}_0 \, {\rm R}_k \biggr] \,\nonumber\\
& & \hskip0.5cm - \, {\rm U}_\lambda \, \biggl[\ov{\rm H}^\lambda + {\rm T}^a \, \ov{w}_{a}{}^\lambda\, + \frac{1}{2} \kappa_{abc} {\rm T}^b {\rm T}^c \, {\rm Q}^{a \lambda} + \frac{1}{6}\, \kappa_{abc} {\rm T}^a {\rm T}^b {\rm T}^c  \, {\rm R}^\lambda - i \, {\rm p}_0 \, {\rm R}^\lambda \biggr], \nonumber
\eea
where we have introduced a shifted version of the flux parameters to absorb the effects from ${\rm p}_{ab}, {\rm p}_a$ in the following manner, e.g. see \cite{Shukla:2019wfo,Biswas:2024ewk},
\bea
\label{eq:IIA-W-fluxshift}
& & \hskip-1cm \ov{e}_0 = e_0 - {\rm p}_a \, m^a, \qquad \, \, \, \, \, \, \ov{e}_a = e_a \, - {\rm p}_{ab} \, m^b + {\rm p}_a \, m^0, \\
& & \hskip-1cm \ov{\rm H}_0 = {\rm H}_0 - {\rm p}_a \, {\rm Q}^a{}_0, \qquad \ov{w}_{a0}= w_{a0}\, - {\rm p}_{ab} \, {\rm Q}^b{}_0 + {\rm p}_a \, {\rm R}_0, \nonumber\\
& & \hskip-1cm \ov{\rm H}_k = {\rm H}_k - {\rm p}_a \, {\rm Q}^a{}_k, \qquad {\ov w}_{ak}= w_{ak}\, - {\rm p}_{ab} \, {\rm Q}^b{}_k + {\rm p}_a \, {\rm R}_k, \nonumber\\
& & \hskip-1cm \ov{\rm H}^\lambda = {\rm H}^\lambda - {\rm p}_a \, {\rm Q}^{a\lambda} , \qquad {\ov w}_{a}{}^\lambda = w_{a}{}^\lambda\, - {\rm p}_{ab} \, {\rm Q}^{b\lambda} + {\rm p}_a \, {\rm R}^\lambda\,.\nonumber
\eea

\subsubsection*{A simple toroidal model}
In particular, we consider the non-geometric type IIA setup based on the toroidal orientifold ${\mathbb T}^6/({\mathbb Z}_2\times{\mathbb Z}_2)$ with the standard involution; e.g. see \cite{Blumenhagen:2013hva, Gao:2017gxk} for more details. In this setting, the untwisted sector corresponds to $h^{1,1}_- = 3, h^{1,1}_+ = 0, h^{2,1} = 3$, which implies that there are three ${\rm U}_\lambda$ moduli and three ${\rm T}^a$ moduli along with a single ${\rm N}^{0}$-modulus. There are no ${\rm N}^k$ moduli present as the even $(1,1)$-cohomology is trivial. Subsequently it turns out that all the fluxes with $k$ index are absent. In addition, let us also note that there will be no $D$-terms generated in the scalar potential as the even $(1,1)$-cohomology is trivial which projects out all the relevant $D$-term fluxes.

Demanding the orientifold requirements, it turns out that there are four components for the ${\rm H}_3$ flux (namely ${\rm H}_0$ and ${\rm H}^\lambda$) and the same for the non-geometric $R$-flux which are denoted as ${\rm R}_0$ and ${\rm R}^\lambda$ for $\lambda \in \{1, 2, 3\}$. In addition, there are 12 flux components for each of the geometric ($w$) flux and the non-geometric (${\rm Q}$) flux, denoted as $\{w_{a0}, \, w_a{}^\lambda\}$ and $\{{\rm Q}^a{}_\lambda, \, {\rm Q}^{a\lambda}\}$ for $\alpha \in \{1, 2, 3\}$ and $\lambda \in \{1, 2, 3\}$. On the RR side, there are eight flux components in total arising from the $p$-form fluxes ($F_p$), namely one from each of the $F_0$ and $F_6$ fluxes denoted as $m^0$ and $e_0$, while three from each of the $F_2$ and $F_4$ fluxes denoted as $m^a, \, e_a$ for $a \in \{1, 2, 3\}$.  This leads to a flux superpotential with 40 flux parameters coupled to 7 complexifield moduli $\{{\rm U}_\lambda, {\rm N}^{0}, {\rm T}^{a}\}$ for $\{\lambda, a\} \in \{1, 2, 3\}$, which further results in a total of 2422 terms in the effective scalar potential \cite{Blumenhagen:2013hva, Gao:2017gxk, Shukla:2019wfo, Marchesano:2020uqz}.

\subsubsection{Type IIB model}

Similarly, the set of ${\cal N}=1$ chiral coordinates $\left\{S, \, T_\alpha, \, G^a, U^i\right\}$ necessary for describing the 4D type IIB supergravity in string-frame are defined as,
\bea
\label{eq:N=1_coordsIIB}
& & \hskip-1cm U^i = v^i  - i \, u^i \, , \qquad S = c_0 + i\, s\,, \qquad G^a =\left(c^a + c_0 \, b^a \right) + \, i \, s \, \, b^a \, ,\nonumber\\
& & \hskip-1cm T_\alpha = \hat{c}_\alpha - i\, s\, \left[\frac{1}{2}\, \ell_{\alpha\beta\gamma} \, t^\beta t^\gamma - \frac{1}{2} \, \hat{\ell}_{\alpha a b} b^a \, b^b\, - \, p_{\alpha a}\, b^a - \, p_\alpha \right],\nonumber
\eea
where $\hat{c}_\alpha$ represents an axionic combination given as $\hat{c}_\alpha = c_\alpha + \hat{\ell}_{\alpha a b} b^a c^b + \frac{1}{2} \, c_0 \, \hat{\ell}_{\alpha a b} b^a \, b^b$, and the T-dual type IIB K\"ahler- and super-potentials take the following forms respectively \cite{Shukla:2019wfo},
\bea
\label{eq:KIIB}
& & \hskip-1cm K_{\rm IIB} = -\ln \left(\frac{4}{3} \, l_{ijk}\, u^i u^j u^k + 2\, \tilde{p}_0 \right)\, - \, 4\, \ln s - 2 \ln\left(\frac{1}{6} \,{\ell_{\alpha \beta \gamma} \, t^\alpha\, t^\beta \, t^{\gamma}} + \frac{p_0}{4} \right),
\eea
and
\bea
\label{eq:WgenIIB}
& & \hskip-0.5cm \sqrt{2}\, W_{\rm IIB} = \biggl[\ov{F}_0 + \, U^i \, \ov{F}_i + \frac{1}{2} \, l_{ijk} U^i U^j\, F^k - \frac{1}{6} \, l_{ijk} U^i U^j U^k\, F^0 - i\, \tilde{p}_0 \, F^0 \biggr] \\
& & - \, S \biggl[\ov{H}_0 + \, U^i \, \ov{H}_i + \frac{1}{2} \, l_{ijk} U^i U^j \, H^k - \frac{1}{6} \, l_{ijk} U^i U^j U^k \, H^0 - i\, \tilde{p}_0 \, H^0 \biggr] \nonumber\\
& & - \, G^a \biggl[\ov{\omega}_{a0} + \, U^i \, \ov{\omega}_{ai} + \frac{1}{2} \, l_{ijk} U^i U^j \,\omega_{a}{}^k - \frac{1}{6} \, l_{ijk} U^i U^j U^k\, \omega_{a}{}^0 - i\, \tilde{p}_0 \, \omega_a{}^0 \biggr] \nonumber\\
& & -  \, T_\alpha \biggl[\ov{\hat{Q}}^\alpha{}_0 + \, U^i \, \ov{\hat{Q}}^\alpha{}_i + \frac{1}{2} \, l_{ijk} U^i U^j \hat{Q}^{\alpha \, k} - \frac{1}{6} l_{ijk} U^i U^j U^k \, \hat{Q}^{\alpha 0} - i\, \tilde{p}_0 \, \hat{Q}^{\alpha 0} \biggr], \nonumber
\eea
 the rational shifts in the usual flux components are given as under,
\bea
\label{eq:IIB-W-fluxshift}
& & \ov{F}_0 = F_0 - \tilde{p}_i \, F^i\,, \qquad \qquad \ov{F}_i = F_i - \tilde{p}_{ij}\, F^j - \tilde{p}_i\, F^0\,, \\
& & \ov{H}_0 = H_0 - \tilde{p}_i \, H^i\,, \qquad \quad \, \, \, \ov{H}_i = H_i - \tilde{p}_{ij}\, H^i - \tilde{p}_i H^0\,,\nonumber\\
& & \ov\omega_{a0} = \omega_{a0} - \tilde{p}_i\, \, \omega_a{}^i \,, \qquad \quad  \ov\omega_{ai} = \omega_{ai} - \tilde{p}_{ij}\, \omega_a{}^j - \tilde{p}_i\, \omega_a{}^0\,,\nonumber\\
& & \ov{\hat{Q}}^\alpha{}_0= \hat{Q}^\alpha{}_0 - \tilde{p}_i \, \hat{Q}^{\alpha i} \,, \qquad \, \, \ov{\hat{Q}}^\alpha{}_i = \hat{Q}^\alpha{}_i - \tilde{p}_{ij}\, \hat{Q}^{\alpha j} - \tilde{p}_i \, \hat{Q}^{\alpha 0}\,.\nonumber
\eea

\subsubsection*{A simple toroidal model}
In particular, we consider the non-geometric type IIB setup based on the toroidal orientifold ${\mathbb T}^6/({\mathbb Z}_2\times{\mathbb Z}_2)$ with the standard involution; e.g. see \cite{Blumenhagen:2013hva, Gao:2015nra, Shukla:2015hpa, Shukla:2016hyy} for more details. In this setting, the untwisted sector corresponds to $h^{1,1}_+ = 3, h^{1,1}_- = 0, h^{2,1}_+ = 0\,, h^{2,1}_- = 3$. This implies that there would be three $T_\alpha$ moduli and three $U^i$ moduli along with the universal axio-dilaton $S$ in this setup. There are no odd-moduli $G^a$ present in this setup as the odd $(1,1)$-cohomology is trivial. It turns out that the geometric flux $\omega$ and the non-geometric $R$ flux do not survive the orientifold projection in this setup, and the only allowed NS-NS fluxes are the three-form $H_3$ flux and the non-geometric $Q$ flux. There are eight components for the $H_3$ flux while 24 components for the $Q$ flux, denoted as $H_\Lambda, H^\Lambda, \hat{Q}^\alpha{}_\Lambda, \hat{Q}^{\alpha \Lambda}$ for $\alpha \in \{1, 2, 3\}$ and $\Lambda \in \{0, 1, 2, 3\}$. On the RR side, there are eight flux components of the three-form $F_3$ flux. In addition, there are no $D$-terms generated in the scalar potential as the even $(2,1)$-cohomology is trivial which projects out all the $D$-term fluxes. This leads to a holomorphic flux superpotential with 40 flux parameters coupled to 7 complexifield variables $\{U^i, S, T_\alpha\}$ for $i, \alpha =\{1, 2, 3\}$, which further results in a total of 2422 terms in the effective scalar potential \cite{Blumenhagen:2013hva,Gao:2015nra,Shukla:2015hpa,Shukla:2016hyy,Leontaris:2023lfc,Leontaris:2023mmm,Biswas:2024ewk}.

The main aim of the current work is to use the flux induced scalar potential corresponding to the standard isotropic toroidal model for exploring the possibility of finding stable de-Sitter solutions, while respecting the Bianchi identities and having all the moduli/axions stabilized. For this analysis we assume that the fluxes are small fluctuations in the background and the ${\cal N} = 1$ scalar potential formula \ref{eq:Vtot} remains applicable. In addition, we work with the tree level effects induced via the flux superpotential, and neglect the possible (sub-leading) corrections induced via a series of the higher derivative $\alpha^\prime$-corrections as well as the string-loop ($g_s$) corrections. 

\subsection{A unified framework for the Type II isotropic toroidal models}
Let us now recollect the necessary pieces of information and ingredients for the type IIB isotropic toroidal model while we present the analogous details for the type IIA setup in the appendix \ref{sec_appendix1}.

We start by imposing the isotropic limit defined by the following identifications among the various respective moduli and the fluxes,
\bea
\label{eq:IIBiso}
& & {U}^1 = {U}^2 = {U}^3 = {U}, \qquad {T}_1 = {T}_2= {T}_3 = {T}, \\
& & {\tau}_1 = {\tau}_1 = {\tau}_1 \equiv {\tau}, \quad c_1 = c_2 = c_3 \equiv c, \quad {u}^1 = {u}^2 = {u}^3 = u, \quad {v}^1 = {v}^2 = {v}^3 \equiv v\,, \nonumber\\
& & \nonumber\\
& & F_1 = F_2 = F_3, \qquad F^1 = F^2 = F^3, \qquad {H}_1 = {H}_2 = {H}_3, \qquad {H}^1 = {H}^2 = {H}^3, \nonumber\\
& & {\hat Q}^1{}_0 ={\hat Q}^2{}_0 = {\hat Q}^3{}_0, \quad {\hat Q}^{1}{}_{1} = {\hat Q}^{2}{}_{2} = {\hat Q}^{3}{}_{3}, \quad {\hat Q}^{1}{}_{2} = {\hat Q}^{1}{}_{3} = {\hat Q}^{2}{}_{1} = {\hat Q}^{2}{}_{3} = {\hat Q}^{3}{}_{1} = {\hat Q}^{3}{}_{2}, \nonumber\\
& & {\hat Q}^{10} ={\hat Q}^{20} = {\hat Q}^{30}, \quad {\hat Q}^{11} = {\hat Q}^{22} = {\hat Q}^{33}, \quad {\hat Q}^{12} = {\hat Q}^{13} = {\hat Q}^{21} = {\hat Q}^{23} = {\hat Q}^{31} = {\hat Q}^{32}\,. \nonumber
\eea
Thus, taking the isotropic limit implies that one has 14 flux components instead of 40 to begin with, out of which 10 components belong to the NS-NS sector, i.e. $(H, Q)$ fluxes, while 4 of them belong to the $F$-flux in the RR sector. Subsequently, the 14 flux components which we consider for our analysis are summarized as below,
\bea
& & F_0, \quad F_1, \quad F^1, \quad F_0, \qquad {H}_0, \quad {H}_1, \quad {H}^1, \quad {H}^0,  \\
& & {\hat Q}^1{}_0, \quad {\hat Q}^{1}{}_{1}, \quad {\hat Q}^{1}{}_{2}, \qquad {\hat Q}^1{}^0, \quad {\hat Q}^{11}, \quad {\hat Q}^{12}\,. \nonumber
\eea
In addition, we have a total of six axion/moduli fields which we denote as,
\bea
& & s, \quad c_0, \qquad \tau, \quad c, \qquad u, \quad v.
\eea
Subsequently, the flux superpotential with isotropic conditions leads to the following form,
\bea
\label{eq:WgenIIB-iso}
& & \hskip-0.5cm W_{\rm IIB}^{\rm iso} = \Bigl({F}_0 + 3\, U \, {F}_1 + 3\, U^2\, F^1 - U^3\, F^0 \Bigr)- \, S \Bigl({H}_0 + 3 \, U\, {H}_1 + 3\, U^2 \, H^1 - U^3 \, H^0 \Bigr) \nonumber\\
& & \qquad \qquad + 3 \, T \Bigl({\hat{Q}}^1{}_0 + \, U \, ({\hat{Q}}^1{}_1 + 2\, {\hat{Q}}^1{}_2) + U^2\, (\hat{Q}^{11} + 2\, \hat{Q}^{12}) - U^3 \, \hat{Q}^{10} \Bigr)\,.
\eea

\subsubsection*{Bianchi identities and tadpole cancellation conditions}
It turns out that the 10 NS-NS fluxes need to satisfy the following seven constraints arising from the Bianchi identities \cite{Aldazabal:2006up},
\bea
\label{eq:BIs-IIB}
& & H^1\, \hat{Q}^1{}_0 + H_0 \, \hat{Q}^{12} - H_1\, (\hat{Q}^1{}_1 + \hat{Q}^1{}_2) = 0, \\
& & H^0\, \hat{Q}^1{}_0 - H_1\, \hat{Q}^{12} + H^1\, (\hat{Q}^1{}_1 + \hat{Q}^1{}_2) = 0, \nonumber\\
& & H_0\, \hat{Q}^{10} - H^1\, \hat{Q}^1{}_2 + H_1 \, (\hat{Q}^{11} + \hat{Q}^{12}) = 0, \nonumber\\
& & H_1\, \hat{Q}^{10} + H^0\, \hat{Q}^1{}_2 + H^1\, (\hat{Q}^{11} + \hat{Q}^{12}) = 0, \nonumber\\
& & \hat{Q}^1{}_2\, (\hat{Q}^1{}_1 + \hat{Q}^1{}_2) - \hat{Q}^1{}_0 \, (\hat{Q}^{11} + \, \hat{Q}^{12}) = 0, \nonumber\\
& & \hat{Q}^{10}\, (\hat{Q}^1{}_1 + \, \hat{Q}^1{}_2) + \hat{Q}^{12}\, (\hat{Q}^{11} + \, \hat{Q}^{12}) = 0, \nonumber\\
& & \hat{Q}^1{}_0\, \hat{Q}^{10} + \, \hat{Q}^1{}_2\, \hat{Q}^{12} = 0. \nonumber
\eea
Note that using the above collection of the seven constraints, one can recover the only Bianchi identity arising from the cohomology formulation \cite{Shukla:2016xdy},
\bea
& & {H}^0 \, \hat{Q}^1{}_0 + \,{H}^1 \, (\hat{Q}^1{}_1 + 2\, \hat{Q}^1{}_2 ) - {H}_1 \, (\hat{Q}^{11} + 2\, \hat{Q}^{12}) - {H}_0 \, \hat{Q}^{10} = 0\,.
\eea
In addition to satisfying the quadratic flux constraints arising from the Bianchi identities, one has to satisfy the following tadpole constraints \cite{Aldazabal:2006up},
\bea
\label{eq:tadpole-IIB}
& & N_3 \equiv {F}^0 \, {H}_0 + 3\, {F}^1 \, {H}_1 - 3\, {F}_1 \, {H}^1 - {F}_0 \, {H}^0 = 32 - N_{D3},\\
& & N_7 \equiv - \, {F}^0 \, \hat{Q}^1{}_0 - \,{F}^1 \, (\hat{Q}^1{}_1 + 2\, \hat{Q}^1{}_2 )+ {F}_1 \, (\hat{Q}^{11} + 2\, \hat{Q}^{12}) + {F}_0 \, \hat{Q}^{10} = -\, 32 + N_{D7}\,. \nonumber
\eea
To summarize, the relevant pieces of information about these isotropic four-dimensional models with generalized fluxes and the respective T-dual exchanges are presented in Table \ref{tab_summaryTdual}.
\noindent
\begin{table}[H]
\begin{center}
\begin{tabular}{|c||c|c|}
\hline
& & \\
& Type IIA with $D6/O6$  \quad  & \quad Type IIB with $D3/O3$ and $D7/O7$ \\
& & \\
\hline
\hline
& & \\
Complex & \, \, ${\rm N}^0$, \, \,  ${\rm U}$, \, \, ${\rm T}$, & $S$, \, \, $T$, \, \, $U$, \\
Moduli& & \\
& ${\rm N}^0 = \, \xi^0 + \, i \, ({\rmz}^0)^{-1}$ & $S = c_0 + i\, s\,$ \\
& ${\rm U} = \xi - \, i \, ({\rm z}^0)^{-1} \, \, {\rm u}^2$ & $T = c - \, i \, s  \, \, \hat\tau$ \\
& ${\rm T} = \, {\rm b} - i\, \, {\rm t}$ & $U = v  - i \, u$,\\
\hline
& & \\
Fluxes & ${\rm H}_0$,  \quad ${\rm H}^1$, & $H_0$, \quad $\hat{Q}^1{}_0$, \\
& $w_{10}$, \quad $w_1{}^1$, \quad $w_1{}^2$, & $H_1$, \quad $\hat{Q}^1{}_{1}$, \quad $\hat{Q}^1{}_{2}$, \\
& ${\rm Q}^1{}_0$, \quad ${\rm Q}^{11}$, \quad ${\rm Q}^{12}$, & $H^1$, \quad $\hat{Q}^{11}$, \quad $\hat{Q}^{12}$,\\
& ${\rm R}_0$, \quad ${\rm R}^1$, & $- H^0$, \quad $- \hat{Q}^{10}$, \\
& & \\
& $e_0$,  \quad $e_1$, \quad $m^1$, \quad $m^0$. & $F_0$,  \quad $F_1$, \quad $F^1$, \quad $- F^0$. \\
\hline
\end{tabular}
\end{center}
\caption{T-duality transformations for the toroidal isotropic models based on ${\mathbb T}^6/({\mathbb Z}_2 \times {\mathbb Z}_2)$ orbifold.}
\label{tab_summaryTdual}
\end{table}
\noindent
The isotropic toroidal models arising from both the type IIA and IIB superstring compactifications on the respective {\it standard} orientifolds of a ${\mathbb T}^6/({\mathbb Z}_2 \times {\mathbb Z}_2)$ orbifold correspond to a simple STU-type model which, apart from the (appropriately normalized and) complexified dilaton, has a single complex-structure modulus along with a K\"ahler modulus. In addition, both these (type IIA and the type IIB) isotropic toroidal models have 4 RR fluxes and 10 NS-NS fluxes restricted by a set of 7 Bianchi identities. 


\section{Analytic exploration of the flux landscape}
\label{sec_analytic}
In this section we present an analytic approach to look for the physical vacua. The main idea is to narrow down the full set of possible candidate flux configurations by checking the consistency of the required constraints without actually solving them explicitly. In this regard we need to check the consistency among the following requirements,
\bea
& & \hskip-1cm {\rm BIs}, \qquad \partial_i V = 0, \qquad \langle V \rangle \geq 0, \qquad \langle V_{ij} \rangle \equiv \, \, {\rm positive \, \, definite}.
\eea
Using the general formula (\ref{eq:Vtot}) for the ${\cal N} =1$ four-dimensional supergravity, the isotropic flux superpotential (\ref{eq:WgenIIB-iso}) results in a total of 309 terms in the scalar potential as also reported in \cite{Biswas:2024ewk}. This isotropic toroidal model has been studied following a numerical approach with the claim of finding stable dS vacua, e.g. see \cite{Guarino:2008ik,deCarlos:2009fq,deCarlos:2009qm,Dibitetto:2011qs,Danielsson:2012by,Damian:2013dq,Blaback:2013ht,Blaback:2015zra}; see \cite{Plauschinn:2020ram} also for a revisit of (some of) these claims.

Our current strategy is different from the point of view of taking the analytic procedure as far as possible, with the aim of making an exhaustive scan for the possible Minkowskian and de-Sitter vacua for the isotropic model. In this regard, given that the scalar potential is huge and results in a set of complicated extremization conditions, we take a route with many intermediate steps to narrow down the region of interest in the full flux parameter space where one could hope to find Minkowskian/de-Sitter vacua. On the way we find several interesting No-Go scenarios which one can avoid in a systematic search or classification of surviving Minkowskian/de-Sitter candidates.
In this approach our methodology is along the following steps:
\begin{itemize}

\item
{\bf Step-1:} There are 14 flux parameters which appear in the set of Bianchi identities (\ref{eq:BIs-IIB}). Without loss of generality one can assume that these flux parameters take either zero or non-zero values, and therefore we begin by considering such $2^{14} = 16384$ flux configurations as our initial input data in the form of what we call `candidate configurations'.

For this purpose, we define the class ${\bf {\cal S}_n}$ of flux configurations to be the one in which $n$ out of 14 fluxes are set to zero, while the remaining $(14-n)$ fluxes are non-zero. For the case of  ${\bf {\cal S}_0}$ we mean that all the 14 fluxes are non-zero while ${\bf {\cal S}_{14}}$ corresponds to the trivial case when all the 14 fluxes are zero and the superpotential identically vanishes.

\item
{\bf Step-2:} We test each of the flux configurations of {\bf Step-1} to see if they satisfy the Bianchi identities. This compatibility test can be performed by using ``{\rm Reduce}" module of Mathematica. Those which do not satisfy the constraints (\ref{eq:BIs-IIB}) will be collected in the set of No-Go configurations.

\item
{\bf Step-3:} The flux configurations which pass the test of satisfying the Bianchi identities (\ref{eq:BIs-IIB}) will be imposed on the extremization conditions. Those which fail to pass this test will be further added to the set of No-Go configurations.

\item
{\bf Step-4:} The flux configurations which pass the test of {\bf Step-3} will be checked for $\langle V \rangle = 0$ and $\langle V \rangle > 0$ to reach the possible candidate configurations which may result in stable Minkowskian or dS vacua respectively. Those which fail to pass this test will be added to the respective set of Minkowskian/dS No-Go configurations.

\item
{\bf Step-5:} The candidate flux configurations resulting from {\bf Step-4} maybe explicitly checked to find some more No-Go scenarios or some physical Minkowskian/dS vacua as well, after checking the Hessian test of having no tachyonic direction along with none of the six fields remaining massless. This positive definiteness test of the Hessian will be performed in an analytic (as well as a follow-up numerical) approach using the analytic results till {\bf Step-4}.

\end{itemize}

\subsection{Main challenges using the standard flux}
Given the huge size of the scalar potential it is reasonable to implement the Step-2 and Step-3 in the given order. In fact, after imposing the Bianchi identities (\ref{eq:BIs-IIB}) on the set of candidate flux configurations we find that 14272 out of 16384 are ruled out. So this helps us in ruling out a significant fraction of the candidate flux configurations to begin with. The details are as below:
\bea
\label{eq:CisNo-dS}
& & \hskip-1cm {\bf {\cal S}_{14}:} = 1({\rm Trivial!}), \quad {\bf {\cal S}_{13}:} = 14(4), \quad {\bf {\cal S}_{12}:} = 91(45), \quad {\bf {\cal S}_{11}:} = 364(242),\\
& & \hskip-1cm {\bf {\cal S}_{10}:} = 1001(798),\quad {\bf {\cal S}_9:} = 2002(1772), \quad {\bf {\cal S}_8:} = 3003(2790), \quad {\bf {\cal S}_7:} = 3432(3210),\nonumber\\
& & \hskip-1cm {\bf {\cal S}_6:} = 3003(2728), \quad {\bf {\cal S}_5:} = 2002(1694), \quad {\bf {\cal S}_4:} = 1001(741), \quad {\bf {\cal S}_3:} = 364(212), \nonumber\\
& & \hskip-1cm {\bf {\cal S}_2:} = 91(34), \quad {\bf {\cal S}_1:} = 14(2), \quad {\bf {\cal S}_0:} = 1(0), \nonumber
\eea
In Eq.(\ref{eq:CisNo-dS}), the numbers mentioned in the respective brackets denote the number of flux configurations which do not satisfy the Bianchi identities (\ref{eq:BIs-IIB}). Subsequently we find that there are 2112 cases which pass Step-2, and are to be tested for satisfying the extremization conditions and other requirements for ensuring the Minkowskian/dS vacua. We summarize the results of this methodology used with standard fluxes in Table \ref{tab_16384-classification1}.

\begin{table}[H]
\centering
\begin{tabular}{ |c||c||c|c||c|c|}
\hline
Flux & Step-1 & Step-2 & Survived/ & Step-3 & Survived/ \\
conf. & & No-Go & Undecided & No-Go & Undecided \\
 \hline
${\bf {\cal S}_{14}}$ & 1 &  0 & 1 & 0 & 1 \\
${\bf {\cal S}_{13}}$ & 14 & 2 & 12 & 8 & 6  \\
${\bf {\cal S}_{12}}$ & 91 & 34 & 57 & 69 & 22  \\
${\bf {\cal S}_{11}}$ & 364 & 212 & 152 & 316 & 48  \\
${\bf {\cal S}_{10}}$ & 1001 & 741 & 260 & 850 & 151 \\
${\bf {\cal S}_{9}}$ & 2002 & 1694 & 308 & 1746 & 256  \\
${\bf {\cal S}_{8}}$ & 3003 & 2728 & 275 & 2737 & 266  \\
\hline
${\bf {\cal S}_{7}}$ & 3432 & 3210 & 222 & 3211 & 221  \\
${\bf {\cal S}_{6}}$ & 3003 & 2790 & 213 & 2790 & 213  \\
${\bf {\cal S}_{5}}$ & 2002 & 1772 & 230 & 1772 & 230  \\
${\bf {\cal S}_{4}}$ & 1001 & 798 & 203 & 798 & 203  \\
${\bf {\cal S}_{3}}$ & 364 & 242 & 122 & 242 & 122  \\
${\bf {\cal S}_{2}}$ & 91 & 45 & 46 & 45 & 46  \\
${\bf {\cal S}_{1}}$ & 14 & 4 & 10 & 4 & 10 \\
${\bf {\cal S}_{0}}$ & 1 & 0 & 1 & 0 & 1  \\
\hline
Total & 16384 & 14272 & 2112 & 14588 & 1796 \\
 \hline
\end{tabular}
\caption{A systematic classification of the No-Go flux configurations using standard fluxes}
\label{tab_16384-classification1}
\end{table}

\noindent
Let us note that {\bf Step-3} involves dealing with complicated expressions of the derivatives of the scalar potential which has 309 terms, and therefore checking whether the Bianchi identities and the extermization conditions are simultaneously compatible for a given candidate flux configuration ${\cal S}_n$ is an extremely challenging task, and one has to set some ``time constraint" in the ``Reduce" module (of the Mathematica) in order to abort the computation if it does not result in an output in a reasonable amount of time. Therefore this challenge may also depend on the capacity of the machine one is using for the computation. Along these lines, using the standard fluxes we find that it is hard to find additional No-Go configurations for ${\cal S}_n$ with $n \leq 7$. For these cases, {\bf Step-3} does not add much to {\bf Step-2} and going beyond {\bf Step-3} is even harder.

Nevertheless, from Table \ref{tab_16384-classification1} we see that the above mentioned three steps are quite significant in our methodology as they have already excluded more than 89\% of the candidate flux configurations as those do not even solve the extremization conditions when considered to be satisfying the Bianchi identities, and therefore at the very outset one can say that such configurations will not result in physical solutions. This analysis serves the following two purposes:
\begin{itemize}

\item
To show that the methodology which we have proposed is a good one to explore the flux landscape, and it can be more efficient if the constraints are compactly expressed in some simple forms. As reflected from the Table \ref{tab_16384-classification1}, one finds that more than 89\% candidate flux configurations are filtered out.

\item
To highlight the main challenge about the huge size of the scalar potential, its derivatives and the Hessian which make it hard to perform the proposed consistency checks in Step-3-Step5 for a given set of (in-)equalities. We attempt to overcome this challenge via using the set of so-called ``axionic flux polynomials".

\end{itemize}


\subsection{Reformulating the various constraints using Axionic fluxes}

Now we present an alternate prescription to look for the physical vacua, however our methodology will be broadly the same 5-step strategy as we have previously adopted. As we have witnessed in the first approach, the main challenge one faces is due to the huge size of the scalar potential leading to a very complicated form of the derivatives which makes it harder even for the purpose of  analytically solving the extremization conditions. Our second approach to tackle this problem is based on rewriting the scalar potential and its derivatives as well as the Hessian in a concise and simpler form by using the ``axionic fluxes", and subsequently use their compact forms to narrow down the set of possible candidate flux configurations which could solve the necessary conditions. These tests can be performed just by checking the compatibility of the set of constraints without looking into the exact solutions of either extremization conditions or the Bianchi identities.

Using the general formula (\ref{eq:Vtot}) for the ${\cal N} =1$ four-dimensional supergravity, the flux superpotential (\ref{eq:WgenIIB-iso}) results in a total of 309 terms ! In this context, the so-called axions flux polynomials which are combinations of fluxes and axions have been found to be extremely useful \cite{Blumenhagen:2013hva, Shukla:2015bca, Shukla:2015rua, Shukla:2015hpa}, and this number 309 reduces to 112 by absorbing the RR axions $\{c_0, c\}$ in a certain useful axionic flux polynomial \cite{Biswas:2024ewk}. Moreover, one can further generalize the RR-axionic polynomials by including the CS axions as well, and such generalized axionic fluxes which include all the axions are given as below \cite{Shukla:2019wfo, Biswas:2024ewk},
\bea
\label{eq:AxionicFluxOrbitsIIB-Z2xZ2-iso1}
& & h_0 = {H}_0 + 3\, v\, {H}_1 + 3\, (v)^2 \, {H}^1\, - \, (v)^3  \, {H}^0, \\
& & h_1 = {H}_1 + 2\, v \, {H}^1 -\, (v)^2 \, {H}^0 \,, \quad h^1 = {H}^1 - v \,{H}^0\,, \quad h^0 = -\, {H}^0 \,, \nonumber\\
& & \nonumber\\
& & q^1{}_0 = \hat{Q}^1{}_0 + \, v\, (\hat{Q}^1{}_1 +  2\, \hat{Q}^1{}_2) +  (v)^2\, (\hat{Q}^{11} + 2\, \hat{Q}^{12})\, - (v)^3 \, \hat{Q}^{10}, \nonumber\\
& & q^1{}_1 = \hat{Q}^1{}_1 + 2\, v \, \hat{Q}^{12} - (v)^2 \, \hat{Q}^{10} \,, \quad  q^1{}_2 = \hat{Q}^1{}_2 + v \, (\hat{Q}^{11} + \, \hat{Q}^{12}) - (v)^2 \, \hat{Q}^{10} \,, \nonumber\\
& & q^{11} = \hat{Q}^{11} - v \,\hat{Q}^{10}\,, \quad  q^{12} = \hat{Q}^{12} - v \,\hat{Q}^{10}\,, \quad q^{10} = -\, \hat{Q}^{10} \,, \nonumber\\
& & \nonumber\\
& & f_0 = {\mathbb F}_0 + 3\, v\, {\mathbb F}_1 + 3\, (v)^2 \, {\mathbb F}^1\, - (v)^3  \, {\mathbb F}^0, \nonumber\\
& & f_1 = {\mathbb F}_1 + 2\, v \, {\mathbb F}^1 - (v)^2 \, {\mathbb F}^0 \,, \quad f^1 = {\mathbb F}^1 - v \,{ \mathbb F}^0\,, \qquad f ^0 = - \,{\mathbb F}^0 \,, \nonumber
\eea
where
\bea
& & {\mathbb F}_0 = {F}_0 - 3\, c\, {\hat{Q}}^1{}_0 \, - \, c_0 \, {H}_0, \qquad {\mathbb F}_1 = {F}_1 - \, c\, ({\hat{Q}}^1{}_1 + 2\, {\hat{Q}}^1{}_2) \, - \, c_0 \, {H}_1,\\
& & {\mathbb F}^0 = F^0 - 3\, c\, \hat{Q}^{10}\, - \, c_0 \, {H}^0, \qquad {\mathbb F}^1 = F^1 - \, c\, (\hat{Q}^{11} + \, 2\, \hat{Q}^{12})\, - \, c_0 \, {H}^1\,. \nonumber
\eea
Now the 309 terms arising from the flux superpotential (\ref{eq:WgenIIB-iso}) can be equivalently expressed by 40 terms if one uses a generalized version of the axionic fluxes as given in (\ref{eq:AxionicFluxOrbitsIIB-Z2xZ2-iso1}) \cite{Biswas:2024ewk}. Moreover, it is important to mention that the set of Bianchi identities in Eq.~(\ref{eq:BIs-IIB}) continue to hold after promoting the standard flux components in to the axionic flux orbits given in Eq.~(\ref{eq:AxionicFluxOrbitsIIB-Z2xZ2-iso1}). This is equivalent to the following set of constraints,
\bea
\label{eq:BIs-IIB-1}
& & \hskip-1.5cm {q}^{1}{}_0\, {h}^{1} + {h}_{0}\, {q}^{12} - {h}_{1}\, \, ({q}^1{}_1+ {q}^1{}_2) = 0, \\
& & \hskip-1.5cm {q}^{1}{}_0 \, {h}^0 - {h}_{1}\, \, {q}^{12} + {h}^1\, ({q}^1{}_1+ {q}^1{}_2) = 0, \nonumber\\
& & \hskip-1.5cm {h}_{0} \, \, {q}^{10} - {q}^1{}_2 \, \, {h}^1 + {h}_{1}\, ({q}^{11} + {q}^{12}) = 0, \nonumber\\
& & \hskip-1.5cm {q}^{10}\, {h}_{1} + {h}^0\, \, {q}^1{}_2 + {h}^{1}\, ({q}^{11} + {q}^{12}) = 0, \nonumber\\
& & \nonumber\\
& & \hskip-1.5cm {q}^{10}\, ({q}^1{}_1+ {q}^1{}_2) + {q}^{12}\, ({q}^{11} + {q}^{12}) = 0, \nonumber\\
& & \hskip-1.5cm {q}^1{}_2\, \, ({q}^1{}_1+ {q}^1{}_2) - {q}^{1}{}_0\, ({q}^{11} + {q}^{12}) = 0, \nonumber\\
& & \hskip-1.5cm {q}^1{}_0\, {q}^{10} + {q}^1{}_2\, \, {q}^{12} = 0, \nonumber
\eea
along with the only cohomology formulation identity being given as,
\bea
& & \hskip-1.5cm {h}_{0} \, \, {q}^{10} + {h}_{1}\, ({q}^{11} + 2\, {q}^{12})  \, \, {h}^0 - {h}^1\, \, ({q}^1{}_1+ 2\, {q}^1{}_2) - {q}^{1}{}_0  = 0.
\eea
As mentioned earlier, the scalar potential arising from the flux superpotential (\ref{eq:WgenIIB-iso}) as well as its derivatives can be expressed in a quite simple from by using the following 14 real quantities $(\gamma_i, \lambda_i)$ defined in terms of axionic fluxes (\ref{eq:AxionicFluxOrbitsIIB-Z2xZ2-iso1}),
\bea
\label{eq:lambda-gamma-def-1}
& &  \gamma_1 = f_0, \quad  \gamma_2 = u\, f_1, \quad  \gamma_3 = u^2\, f^1, \quad \gamma_4 =  u^3\, f^0,\\
& &  \lambda_1 = s\, h_0, \quad \lambda_2 = s\, u\, h_1, \quad \lambda_3 =  s\, u^2\, h^1, \quad \lambda_4 =s\, u^3\, h^0, \nonumber\\
& &  \lambda_5 = \tau\, q^1_0, \quad  \lambda_6 = \tau\, u\, q^1{}_1, \quad \lambda_7 = \tau\, u^2\, q^{11}, \quad \lambda_8 = \tau\, u^3\, q^{10}, \nonumber\\
& &  \lambda_9 = \tau\, u\, q^1{}_2, \quad  \lambda_{10} = \tau\, u^2\, q^{12}. \nonumber
\eea
Subsequently, using (\ref{eq:AxionicFluxOrbitsIIB-Z2xZ2-iso1}) and (\ref{eq:lambda-gamma-def-1}), the scalar potential takes the following simple form,
\bea
\label{eq:V-lambda-gamma-1}
& & \hskip-1cm V = \frac{1}{4 \, s \, u^3\, \tau^3} \biggl[\gamma_1^2+3 \gamma_2^2+3 \gamma_3^2+\gamma_4^2+\lambda_1^2+3 \lambda_2^2+3 \lambda_3^2+\lambda_4^2+3 \lambda_5^2 -3 \lambda_6^2-3 \lambda_7^2+3 \lambda_8^2\\
& & \hskip1cm +2 \gamma_4 \lambda_1-6 \gamma_3 \lambda_2+6 \gamma_4 \lambda_5+18 \lambda_3 \lambda_5-6 \gamma_3 \lambda_6-12 \lambda_2 \lambda_6+6 \lambda_4 \lambda_6 +6 \lambda_1 \lambda_7 \nonumber\\
& & \hskip1cm -12 \lambda_3 \lambda_7 +12 \lambda_5 \lambda_7 -2 \left(\lambda_4+3 \lambda_8\right) \gamma_1+6 \gamma_2 \left(\lambda_3+\lambda_7\right)+18 \lambda_2 \lambda_8+12 \lambda_6 \lambda_8 \nonumber\\
& & \hskip1cm -12 \gamma_3 \lambda_9+12 \gamma_2 \lambda_{10}-12 \lambda_9^2-12 \lambda_{10}^2-24 \lambda_2 \lambda_9+12 \lambda_4 \lambda_9-12 \lambda_6 \lambda_9+24 \lambda_8 \lambda_9\nonumber\\
& & \hskip1cm +12 \lambda_1 \lambda_{10}-24 \lambda_3 \lambda_{10}+24 \lambda_5 \lambda_{10}-12 \lambda_7 \lambda_{10}\biggr], \nonumber
\eea
which has a total of 40 terms. In fact, using (\ref{eq:lambda-gamma-def-1}) will simply translate the task of finding No-Go configurations in terms of checking the compatibility of a set of constraints with real choices of $(\gamma_i, \lambda_i)$ parameters. For example, the set of Bianchi identities (\ref{eq:BIs-IIB-1}) are equivalently expressed as
\bea
\label{eq:BIs-IIB-2}
& & \hskip-0.5cm \lambda_4 \lambda_5+\lambda_2 \lambda_{10}=\lambda_3 \lambda_6+\lambda_3 \lambda_9, \quad \lambda_3 \lambda_5+\lambda_1 \lambda_{10}=\lambda_2 \lambda_6+\lambda_2 \lambda_9, \quad \lambda_5 \lambda_8=\lambda_9 \lambda_{10}, \\
& & \hskip-0.5cm \lambda_1 \lambda_8+\lambda_3 \lambda_9=\lambda_2 \lambda_7+\lambda_2 \lambda_{10}, \quad \lambda_9^2+\lambda_6 \lambda_9=\lambda_5 \lambda_7+\lambda_5 \lambda_{10}, \quad \lambda_6 \lambda_8+\lambda_9 \lambda_8=\lambda_{10}^2+\lambda_7 \lambda_{10},\nonumber\\
& & \hskip-0.5cm \lambda_2 \lambda_8+\lambda_4 \lambda_9=\lambda_3 \lambda_7+\lambda_3 \lambda_{10}, \quad \lambda_4 \lambda_5+\lambda_2 \lambda_7+2 \lambda_2 \lambda_{10}=\lambda_3 \lambda_6+\lambda_1 \lambda_8+2 \lambda_3 \lambda_9,\nonumber
\eea
where the last identity is the only one which can arise from the known version of the cohomology formulation \cite{Ihl:2006pp,Ihl:2007ah,Robbins:2007yv,Shukla:2016xdy,Gao:2018ayp}, and this can be derived from the set of 7 identities. Other identities are known as `missing identities'. In fact one can see that 40 terms of the scalar potential as given in (\ref{eq:V-lambda-gamma-1}) reduces  to 31 terms after eliminating some terms through the Bianchi identities (\ref{eq:BIs-IIB-2}). The effective scalar potential takes the following form:
\bea
\label{eq:V-lambda-gamma-2}
& & \hskip-1cm V = \frac{1}{4 \, s \, u^3\, \tau^3} \biggl[\gamma_1^2+3 \gamma_2^2+3 \gamma_3^2+\gamma_4^2+\lambda_1^2+3 \lambda_2^2+3 \lambda_3^2+\lambda_4^2+3 \lambda_5^2 -3 \lambda_6^2-3 \lambda_7^2+3 \lambda_8^2  \\
& & +2 \gamma_4 \lambda_1-2 \lambda_4 \gamma_1-6 \lambda_8 \gamma_1-6 \gamma_3 \lambda_2+6 \gamma_2 \lambda_3+6 \gamma_4 \lambda_5-6 \gamma_3 \lambda_6+6 \lambda_2 \lambda_6+6 \lambda_4 \lambda_6+6 \gamma_2 \lambda_7 \nonumber\\
& & +6 \lambda_1 \lambda_7+6 \lambda_2 \lambda_8-6 \lambda_1 \lambda_{10}-6 \lambda_2 \lambda_9-12 \gamma_3 \lambda_9+12 \lambda_8 \lambda_9+12 \gamma_2 \lambda_{10}-12 \lambda_3 \lambda_{10}+12 \lambda_5 \lambda_{10}\biggr], \nonumber
\eea
where we note that so far we have not ``solved" the Bianchi identities and we have simply eliminated those quadratic pieces from (\ref{eq:V-lambda-gamma-1}) which can be cancelled via considering the combinations of the various constraints in the set of Bianchi identities. 
Following this strategy, the explicit form of the scalar potential derivatives w.r.t.~the three axionic and three saxionic moduli are given by the following simple and compact relations,
\bea
\label{eq:derV-lambda-gamma-1}
& & \hskip-0cm\frac{\partial V}{\partial c_0} = - \frac{1}{2\, s^2 \,u^3 \,\tau^3} \biggl[\gamma_1 \lambda_1+3 \gamma_2 \lambda_2+3 \gamma_3 \lambda_3+\gamma_4 \lambda_4\biggr],\\
& & \hskip-0cm \frac{\partial V}{\partial c} = \frac{3}{2 \,s \,u^3 \,\tau^4} \bigg[\gamma_1 \lambda_5+\gamma_2 \lambda_6+\gamma_3 \lambda_7+\gamma_4 \lambda_8+2 \gamma_2 \lambda_9+2 \gamma_3 \lambda_{10}\biggr], \nonumber\\
& & \hskip-0cm \frac{\partial V}{\partial v} = \frac{3}{2 \,s \,u^4 \,\tau^3} \biggl[\gamma_1 \gamma_2+2 \gamma_3 \gamma_2+\gamma_3 \gamma_4+\lambda_1 \lambda_2+2 \lambda_2 \lambda_3+\lambda_3 \lambda_4+2 \lambda_3 \lambda_6+\lambda_5 \lambda_6+2 \lambda_1 \lambda_8\nonumber\\
& & \hskip1cm +\lambda_7 \lambda_8+2 \lambda_5 \lambda_9-4 \lambda_2 \lambda_{10}+2 \lambda_8 \lambda_{10}+6 \lambda_9 \lambda_{10} \biggr], \nonumber\\
& & \hskip-0cm \frac{\partial V}{\partial s} = -\frac{1}{4 \,s^2 \,u^3 \,\tau^3} \biggl[\gamma_1^2+3 \gamma_2^2+3 \gamma_3^2+\gamma_4^2-\lambda_1^2-3 \lambda_2^2-3 \lambda_3^2-\lambda_4^2+3 \lambda_5^2-3 \lambda_6^2-3 \lambda_7^2 +3 \lambda_8^2\nonumber\\
& & \hskip1cm +6 \gamma_4 \lambda_5-6 \lambda_8 \gamma_1-6 \gamma_3 \lambda_6+6 \gamma_2 \lambda_7-12 \gamma_3 \lambda_9+12 \lambda_8 \lambda_9+12 \gamma_2 \lambda_{10}+12 \lambda_5 \lambda_{10}\biggr], \nonumber\\
& & \hskip-0cm \frac{\partial V}{\partial \tau} = - \frac{3}{2 \,s \,u^3 \,\tau^4} \biggl[\gamma_1^2+3 \gamma_2^2+3 \gamma_3^2+\gamma_4^2+\lambda_1^2+3 \lambda_2^2+3 \lambda_3^2+\lambda_4^2+\lambda_5^2-\lambda_6^2-\lambda_7^2+\lambda_8^2 \nonumber\\
& & \hskip1cm -2 \lambda_4 \gamma_1-4 \lambda_8 \gamma_1+2 \gamma_4 \lambda_1-6 \gamma_3 \lambda_2+6 \gamma_2 \lambda_3+4 \gamma_4 \lambda_5-4 \gamma_3 \lambda_6+4 \lambda_2 \lambda_6+4 \lambda_4 \lambda_6+4 \gamma_2 \lambda_7\nonumber\\
& & \hskip1cm +4 \lambda_1 \lambda_7+4 \lambda_2 \lambda_8-8 \gamma_3 \lambda_9-4 \lambda_2 \lambda_9+4 \lambda_8 \lambda_9+8 \gamma_2 \lambda_{10}-4 \lambda_1 \lambda_{10}-8 \lambda_3 \lambda_{10}+4 \lambda_5 \lambda_{10}\biggr], \nonumber\\
& & \hskip-0cm \frac{\partial V}{\partial u} = - \frac{3}{2 \,s \,u^4 \,\tau^3} \biggl[-\gamma_1^2-\gamma_2^2+\gamma_3^2+\gamma_4^2-\lambda_1^2-\lambda_2^2+\lambda_3^2+\lambda_4^2-3 \lambda_5^2+\lambda_6^2-\lambda_7^2+3 \lambda_8^2-2 \lambda_2 \lambda_6 \nonumber\\
& & \hskip1cm +2 \lambda_4 \lambda_6-2 \lambda_1 \lambda_7+2 \lambda_2 \lambda_8+2 \lambda_2 \lambda_9+4\lambda_8 \lambda_9+2 \lambda_1 \lambda_{10}-4 \lambda_3 \lambda_{10}-4 \lambda_5 \lambda_{10}\biggr]. \nonumber
\eea
Let us emphasize again that we have used the Bianchi identities (\ref{eq:BIs-IIB-2}) to eliminate some terms from the expressions of the scalar potential derivatives to arrive at the collection presented in Eq.~(\ref{eq:derV-lambda-gamma-1}). Following this strategy the generic Hessian components can also receive some simplifications, however these are still too complicated to be presented here, and therefore we collect them in Eq.~(\ref{eq:Hess-eqn}) of the appendix \ref{sec_appendix2}.

\subsection{Capturing the No-Go flux configurations}

Let us appreciate the fact that the scalar potential arising from the  superpotenial (\ref{eq:WgenIIB-iso}) consists of 309 terms when expressed in terms of standard $F, H$ and $Q$ fluxes, however using the axionic fluxes  (\ref{eq:AxionicFluxOrbitsIIB-Z2xZ2-iso1}) it can be written by using 40 terms only as seen from (\ref{eq:V-lambda-gamma-1})! In fact it can be further reduced to 31 terms by using a partial simplifications due to Bianchi identities as given in (\ref{eq:V-lambda-gamma-2}). The use of redefinition (\ref{eq:lambda-gamma-def-1}) is  the fact that the scalar potential as well as its derivatives can be formulated in a very compact way which can facilitate the possibility of performing an analytic search of physical vacua. Otherwise starting with a scalar potential having 309 terms, it was not a pragmatic task to proceed along as we have already witnessed in our earlier effort in the previous section, and in fact this has been the reason why earlier studies have followed some fully numerical `black box' kind of approach to find the physical vacua.

To begin with, there are 16384 candidate flux configurations corresponding to 14 flux parameters being zero or non-zero. These are equivalently encoded in $(\lambda_i, \gamma_i)$ parameters where $\gamma_i$'s are generalized version of the four RR fluxes while $\lambda_i$'s denote the generalized version of the ten NS-NS fluxes. In {\bf Step-2} we find that imposing Bianchi identities (\ref{eq:BIs-IIB-2}) one can rule out 14272 configurations which do not satisfy the quadratic flux constraints, and subsequently one is left with 2112 possible configurations. Following {\bf Step-3} which includes imposing the constraints from Bianchi identities (as in {\bf Step-2}) and the six extremization conditions using (\ref{eq:derV-lambda-gamma-1}) we find that there are only 393 undecided configurations while a total of 15991 are ruled out. The undecided cases are to be tested for Minkowskian/dS vacua as a next step. This way we can see how crucial role our axionic fluxes have played in this analytic approach by simply ruling more than 97\% of the candidate flux configurations which will not result in physical vacua ! The main pieces of information are summarized in the Table \ref{tab_16384-classification-2} which consists of the following results:

\begin{itemize}
\item
Out of 16384 candidate flux configurations, there are at least 15991 of those which do not simultaneously satisfy the Bianchi identities and the extremization conditions for real values of $(\gamma_i, \lambda_i)$, leaving at most 393 configurations which may (or may not) have the physical vacua. This makes a reduction of 97.6\% of the total candidate configurations we began with.

The ``Reduce" module in Mathematica is quite slow when the reality condition for $\gamma_i$ and $\lambda_i$ parameters is imposed. In this regard, first we have taken an intermediate step to get an equivalent set of equalities from the ``Reduce" module corresponding to those outputs which give a solution\footnote{Note that with a given time constraint, one does not always have an output solution using ``Reduce" as sometimes if the running time is more than what is assigned, the Mathematica program is automatically aborted for that particular flux configuration, and subsequently such a configuration remain ``undecided".}, and then we discard the solutions which correspond to the imaginary values of $\gamma_i$ and $\lambda_i$ parameters.

\item
These 393 `survived/undecided' configurations when tested for the compatibility with the Minkowsian condition $\langle V \rangle = 0$ or the dS condition $\langle V \rangle > 0$ result in additional respective No-Go scenarios. This process, subsequently reduces the number of possible candidate configurations to 204 for Minkowskian case, and 288 for the dS case.

\item
In Table \ref{tab_16384-classification-2}, the Step-4 results for the Minkowskian and dS cases represent the number of flux configurations (in respective cases) which are NOT ruled out in our analytic exploration. However, there can certainly be some additional No-Go configurations among these so-called `survived/undecided' cases.

\end{itemize}

\begin{table}[H]
\centering
\begin{tabular}{ |c||c||c|c||c|c||c|c|}
\hline
Flux & Step-1 & Step-2 & Survived/ & Step-3 & Survived/ & Step-4 & Step-4\\
conf. & & No-Go & Undecided & No-Go & Undecided & (Mink) & (dS) \\
 \hline
 \hline
${\bf {\cal S}_{14}}$ & 1 &  0 & 1 & 0 & 1 & 1 & 0\\
${\bf {\cal S}_{13}}$ & 14 & 2 & 12 & 14 & 0 & 0 & 0\\
${\bf {\cal S}_{12}}$ & 91 & 34 & 57 & 87 & 4 & 4 & 0\\
${\bf {\cal S}_{11}}$ & 364 & 212 & 152 & 364 & 0 & 0 & 0\\
${\bf {\cal S}_{10}}$ & 1001 & 741 & 260 & 989 & 12 & 6 & 0\\
${\bf {\cal S}_{9}}$ & 2002 & 1694 & 308 & 1991 & 11 & 2 & 0\\
\hline
${\bf {\cal S}_{8}}$ & 3003 & 2728 & 275 & 2976 & 27 & 8 & 5\\
${\bf {\cal S}_{7}}$ & 3432 & 3210 & 222 & 3396 & 36 & 28 & 10\\
${\bf {\cal S}_{6}}$ & 3003 & 2790 & 213 & 2967 & 36 & 28 & 16\\
${\bf {\cal S}_{5}}$ & 2002 & 1772 & 230 & 1968 & 34 & 14 & 27\\
${\bf {\cal S}_{4}}$ & 1001 & 798 & 203 & 924 & 77 & 7 & 75\\
${\bf {\cal S}_{3}}$ & 364 & 242 & 122 & 266 & 98 & 49 & 98\\
${\bf {\cal S}_{2}}$ & 91 & 45 & 46 & 45 & 46 & 46 & 46\\
${\bf {\cal S}_{1}}$ & 14 & 4 & 10 & 4 & 10 & 10 & 10\\
${\bf {\cal S}_{0}}$ & 1 & 0 & 1 & 0 & 1 & 1 & 1\\
\hline
\hline
Total & 16384 & 14272 & 2112 & 15991 & 393 & 204 & 288\\
 \hline
\end{tabular}
\caption{Analytic classification of the number of flux configurations using the ``Reduce" module.}
\label{tab_16384-classification-2}
\end{table}

\noindent
For a given real symmetric $n\times n$ matrix $A= a_{ij}$, the principle sub-matrices $A^{(n)}$ are defined as,
\bea
\label{eq:principle-submatrix}
& & \hskip-1.5cm A^{(1)} = [a_{11}], \quad A^{(2)} = \begin{bmatrix} a_{11} & a_{12}\\ a_{12} & a_{22} \end{bmatrix}, \quad A^{(3)} = \begin{bmatrix} a_{11} & a_{12} & a_{13} \\ a_{12} & a_{22} & a_{23}\\ a_{13} & a_{23} & a_{33} \end{bmatrix}, \quad \dots , \quad A^{(n)} = A.
\eea
Subsequently, the positive definiteness of the matrix $A$ is guaranteed by considering the positive definiteness of the determinants of all the principle sub-matrices, i.e.
\bea
\label{eq:Hessian-check}
& & {\rm det}A^{(1)} > 0, \quad {\rm det}A^{(2)} > 0, \quad \dots , \quad {\rm det}A^{(n)} > 0.
\eea
For our current case, the Hessian is a $6\times6$ matrix with entries having a very complicated dependence on the $(\gamma_i, \lambda_i)$ parameters\footnote{In fact, each of the principle sub-matrices $A^{(n)}$ have some dependences on the saxion moduli $\{s, \tau, u\}$. However, all such dependences boil down to an overall factor in their respective determinants, and therefore $\{s, \tau, u\}$ do not effectively appear in the Hessian check condition (\ref{eq:Hessian-check}).}, and therefore following the analytic approach, it is hard to analytically check the positive definiteness of the most generic Hessian for the purpose of excluding the tachyonic and massless cases. However we can still rule out some configurations for which the respective expressions of the various determinants in (\ref{eq:principle-submatrix}) remain a bit simpler to perform a consistency check of conditions in (\ref{eq:Hessian-check}). This leads to the results in Table \ref{tab_16384-classification-3}.

\begin{table}[H]
\centering
\begin{tabular}{ |c||c|c|c||c|c|c|c|}
\hline
Flux & Step-3 & Step-4 & Step-4 & Step-5 & Step-5 & Step-5 \\
conf. & Extrema & Mink & dS & Extrema & Mink & dS \\
 \hline
 \hline
${\bf {\cal S}_{14}}$  & 1 & 1 & 0 & 0 & 0 & 0\\
${\bf {\cal S}_{13}}$  & 0 & 0 & 0 & 0 & 0 & 0\\
${\bf {\cal S}_{12}}$  & 4 & 4 & 0 & 0 & 0 & 0\\
${\bf {\cal S}_{11}}$  & 0 & 0 & 0 & 0 & 0 & 0\\
${\bf {\cal S}_{10}}$  & 12 & 6 & 0 & 0 & 0 & 0\\
${\bf {\cal S}_{9}}$  & 11 & 2 & 0 & 0 & 0 & 0\\
\hline
${\bf {\cal S}_{8}}$  & 27 & 8 & 5 & 16 & 0 & 5\\
${\bf {\cal S}_{7}}$  & 36 & 28 & 10 & 15 & 7 & 10 \\
${\bf {\cal S}_{6}}$  & 36 & 28 & 16 & 13 & 8 & 13 \\
${\bf {\cal S}_{5}}$  & 34 & 14 & 27 & 28 & 8 & 27 \\
\hline
${\bf {\cal S}_{4}}$  & 77 & 7 & 75 & 76 & 6 & 75 \\
${\bf {\cal S}_{3}}$  & 98 & 49 & 98 & 98 & 49 & 98 \\
${\bf {\cal S}_{2}}$  & 46 & 46 & 46 & 46 & 46 & 46 \\
${\bf {\cal S}_{1}}$  & 10 & 10 & 10 & 10 & 10 & 10 \\
${\bf {\cal S}_{0}}$  & 1 & 1 & 1 & 1 & 1 & 1 \\
\hline
\hline
Total & 393 & 204 & 288 & 303 & 135 & 285 \\
 \hline
\end{tabular}
\caption{Number of survived/undecided configurations after imposing the Hessian condition (\ref{eq:Hessian-check}).}
\label{tab_16384-classification-3}
\end{table}

\noindent
Let us recall that ${\cal S}_{14}$ corresponds to the trivial fluxless case with all the 14 fluxes set to zero. This simply means that the superpotential identically vanishes, and so no moduli stabilization is possible with this configuration. It survived the Step-1, Step-2, Step-3 and Step-4, as seen from the top left part of the Table \ref{tab_16384-classification-3}, simply because it has trivially satisfied the Bianchi identities and the various extremization conditions as well as Minkowskian condition $\langle V \rangle = 0$. However Hessian checks  of the condition (\ref{eq:Hessian-check}) in Step-5 certainly rules out this case which has kept all the moduli/axions still massless. From Table \ref{tab_16384-classification-3} we can see that the number of candidate configurations has been reduced a lot by following the well designed steps we have framed in our methodology. This subsequently has resulted in a reduced flux landscape as compared to what was generically available to begin with.

Finally let us emphasize again that the number of ``survived/undecided" cases mentioned in Table \ref{tab_16384-classification-3} correspond to those configurations for which we could not manage to find the No-Go in our analytic analysis. This does not mean that these cases will certainly have stable Minkowskian or dS vacua. Our strategy so far has been focused on ruling out as many candidate configurations as possible via following our earlier described methodology having some concrete steps to check. Next we will further explore a subset of the remaining candidate configurations in a numerical approach.


\section{Numerical exploration of the reduced flux landscape}
\label{sec_numeric}

In this section we aim to numerically explore the ``reduced flux landscape" which we have obtained by narrowing down the full parameter space by taking an analytical approach. From Table \ref{tab_16384-classification-3} we observe that one can not have stable Minkowskian or de-Sitter solutions corresponding to the first 6 classes of configurations from top, namely ${\cal S}_{14}$, ${\cal S}_{13}$, ${\cal S}_{12}$, ${\cal S}_{11}$, ${\cal S}_{10}$ and ${\cal S}_{9}$. Similar observations were made for de-Sitter vacua in Table \ref{tab_16384-classification-2} where Hessian checks were not included. To begin with these are quite non-trivial results in the sense that these six classes ${\cal S}_{14}$-${\cal S}_{9}$ constitute a total of 3473 candidate configurations, and all of these result in a No-Go scenario for searching Minkowskian/dS vacua.

Now using the classification summarized in Table \ref{tab_16384-classification-2} and Table \ref{tab_16384-classification-3}, we will use the ``NSolve" module of Mathematica for investigating the 135 candidate cases of the Minkowskian solutions, and 285 candidate cases of the dS solutions. While looking for the numerical solutions we will also demand that $(\gamma_i, \lambda_i)$ parameters are real, and subsequently as part of the Step-5 we will check the positive definiteness of the Hessian for a given candidate flux configuration. 


\subsection{Analyzing the Minkowskian vacua}

Similar to the Bianchi identities (\ref{eq:BIs-IIB-2}) and the extremization conditions (\ref{eq:derV-lambda-gamma-1}), using the scalar potential expression (\ref{eq:V-lambda-gamma-2}), the Minkowskian condition $\langle V \rangle  = 0$ represents an equality constraint which is purely determined in terms of $(\gamma_i, \lambda_i)$ parameters, and therefore Minkowskian solutions can be analyzed using the ``NSolve" module for looking at the $(\gamma_i, \lambda_i)$ values which numerically solve all the constraints in Step-4. Subsequently as part of Step-5, we will use the $(\gamma_i, \lambda_i)$ numerical solutions arising from Step-4 to check if the Hessian positive definiteness condition (\ref{eq:Hessian-check}) is satisfied or not. Let us also note that Step-4 may lead to multiple solutions for a single candidate configuration, and one needs to check the Hessian positive definiteness for each of those numerical solutions. Once such $(\gamma_i, \lambda_i)$ values are known, one can trace back the VEVs of various saxionic/axionic moduli via using the relations (\ref{eq:lambda-gamma-def-1}).

From Table \ref{tab_16384-classification-2} and Table \ref{tab_16384-classification-3} we observe that there are only 29 Minkowskian candidates left out of 15914 configurations from the upper 11 classes, namely ${\cal S}_{14}$-${\cal S}_4$. In fact the top 7 classes ${\cal S}_{14}$-${\cal S}_8$ completely result in No-Go scenarios for stable Minkowskian vacua. This is quite a  significant reduction of the vast flux parameter space. Now we elaborate on what we find for the remaining classes ${\cal S}_7$-${\cal S}_0$.

\subsubsection*{${\cal S}_7$ models}
This class has 7 configurations and we find that 5 of those result in ``{\rm No Solution}" cases while there are 2 configurations which solve the BIs+Extremization+Minkowskian conditions. However we find that these two numerical solutions fail to satisfy the Hessian check (\ref{eq:Hessian-check}).

\subsubsection*{${\cal S}_6$ models}

This class has 8 configurations and we find that 7 of those result in ``{\rm No Solution}" cases while there is one configuration which solves the BIs+Extremization+Minkowskian conditions. However we find that this numerical solution fails to satisfy the Hessian positive definiteness condition (\ref{eq:Hessian-check}).

\subsubsection*{${\cal S}_5$ models}

This class has 8 configurations and we find that all of these result in ``{\rm No Solution}" cases while demanding the solutions of the BIs+Extremization+Minkowskian conditions.

\subsubsection*{${\cal S}_4$ models}

This class has 6 configurations and we find that all of these result in ``{\rm No Solution}" cases while demanding the solutions of the BIs+Extremization+Minkowskian conditions.

\subsubsection*{${\cal S}_3$ models}

This class has 49 configurations and we find that 43 of those result in ``{\rm No Solution}" cases while we could not get an output for 6 configurations which remain undecided.

\subsubsection*{${\cal S}_2$-${\cal S}_0$ models}
There are 57 configurations which could not be checked using our approach. 
\vskip0.15cm
To summarize, all the results corresponding to the search of stable Minkowskain vacua are presented in Table \ref{tab_16384-classification-4} which shows that there are only 63 cases out of 16384 which remain undecided while all the decided ones result in the No-Go scenarios.


\subsection{Analyzing the de-Sitter vacua}

For de-Sitter scan, we have only 55 de-Sitter candidates left out of 14913 configurations from the upper ten classes listed in Table \ref{tab_16384-classification-3}, namely ${\cal S}_{14}$-${\cal S}_5$. Note that this means that there are only 55 candidates undecided out of the 91\% of the collection of candidate configurations, which is a very significant reduction. In fact the top 6 classes ${\cal S}_{14}$-${\cal S}_9$ completely result in No-Go scenarios for stable de-Sitter vacua. This is quite a  significant reduction of the vast flux parameter space. Now we elaborate on what we find for the remaining classes ${\cal S}_8$-${\cal S}_0$.

\subsubsection*{${\cal S}_8$ models}
This class has 5 configurations and we find that 2 of those result in ``{\rm No Solution}" cases while there are 3 configurations which solve the BIs+Extremization conditions. However we find that these 3 numerical solutions fail to simultaneously satisfy  $\langle V \rangle > 0$ and the Hessian check.

\subsubsection*{${\cal S}_7$ models}

This class has 10 configurations and all of those solve the BIs+Extremization conditions. However all these solutions fail to simultaneously satisfy the $\langle V \rangle > 0$ and the Hessian check.

\subsubsection*{${\cal S}_6$ models}

This class has 13 configurations and we find that 6 of those result in ``{\rm No Solution}" cases while there are 7 configurations which solve the BIs+Extremization conditions. However we find that these 7 numerical solutions fail to simultaneously satisfy the $\langle V \rangle > 0$ and the Hessian check.

\subsubsection*{${\cal S}_5$ models}

This class has 27 configurations and we find that 25 of those result in ``{\rm No Solution}" cases while there are 2 configurations which solve the BIs+Extremization conditions. However we find that these 2 numerical solutions fail to simultaneously satisfy the $\langle V \rangle > 0$ and the Hessian positive definiteness conditions.

\subsubsection*{${\cal S}_4$ models}

This class has 75 configurations and we find that 68 of those fail to simultaneously satisfy the $\langle V \rangle > 0$ and the Hessian positive definiteness conditions, while we are left with 7 undecided cases.

\subsubsection*{${\cal S}_3$-${\cal S}_0$ models}
There are 155 configurations which could not be checked using our approach.
\vskip0.2cm
To summarize, all the results corresponding to the search of stable stable de-Sitter vacua are presented in Table \ref{tab_16384-classification-4} which shows that there are only 162 cases (out of 16384) which remain undecided while all the decided ones result in the No-Go scenarios. These undecided configurations are collected in the appendix \ref{sec_appendix3}.


\subsection{Status after the numerical analysis}

To summarize the status of the candidate configurations, let us mention that after the numerical analysis of the 11 classes ${\cal S}_{14}$-${\cal S}_4$ having a total of 15914 configurations, we find none of these can result in a tachyon-free Minkowskian or de-Sitter solution, expect for 7 undecided cases.  We recall again that we have imposed the positive definiteness of the Hessian and therefore configurations resulting in at least one (or more) massless moduli/axions are also excluded from this scan. The results are summarized in Table \ref{tab_16384-classification-4} where the last three columns presenting Ext.$^\ast$, Mink$^\ast$ and dS$^\ast$ correspond to the undecided cases left after the numerical analysis.

We also note that the ``NSolve" module works for classes ${\cal S}_{14}$-${\cal S}_4$ which involve at the most 10 non-zero $(\gamma_i, \lambda_i)$ parameters. However, ``NSolve" generically does not work for classes ${\cal S}_3$-${\cal S}_0$. This is because of the fact that we have 6 extremization conditions (\ref{eq:derV-lambda-gamma-1}) while 7 constraints from the Bianchi identities (\ref{eq:BIs-IIB-2}) are not independent. Even for the case when none of the $(\gamma_i, \lambda_i)$ parameters vanish, these 7 identities can be effectively expressed in terms of 4 constraints, and therefore there can be at the most 10 independent conditions after taking care of Bianchi identities and the extremization conditions. Considering the fact that configurations from class ${\cal S}_3$ involve 11 non-zero $(\gamma_i, \lambda_i)$ parameters which may not be necessarily solved using the numerical approach and therefore the Hessian checks (\ref{eq:Hessian-check}) could not be performed. However, for the Minkowskian case we have another equality condition, namely $\langle V \rangle = 0$, making it a total of 11 constraints from BIs+Extremization+Minkowskian requirements, and therefore most of the configurations in ${\cal S}_3$ can be numerically checked for Minkowskian case using ``NSolve", and we have additional No-Go results for the Minkowskian test corresponding to the class ${\cal S}_3$ as shown in Table \ref{tab_16384-classification-4}.

Thus, we find that out of 16384 candidate configurations to begin with, we are left with only 63 undecided cases for stable Minkowskian solutions and 162 cases for stable de-Sitter solutions while for around 99\% of those we end up confirming the No-Go case. The configurations corresponding to the undecided de-Sitter case are collected in appendix \ref{sec_appendix3}.

\begin{table}[h!]
\centering
\begin{tabular}{ |c||c|c|c||c|c|c||c|c|c|}
\hline
Flux & Step-3 & Step-4 & Step-4 & Step-5 & Step-5 & Step-5 & Step-5 & Step-5 & Step-5 \\
conf. & Ext. & Mink & dS & Ext. & Mink & dS & Ext.$^\ast$ & Mink$^\ast$ & dS$^\ast$ \\
 \hline
 \hline
${\bf {\cal S}_{14}}$  & 1 & 1 & 0 & 0 & 0 & 0 & 0 & 0 & 0\\
${\bf {\cal S}_{13}}$  & 0 & 0 & 0 & 0 & 0 & 0 & 0 & 0 & 0\\
${\bf {\cal S}_{12}}$  & 4 & 4 & 0 & 0 & 0 & 0 & 0 & 0 & 0\\
${\bf {\cal S}_{11}}$  & 0 & 0 & 0 & 0 & 0 & 0 & 0 & 0 & 0\\
${\bf {\cal S}_{10}}$  & 12 & 6 & 0 & 0 & 0 & 0 & 0 & 0 & 0\\
${\bf {\cal S}_{9}}$  & 11 & 2 & 0 & 0 & 0 & 0 & 0 & 0 & 0\\
\hline
${\bf {\cal S}_{8}}$  & 27 & 8 & 5 & 16 & 0 & 5 & 4 & 0 & 0\\
${\bf {\cal S}_{7}}$  & 36 & 28 & 10 & 15 & 7 & 10 & 4 & 0 & 0\\
${\bf {\cal S}_{6}}$  & 36 & 28 & 16 & 13 & 8 & 13 & 4 & 0 & 0\\
${\bf {\cal S}_{5}}$  & 34 & 14 & 27 & 28 & 8 & 27 & 0 & 0 & 0\\
${\bf {\cal S}_{4}}$  & 77 & 7 & 75 & 76 & 6 & 75 & 10 & 0 & 7 \\
\hline
${\bf {\cal S}_{3}}$  & 98 & 49 & 98 & 98 & 49 & 98 & 98 & 6 & 98 \\
${\bf {\cal S}_{2}}$  & 46 & 46 & 46 & 46 & 46 & 46 & 46 & 46 & 46\\
${\bf {\cal S}_{1}}$  & 10 & 10 & 10 & 10 & 10 & 10 & 10 & 10 & 10\\
${\bf {\cal S}_{0}}$  & 1 & 1 & 1 & 1 & 1 & 1 & 1 & 1 & 1 \\
\hline
\hline
Total & 393 & 204 & 288 & 303 & 135 & 285 & 177 & 63 & 162 \\
 \hline
\end{tabular}
\caption{Classification of ``undecided" configurations at various steps showing No-Go for ${\cal S}_{14}$-${\cal S}_{4}$ configurations.}
\label{tab_16384-classification-4}
\end{table}


\subsection{Comments on the previous de-Sitter models}
For our analysis about exploring the Minkowskian or de-Sitter solutions, let us mention here again that we have aimed to find the (axionic-) flux configurations which are not ruled out by imposing the following constraints:

\begin{itemize}

\item
All the Bianchi identities are satisfied.

\item
Except for the numerical test in the final stage which was performed for a significantly reduced number of flux configurations, our test for extremization + Minkowskian/de-Sitter conditions is precise in analytic sense, as it checks the compatibility of the set of constraints.

\item
There are no tachyons present.

\item
There are no moduli/axions which remain massless. 

\end{itemize}

\noindent
We note that the {\it last condition} of the above is sometimes overlooked/relaxed by arguing that the set of remaining massless moduli/axions can be fixed by including additional sub-leading effects. However, we avoided such situations as well, in the sense that a genuine de-Sitter minimum must have all moduli/axions stabilized. Otherwise, as a counter example, one could consider the situation of the tree level no-scale cancellations in models with the standard $F_3/H_3$ fluxes only, which keeps all the K\"ahler moduli massless, and leads to a positive semi-definite scalar potential for the complex-structure moduli and the axio-dilaton fields, and any minimum (if it exists) will be Minkowskian or de-Sitter type. But we know that the main challenge starts with incorporating the K\"ahler moduli stabilization into the overall moduli dynamics via the sub-leading corrections. For this reason we ensured the positive definiteness of the Hessian, and have ruled out the configurations leading to one or more massless moduli/axions. Such a condition is applicable to the tachyon free de-Sitter solutions of \cite{Blumenhagen:2015xpa, Shukla:2022srx} in which a combination of axions remain flat. 

Finally let us make some comments on the possible correlation among our results and the previous studies/claims on finding de-Sitter vacua in the context of non-geometric setups. The STU-like non-geometric models have been studied following a numerical approach with some of those claiming to find stable dS vacua, e.g. see \cite{Guarino:2008ik,deCarlos:2009fq,deCarlos:2009qm,Dibitetto:2011qs,Danielsson:2012by,Damian:2013dq,Blaback:2013ht,Blaback:2015zra,Plauschinn:2020ram}. These studies correspond to STU-like models having a set of (non-)geometric fluxes and other superpotential contributions. For example, 

\begin{itemize}

\item
The models studied in \cite{Guarino:2008ik,Danielsson:2012by,Blaback:2013ht,Damian:2013dwa,Blaback:2015zra} include an additional non-geometric fluxes, namely the so-called $P$ flux, which is S-dual to the non-geometric $Q$ flux. 

\item 
Similarly, a combination of (non-)geometric flux and the non-perturbative superpotential contributions have been used to realize de-Sitter solutions in \cite{Blaback:2015zra}, however it is not clear if the Bianchi identities have been (properly) incorporated in this analysis. 

\item
In addition, some non-geometric flux models have included additional uplifting terms, like anti-D3 uplifting or D-term contributions, to realize de-Sitter solutions \cite{Blumenhagen:2015kja,Blumenhagen:2015xpa,Damian:2018tlf,Shukla:2022srx,Damian:2023ote}. 

\end{itemize}

\noindent
Finding a direct correlation with such STU-like non-geometric models is beyond the scope of current analysis which is limited to an isotropic limit of a particular ${\mathbb T}^6/({\mathbb Z}_2\times{\mathbb Z}_2)$ orientifold. However, we can compare or correlate the implications of the current analysis with a couple of models, e.g. those presented in \cite{Damian:2013dq} and \cite{deCarlos:2009fq} which have the same underlying ingredients.

\begin{itemize}

\item
First, let us consider the de-Sitter model presented in \cite{Damian:2013dq}. Comparing the superpotential in Eq.~(A.1) of \cite{Damian:2013dq} with our isotropic superpotential in Eq.~(\ref{eq:WgenIIB-iso}), one finds that this model corresponds to having $\{H_0 = 0, \, H_1 = 0, H^1 = 0, Q^1{}_0 = 0, Q^1{}_1 + 2 Q^1{}_2 = 0 \}$, i.e. it belongs to the class ${\cal S}_4$ of Table \ref{tab_16384-classification1}. Subsequently, using axionic fluxes as defined in Eq.~(\ref{eq:AxionicFluxOrbitsIIB-Z2xZ2-iso1}), this model corresponds to ${\cal S}_0$ of Table \ref{tab_16384-classification-2} as none of the axionic fluxes are zero. 

\item
Next, let us consider the de-Sitter model presented in \cite{deCarlos:2009fq}. Comparing the superpotential in Eq.~(4.21) of \cite{deCarlos:2009fq} with our isotropic superpotential in Eq.~(\ref{eq:WgenIIB-iso}), one finds that this model corresponds to having $\{Q^1{}_0 = 0, Q^1{}_1 + 2 Q^1{}_2 = 0 \}$, i.e. it belongs to the class ${\cal S}_2$ of Table \ref{tab_16384-classification1}. Subsequently, using axionic fluxes as defined in Eq.~(\ref{eq:AxionicFluxOrbitsIIB-Z2xZ2-iso1}), this model corresponds to ${\cal S}_0$ of Table \ref{tab_16384-classification-2} as none of the axionic fluxes are zero.

\end{itemize}


\section{Summary and conclusions}
\label{sec_conclusions}
In this article, we have performed a systematic scan of the isotropic non-geometric flux vacua arising in the four-dimensional type II supergravity models based on toroidal orientifold. Using the scan results, we have presented a classification of the No-Go scenarios corresponding to various possible candidate configurations, regarding the possibility of those resulting in stable Minkowskian or de-Sitter vacua. The main attractive feature of our approach is the analytic nature of the flux vacua scan unlike the earlier approaches of using a black-box kind of numerical analysis. This analytic approach has been possible because of the so-called `axionic flux polynomials' which have helped us in rewriting the scalar potential, its various derivatives as well as the Hessian in a relatively very compact form.

It has been demonstrated in a series of works that scalar potentials arising from the concrete non-geometric toroidal setups are typically very huge in size \cite{Blumenhagen:2013hva, Gao:2015nra,Shukla:2015bca,Shukla:2015rua,Leontaris:2023mmm,Biswas:2024ewk}; for example, the type IIB supergravity model arising from ${\mathbb T}^6/({\mathbb Z}_2 \times {\mathbb Z}_2)$ orientifold with three-form fluxes $(F_3, H_3)$ and the non-geometric $Q$-flux results in 2422 terms depending on 14 moduli/axions and 40 flux parameters \cite{Blumenhagen:2013hva}. Even after imposing the isotropy conditions, there still remain 309 terms in the scalar potential which depend on 14 flux parameters and 6 moduli/axionic variables \cite{Biswas:2024ewk}. However, it turns out that using the axionic flux polynomials (\ref{eq:AxionicFluxOrbitsIIB-Z2xZ2-iso1}),  the scalar potential can be re-expressed in terms of just 40 terms \cite{Biswas:2024ewk}. We also observe that the set of Bianchi identities (\ref{eq:BIs-IIB}) continues to hold true even after promoting all the standard fluxes to their axionic generalizations as seen from Eq.~(\ref{eq:BIs-IIB-1}). This observation suggests us to define a new set of 14 parameters $(\gamma_i, \lambda_i)$ in Eq.~(\ref{eq:lambda-gamma-def-1}) corresponding to the 14 axionic flux polynomials such that the six extremization conditions (\ref{eq:derV-lambda-gamma-1}) and the Bianchi identities (\ref{eq:BIs-IIB-2}) can be solely expressed in terms of these parameters. Subsequently, aiming for an analytic analysis we  followed a five-step methodology:
\begin{itemize}

\item
First, without any loss of generality, we assumed that as 14 real parameters, $(\gamma_i, \lambda_i)$ can take either zero or non-zero values. This led us to having a collection of $2^{14} = 16384$ possible candidate configurations to analyze.

\item
Using the ``Reduce" module of Mathematica we subsequently found that 14272 configurations are incompatible with constraints from the set of Bianchi identities (\ref{eq:BIs-IIB-1}), and only 2112 candidates remain undecided after Step-2.

\item
In Step-3, after including the six extremization conditions $\partial_i V = 0$ using (\ref{eq:derV-lambda-gamma-1}) along with the Bianchi identities (\ref{eq:BIs-IIB-1}), the compatibility test showed that there are 15991 configurations which result in No-Go while only 393 remain undecided.

\item
We further tested these 393 configurations for satisfying the Minkowskian condition $\langle V \rangle = 0$ as well as the de-Sitter condition $\langle V \rangle > 0$ separately as Step-4, and found that there are only 204 undecided configurations for the Minkowskian case while there remain 288 undecided configurations for the de-Sitter case.

\item
We included the Hessian checks (\ref{eq:Hessian-check}) in Step-5 as the final stage of our analytic procedure. Note that using the analytic analysis so far, we have not explicitly solved any of the Bianchi identities or extremization conditions, and we have only checked whether the full set of constraints are simultaneously ``compatible" or not. As summarized in Table \ref{tab_16384-classification-3}, after our analytic exploration we are left with only 135 undecided configurations for Minkowskian case, and 285 undecided configuration for de-Sitter case.

\end{itemize}

\noindent
The undecided cases listed in Table \ref{tab_16384-classification-3} are those for which we could not manage to ensure a No-Go, however this does not necessarily mean that such leftover configurations will result in physical Minkowskian/de-Sitter solutions. As a final step of our analysis, we attempted to numerically explore what we call the reduced flux landscape. As we have summarized the main results in Table \ref{tab_16384-classification-4}, the underlying strategy for our numerical scan has been along the following points:
\begin{itemize}

\item
We note that although there are 7 Bianchi identities (\ref{eq:BIs-IIB-1}) and 6 extremization condition from (\ref{eq:derV-lambda-gamma-1}), it turns out that at the most only 4 Bianchi identities can be independent. This sums up to a total of 10 effective quadratic flux constraints expressed in terms of 14 $(\gamma_i, \lambda_i)$ parameters. Now, considering a particular type of configurations ${\cal S}_n$ at a time, we use `NSolve' module to look for numerical solutions, and subsequently check if those numerical solutions satisfy the Hessian positive definiteness conditions (\ref{eq:Hessian-check}) or not.

We find that generically we can analyze almost all the configurations of type ${\cal S}_{14}$-${\cal S}_4$, i.e. those having at the most 10 $(\gamma_i, \lambda_i)$ parameters being non-zero. In fact, we find a complete No-Go case for stable de-Sitter solutions corresponding to the classes ${\cal S}_{14}$-${\cal S}_5$, while ${\cal S}_4$ have resulted in 7 configurations which we could not decide due to a complicated form of the derivatives/Hessian. For the remaining configurations ${\cal S}_n$ with $n\leq3$, which have 3 or less number of $(\gamma_i, \lambda_i)$ parameters being set to zero, it is hard to reach a conclusion for the existence of stable de-Sitter vacua.

\item
Given that Bianchi identities, the extremization conditions and the Minkowskian condition $\langle V \rangle =0$ constitute a set of equalities, now one can generically have 11 independent constraints. So we can analyze all the configurations of type ${\cal S}_{14}$-${\cal S}_3$, i.e. those having at the most 11 $(\gamma_i, \lambda_i)$ parameters being non-zero.

In this regard, we find that the configurations in ${\cal S}_{14}$-${\cal S}_4$ result in a complete No-Go for having stable Minkowskian solutions. In addition, although almost all the configurations of type ${\cal S}_3$ result in Minkowskian No-Go case, we found 6 undecided cases. Finally, for the remaining configurations ${\cal S}_n$ with $n\leq2$, in which we have 2 or less number of $(\gamma_i, \lambda_i)$ parameters being set to zero, it is hard to reach a conclusion for stable Minkowskian vacua.

\end{itemize}

\noindent
To summarize, we have presented a detailed analytic test of the compatibility of the set of constraints needed for having stable Minkowsian and de-Sitter solutions in an isotropic toroidal type IIB model with non-geometric fluxes. Quite surprisingly we find that out of the total of 16384 candidate configurations, 16321 result in No-Go for Minkowskian while 16222 of those results in No-Go for de-Sitter vacua. In the light of the recent interests regarding the swampland conjectures, it would be interesting to explore the leftover configurations as collected in the appendix \ref{sec_appendix3} in order to make an exhaustive claim about the (non-)existence of stable de-Sitter vacua in such models.


\section*{Acknowledgments}
GKL would like to thanks the Physics Department of National and Kapodistrian University of Athens for kind hospitality. PS would like thank the {\it Department of Science and Technology (DST), India} for the kind support.


\appendix


\section{Details of the Type IIA isotropic toroidal model}
\label{sec_appendix1}

Taking the isotropic limit implies that one has 14 flux components instead of 40 to begin with, out of which 10 components belong to the NS-NS sector while 4 of them belong to the RR sector. To be specific, the isotropic limit corresponds to the following identifications among the various respective moduli and the fluxes,
\bea
\label{eq:IIAiso}
& & {\rm U}_1 = {\rm U}_2 = {\rm U}_3 \equiv {\rm U}, \qquad {\rm T}^1 = {\rm T}^2= {\rm T}^3 \equiv {\rm T}, \\
& & {\rm t}^1 = {\rm t}^2 = {\rm t}^3 \equiv {\rm t}, \quad {\rm z}^1 = {\rm z}^2 = {\rm z}^3 \equiv {\rm z}, \quad {\rm b}^1 = {\rm b}^2 = {\rm b}^3 \equiv {\rm b}, \quad \xi_1 = \xi_2 = \xi_3 \equiv \xi\,, \nonumber\\
& & \nonumber\\
& & e_1 = e_2 = e_3, \quad m^1 = m^2 = m^3, \quad {\rm H}^1 = {\rm H}^2 = {\rm H}^3, \qquad {\rm R}^1 = {\rm R}^2 = {\rm R}^3, \nonumber\\
& & w_{10} = w_{20} = w_{30}, \quad w_1{}^1 = w_2{}^2 = w_3{}^3, \quad w_1{}^2 =  w_1{}^3 =  w_2{}^1 =  w_2{}^3 =  w_3{}^1 =  w_3{}^2, \nonumber\\
& & {\rm Q}^1{}_0 ={\rm Q}^2{}_0 = {\rm Q}^3{}_0, \quad {\rm Q}^{11} = {\rm Q}^{22} = {\rm Q}^{33}, \quad {\rm Q}^{12} = {\rm Q}^{13} = {\rm Q}^{21} = {\rm Q}^{23} = {\rm Q}^{31} = {\rm Q}^{32}\,. \nonumber
\eea
Subsequently, the 14 independent flux parameters which we consider are summarized as below,
\bea
& & e_0, \quad e_1, \quad m^1, \quad m^0, \qquad   \\
& & {\rm H}_0, \quad {\rm H}^1, \qquad w_{10}, \quad w_1{}^1, \quad w_1{}^2, \qquad {\rm Q}^1{}_0, \quad {\rm Q}^{11}, \quad {\rm Q}^{12}\,, \qquad {\rm R}_0, \quad {\rm R}^1, \nonumber
\eea
along with the following T-dual set of the six moduli/axions,
\bea
& & D, \quad \xi^0, \qquad \sigma, \quad \xi, \qquad \rho, \quad {\rm b}.
\eea
Subsequently, the flux superpotential with isotropic conditions leads to the following form,
\bea
\label{eq:WgenIIA-iso}
& & \hskip-0.5cm W_{\rm IIA}^{\rm iso} = \frac{1}{\sqrt{2}} \biggl[\Bigl({e}_0 + 3\, {\rm T} \, {e}_1 + 3\, {\rm T}^2 \, m^1  + \, {\rm T}^3\, \, m^0 \Bigr) - \, {\rm N}^0 \Bigl({\rm H}_0 + 3\, {\rm T} \, {w}_{10}\, + 3\, {\rm T}^2 \, {\rm Q}^1{}_0 + \, {\rm T}^3 \, {\rm R}_0 \Bigr) \,\nonumber\\
& & \qquad \qquad - 3\, {\rm U} \, \Bigl({\rm H}^1 + {\rm T} \, ({w}_{1}{}^1 + 2\, {w}_{1}{}^2)\, +  {\rm T}^2 \, ({\rm Q}^{11} + 2\, {\rm Q}^{12}) + {\rm T}^3  \, {\rm R}^1 \Bigr) \biggr]\,.
\eea
The axionic fluxes take the form given as below,
\bea
\label{eq:AxionicFluxOrbitsIIA-Z2xZ2-iso}
& &  {\rm h}_0 \, \, = {\rm H}_{0} + 3\, {w}_{10}\, {\rm b} + 3\, ({\rm b})^2 \, {\rm Q}^1{}_{0} + ({\rm b})^3 \, {\rm R}_{0} \,,\\
& &  {\rm h}_1 = {w}_{10} + 2\, {\rm b} \, {\rm Q}^1{}_0 + \, ({\rm b})^2 \, {\rm R}_{0}, \nonumber\\
& & {\rm h}^1 = {\rm Q}^1{}_{0} + \, {\rm b}\, {\rm R}_{0}\, , \nonumber\\
& &  {\rm h}^0 \, = \, {\rm R}_{0} \,, \nonumber
\eea
\bea
& & {\rm h}^1{}_0 \, \, = {\rm H}^1 +  {\rm b}\, ({w}_{1}{}^1\, + 2\, {w}_{1}{}^2)\, +\, ({\rm b})^2 \, ({\rm Q}^{11} + 2\, {\rm Q}^{12}) + \, ({\rm b})^3 \, {\rm R}^1 \, ,\nonumber\\
& & {\rm h}_1{}^1 = {w}_1{}^1 + 2\, {\rm b} \, {\rm Q}^{12} + \, ({\rm b})^2 \, {\rm R}^1, \nonumber\\
& & {\rm h}_1{}^2 = {w}_1{}^2 + \, {\rm b} \, ({\rm Q}^{11} +  {\rm Q}^{12})+ \, ({\rm b})^2 \, {\rm R}^1,  \nonumber\\
& & {\rm h}^{11} = {\rm Q}^{11}+ \, {\rm b}\, {\rm R}^1, \nonumber\\
& & {\rm h}^{12} = {\rm Q}^{12}+ \, {\rm b}\, {\rm R}^1, \nonumber\\
& & {\rm h}^{10} \, = \, {\rm R}^1 \,,\nonumber\\
& & \nonumber\\
& & {\rm f}_0  = {\mathbb G}_0 - \, \xi^{0} \, {\rm h}_0 - 3\, {\xi}_1 \, {\rm h}^1{}_0 \,, \qquad \quad {\rm f}_1 = {\mathbb G}_1 - \, \xi^{0} \, {\rm h}_1 - \, {\xi}_1 \, ({\rm h}_1{}^1 + 2 \, {\rm h}_1{}^2)\,, \nonumber\\
& & {\rm f}^0 = {\mathbb G}^0 - \, \xi^{0} \, {\rm h}^0 - 3\, {\xi}_1 \,  {\rm h}^{10}, \qquad \quad {\rm f}^1 = {\mathbb G}^1 - \, \xi^{0} \, {\rm h}^1 - {\xi}_1 \, ({\rm h}^{11} + 2 \,{\rm h}^{12})\,, \nonumber
\eea
where
\bea
& & {\mathbb G}_0 = {e}_0 + 3\, {\rm b}\, {e}_1 + 3\, ({\rm b})^2 \,m^1 + ({\rm b})^3 \, \, m^0\, , \nonumber\\
& & {\mathbb G}_1 = {e}_1 + 2\, {\rm b} \,m^1 +\, ({\rm b})^2 \, m^0\,, \nonumber\\
& & {\mathbb G}^1 = m^1 + m^0\,  {\rm b}\, \nonumber\\
& & {\mathbb G}^0 = m^0\,. \nonumber
\eea

\subsubsection*{Type IIA Bianchi identities}
It turns out that the 10 NS-NS fluxes need to satisfy the following seven constraints arising from the Bianchi identities,
\bea
\label{eq:BIs-IIA}
& & \hskip-1.5cm w_{10}\, \, (w_1{}^1+ w_1{}^2) = {\rm H}^{1}\, {\rm Q}^{1}{}_0 + {\rm H}_{0}\, {\rm Q}^{12}, \nonumber\\
& & \hskip-1.5cm w_1{}^2\, \, (w_1{}^1+ w_1{}^2) = {\rm H}^{1}\, ({\rm Q}^{11} + {\rm Q}^{12}), \\
& & \hskip-1.5cm {\rm H}^1\, {\rm R}^1 = w_1{}^2\, \, {\rm Q}^{12}, \nonumber\\
& & \hskip-1.5cm {\rm H}_{0} \, \, {\rm R}^1 + w_1{}^2 \, \, {\rm Q}^1{}_0 = w_{10}\, ({\rm Q}^{11} + {\rm Q}^{12}), \nonumber\\
& & \hskip-1,5cm {\rm H}^{1} \, \, {\rm R}_0 + w_{10}\, \, {\rm Q}^{12} = {\rm Q}^1{}_0\, \, (w_1{}^1+ w_1{}^2), \nonumber\\
& & \hskip-1.5cm {\rm R}^1\, (w_1{}^1+ w_1{}^2) = {\rm Q}^{12}\, ({\rm Q}^{11} + {\rm Q}^{12}), \nonumber\\
& & \hskip-1.5cm {\rm R}^1\, w_{10} + {\rm R}_0\, \, w_1{}^2 = {\rm Q}^{1}{}_0\, ({\rm Q}^{11} + {\rm Q}^{12}). \nonumber
\eea
Let us note that using the forth and fifth identities, one derives the following identity which is the only one arising from what is called as ``cohomology formulation" of identities \cite{},
\bea
& & \hskip-1.5cm {\rm H}_{0} \, \, {\rm R}^1 - {\rm H}^{1} \, \, {\rm R}_0 + {\rm Q}^1{}_0\, \, (w_1{}^1+ 2\, w_1{}^2) - w_{10}\, ({\rm Q}^{11} + 2\, {\rm Q}^{12}) = 0.
\eea
It is important to mention that the set of Bianchi identities in eqn. (\ref{eq:BIs-IIA}) continue to hold after promoting the standard flux components in to the axionic flux orbits given in eqn. (\ref{eq:AxionicFluxOrbitsIIA-Z2xZ2-iso}). This is equivalent to the following set of constraints,
\bea
\label{eq:BIs-IIA-gen}
& & \hskip-1.5cm {\rm h}_{1}\, \, ({\rm h}_1{}^1+ {\rm h}_1{}^2) = {\rm h}^{1}{}_0\, {\rm h}^{1} + {\rm h}_{0}\, {\rm h}^{12}, \\
& & \hskip-1.5cm {\rm h}_1{}^2\, \, ({\rm h}_1{}^1+ {\rm h}_1{}^2) = {\rm h}^{1}{}_0\, ({\rm h}^{11} + {\rm h}^{12}), \nonumber\\
& & \hskip-1.5cm {\rm h}^1{}_0\, {\rm h}^{10} = {\rm h}_1{}^2\, \, {\rm h}^{12}, \nonumber\\
& & \hskip-1.5cm {\rm h}_{0} \, \, {\rm h}^{10} + {\rm h}_1{}^2 \, \, {\rm h}^1 = {\rm h}_{1}\, ({\rm h}^{11} + {\rm h}^{12}), \nonumber\\
& & \hskip-1,5cm {\rm h}^{1}{}_0 \, {\rm h}^0 + {\rm h}_{1}\, \, {\rm h}^{12} = {\rm h}^1\, ({\rm h}_1{}^1+ {\rm h}_1{}^2), \nonumber\\
& & \hskip-1.5cm {\rm h}^{10}\, ({\rm h}_1{}^1+ {\rm h}_1{}^2) = {\rm h}^{12}\, ({\rm h}^{11} + {\rm h}^{12}), \nonumber\\
& & \hskip-1.5cm {\rm h}^{10}\, {\rm h}_{1} + {\rm h}^0\, \, {\rm h}_1{}^2 = {\rm h}^{1}\, ({\rm h}^{11} + {\rm h}^{12}), \nonumber
\eea
along with the only cohomology formulation identity being given as,
\bea
& & \hskip-1.5cm {\rm h}_{0} \, \, {\rm h}^{10} - {\rm h}^{1}{}_0 \, \, {\rm h}^0 + {\rm h}^1\, \, ({\rm h}_1{}^1+ 2\, {\rm h}_1{}^2) - {\rm h}_{1}\, ({\rm h}^{11} + 2\, {\rm h}^{12}) = 0.
\eea


\section{Details of the various Hessian components}
\label{sec_appendix2}
Considering the basis of moduli/axionic fields as $\{c_0, c, v, s, \tau, u\}$, the various Hessian components can be expressed in terms of $(\gamma_i, \lambda_i)$ parameters in the following way,
\bea
\label{eq:Hess-eqn}
& & V_{11} = \frac{1}{2 s^3 \tau ^3 u^3} \biggl[\lambda _1^2+3 \lambda _2^2+3 \lambda _3^2+\lambda _4^2\biggr], \\
& & V_{12} = -\frac{3}{2
   s^2 \tau ^4 u^3} \biggl[\lambda _1 \lambda _5+\lambda _3 \lambda _7+\lambda _4 \lambda _8+\lambda _2 \left(\lambda _6+2 \lambda _9\right)+2 \lambda _3 \lambda _{10}\biggr], \nonumber\\
& & V_{13} = -\frac{3}{2 s^2 \tau
   ^3 u^4} \biggl[\left(\gamma _1+2 \gamma _3\right) \lambda _2+\gamma _4 \lambda _3+\gamma _2 \left(\lambda _1+2 \lambda _3\right)+\gamma _3 \lambda _4\biggr], \nonumber\\
& & V_{14} = \frac{1}{2 s^3 \tau ^3 u^3} \biggl[\gamma _1 \lambda _1+3 \gamma _2 \lambda _2+3 \gamma _3 \lambda _3+\gamma _4 \lambda _4\biggr], \nonumber\\
& & V_{15} = \frac{3}{2 s^2 \tau ^4 u^3} \biggl[\gamma _1 \lambda _1+3 \gamma _2 \lambda _2+3 \gamma _3 \lambda _3+\gamma _4 \lambda _4\biggr], \nonumber\\
& & V_{16} = \frac{3}{2 s^2 \tau ^3 u^4} \biggl[\gamma _1 \lambda _1+\gamma _2 \lambda _2-\gamma _3 \lambda _3-\gamma _4 \lambda _4\biggr], \nonumber\\
& & V_{22} = \frac{3}{2 s
   \tau ^5 u^3} \biggl[3 \lambda _5^2+\lambda _6^2+\lambda _7^2+3 \lambda _8^2+4 \lambda _9^2+4 \lambda _{10}^2+4 \lambda _6 \lambda _9+4 \lambda _7 \lambda _{10}\biggr], \nonumber\\
& & V_{23} = \frac{1}{2 s \tau ^4 u^4} \biggl[9 \gamma _2 \lambda _5+3 \left(\gamma _1+2 \gamma _3\right) \lambda _6+3 \left(2 \gamma _2+\gamma _4\right) \lambda _7+9 \gamma _3 \lambda _8+6 \left(\gamma _1+2
   \gamma _3\right) \lambda _9 \nonumber\\
& & \hskip0.75cm +6 \left(2 \gamma _2+\gamma _4\right) \lambda _{10}\biggr], \nonumber\\
& & V_{24} = -\frac{3}{2 s^2
   \tau ^4 u^3} \biggl[\gamma _1 \lambda _5+\gamma _3 \lambda _7+\gamma _4 \lambda _8+\gamma _2 \left(\lambda _6+2 \lambda _9\right)+2 \gamma _3 \lambda _{10}\biggr], \nonumber\\
& & V_{25} = -\frac{9}{2 s \tau
   ^5 u^3} \biggl[\gamma _1 \lambda _5+\gamma _3 \lambda _7+\gamma _4 \lambda _8+\gamma _2 \left(\lambda _6+2 \lambda _9\right)+2 \gamma _3 \lambda _{10}\biggr],\nonumber\\
& & V_{26} = \frac{1}{2 s \tau ^4 u^4} \biggl[-9 \gamma _1 \lambda _5+3 \gamma _3 \lambda _7+9 \gamma _4 \lambda _8-3 \gamma _2 \left(\lambda _6+2 \lambda _9\right)+6 \gamma _3 \lambda _{10}\biggr], \nonumber\\
& & V_{33} = \frac{3}{2 s \tau ^3 u^5} \biggl[3 \gamma _2^2+2 \gamma _4 \gamma _2+4 \gamma _3^2+\gamma _4^2+2 \gamma _1 \gamma _3+3 \lambda _2^2+4 \lambda _3^2+\lambda _4^2+\lambda _6^2+3 \lambda
   _8^2+6 \lambda _9^2+6 \lambda _{10}^2 \nonumber\\
& & \hskip0.75cm +2 \lambda _1 \lambda _3+2 \lambda _2 \lambda _4+2 \lambda _4 \lambda _6+2 \lambda _2 \lambda _8+6 \lambda _6 \lambda _9+6
   \lambda _8 \lambda _9-4 \lambda _3 \lambda _{10}+2 \lambda _5 \lambda _{10}+6 \lambda _7 \lambda _{10}\biggr], \nonumber\\
& & V_{34} = -\frac{3}{2 s^2 \tau ^3 u^4} \biggl[\gamma _1 \gamma _2+2 \gamma _3 \gamma _2+\gamma _3 \gamma _4-\lambda _1 \lambda _2-2 \lambda _2 \lambda _3-\lambda _3 \lambda _4+\lambda _5 \lambda _6+6
   \lambda _5 \lambda _8+\lambda _7 \lambda _8 \nonumber\\
& & \hskip0.75cm +2 \lambda _5 \lambda _9+2 \lambda _8 \lambda _{10}\biggr], \nonumber\\
& & V_{35} = -\frac{3}{2 s \tau ^4 u^4} \biggl[3 \gamma _1 \gamma _2+6 \gamma _3 \gamma _2+3 \gamma _3 \gamma _4+3 \lambda _1 \lambda _2+6 \lambda _2 \lambda _3+3 \lambda _3 \lambda _4+4 \lambda _3
   \lambda _6+\lambda _5 \lambda _6+4 \lambda _1 \lambda _8 \nonumber\\
& & \hskip0.75cm +\lambda _7 \lambda _8+2 \lambda _5 \lambda _9-8 \lambda _2 \lambda _{10}+2 \lambda _8 \lambda _{10}+6
   \lambda _9 \lambda _{10}\biggr], \nonumber\\
& & V_{36} = -\frac{3}{2 s \tau ^3 u^5} \biggl[3 \gamma _1 \gamma _2+2 \gamma _3 \gamma _2-\gamma _3 \gamma _4+3 \lambda _1 \lambda _2+2 \lambda _2 \lambda _3-\lambda _3 \lambda _4+2 \lambda _3
   \lambda _6+3 \lambda _5 \lambda _6+2 \lambda _1 \lambda _8 \nonumber\\
& & \hskip0.75cm -\lambda _7 \lambda _8+6 \lambda _5 \lambda _9-4 \lambda _2 \lambda _{10}-2 \lambda _8 \lambda _{10}+6
   \lambda _9 \lambda _{10}\biggr], \nonumber\\
& & V_{44} = \frac{1}{2 s^3 \tau ^3
   u^3} \biggl[-6 \gamma _1 \lambda _8+6 \gamma _4 \lambda _5-6 \gamma _3 \lambda _6-12 \gamma _3 \lambda _9+6 \gamma _2 \left(\lambda _7+2 \lambda _{10}\right)+\gamma _1^2+3
   \gamma _2^2+3 \gamma _3^2+\gamma _4^2 \nonumber\\
& & \hskip0.75cm +3 \lambda _5^2-3 \lambda _6^2-3 \lambda _7^2+3 \lambda _8^2+12 \lambda _8 \lambda _9+12 \lambda _5 \lambda _{10}\biggr], \nonumber\\
& & V_{45} = \frac{3}{4 s^2 \tau ^4 u^3} \biggl[-4 \gamma _1 \lambda _8+4 \gamma _4 \lambda _5-4 \gamma _3 \lambda _6-8 \gamma _3 \lambda _9+4 \gamma _2 \left(\lambda _7+2 \lambda _{10}\right)+\gamma
   _1^2+3 \gamma _2^2+3 \gamma _3^2+\gamma _4^2 \nonumber\\
& & \hskip0.75cm -\lambda _1^2-3 \lambda _2^2-3 \lambda _3^2-\lambda _4^2+\lambda _5^2-\lambda _6^2-\lambda _7^2+\lambda _8^2+4 \lambda _8
   \lambda _9+4 \lambda _5 \lambda _{10}\biggr], \nonumber\\
& & V_{46} = \frac{3}{4 s^2 \tau ^3 u^4} \biggl[\gamma _1^2+\gamma _2^2-\gamma _3^2-\gamma _4^2-\lambda _1^2-\lambda _2^2+\lambda _3^2+\lambda _4^2+3 \lambda _5^2-\lambda _6^2+\lambda _7^2-3 \lambda
   _8^2-4 \lambda _8 \lambda _9+4 \lambda _5 \lambda _{10}\biggr], \nonumber\\
& & V_{55} = \frac{3}{2 s
   \tau ^5 u^3} \biggl[2 \gamma _1^2+6 \gamma _2^2+6 \gamma _3^2+2 \gamma _4^2+2 \lambda _1^2+6 \lambda
   _2^2+6 \lambda _3^2+2 \lambda _4^2+\lambda _5^2-\lambda _6^2-\lambda _7^2 +\lambda _8^2  \nonumber\\
& & \hskip0.75cm -2 \gamma _1 \left(2 \lambda _4+3 \lambda _8\right)+4 \gamma _4 \lambda _1-12 \gamma _3 \lambda _2+6 \gamma _4 \lambda _5-6 \gamma _3 \lambda _6-12 \gamma
   _3 \lambda _9 +6 \gamma _2 \left(2 \lambda _3+\lambda _7+2 \lambda _{10}\right) \nonumber\\
& & \hskip0.75cm +6 \lambda _2 \lambda _6+6 \lambda _4 \lambda _6+6 \lambda _1 \lambda _7+6
   \lambda _2 \lambda _8-6 \lambda _2 \lambda _9+4 \lambda _8 \lambda _9-6 \lambda _1 \lambda _{10}-12 \lambda _3 \lambda _{10}+4 \lambda _5 \lambda _{10}\biggr], \nonumber\\
& & V_{56} = \frac{3}{4 s \tau ^4 u^4} \biggl[3 \gamma _1^2+3 \gamma _2^2-3 \gamma _3^2-3 \gamma _4^2+3 \lambda _1^2+3 \lambda _2^2-3 \lambda _3^2-3 \lambda _4^2+3 \lambda _5^2-\lambda _6^2+\lambda
   _7^2-3 \lambda _8^2+4 \lambda _2 \lambda _6 \nonumber\\
& & \hskip0.75cm -4 \lambda _4 \lambda _6+4 \lambda _1 \lambda _7-4 \lambda _2 \lambda _8-4 \lambda _2 \lambda _9-4 \lambda _8 \lambda _9-4
   \lambda _1 \lambda _{10}+8 \lambda _3 \lambda _{10}+4 \lambda _5 \lambda _{10}\biggr], \nonumber\\
& & V_{66} = \frac{3}{2 s \tau ^3 u^5} \biggl[2 \gamma _1^2+\gamma _2^2+\gamma _4^2+2 \lambda _1^2+\lambda _2^2+\lambda _4^2+6 \lambda _5^2-\lambda _6^2+3 \lambda _8^2+2 \lambda _2 \lambda _6+2
   \lambda _1 \lambda _7-2 \lambda _2 \lambda _9 \nonumber\\
& & \hskip0.75cm -2 \lambda _1 \lambda _{10}+4 \lambda _5 \lambda _{10}\biggr]. \nonumber
\eea
Let us recall that Hessian components expressed in terms of the standard fluxes are too complicated, given that scalar potential itself has 309 terms. However, axionic fluxes help us to rewrite those complicated expressions in a rather simple way.

\section{Undecided configurations for de-Sitter vacua}
\label{sec_appendix3}
There are a total of 162 configurations which remain undecided for the existence of stable de-Sitter vacua. Apart from the most general case when none of the 14 parameters are zero, the renaming 161 configurations are listed as follows:
\bea
& & \hskip-1.5cm {\cal S}_1 \equiv \bigl\{\left\{\gamma _1 = 0\right\},\left\{\gamma _2 = 0\right\},\left\{\gamma _3 = 0\right\},\left\{\gamma _4 = 0\right\},\left\{\lambda _1 = 0\right\},\left\{\lambda _2 = 0\right\},\left\{\lambda _3 = 0\right\},\left\{\lambda _4 = 0\right\}, \nonumber\\
& & \hskip1cm \left\{\lambda _6 = 0\right\},\left\{\lambda
   _7 = 0\right\}\bigr\}. \nonumber
\eea
\bea
& & \hskip-0.25cm {\cal S}_2 \equiv \bigl\{\left\{\gamma _1 = 0,\gamma _2 = 0\right\},\left\{\gamma _1 = 0,\gamma _3 = 0\right\},\left\{\gamma _1 = 0,\gamma _4 = 0\right\},\left\{\gamma _1 = 0,\lambda _1 = 0\right\},\left\{\gamma _1 = 0,\lambda _2 = 0\right\}, \nonumber\\
& & \hskip0.75cm \left\{\gamma _1 = 0,\lambda _3 = 0\right\},\left\{\gamma_1 = 0,\lambda _4 = 0\right\},\left\{\gamma _1 = 0,\lambda _6 = 0\right\},\left\{\gamma _1 = 0,\lambda _7 = 0\right\},\left\{\gamma _2 = 0,\gamma _3 = 0\right\},
\nonumber\\
& & \hskip0.75cm \left\{\gamma _2 = 0,\gamma _4 = 0\right\},\left\{\gamma _2 = 0,\lambda _1 = 0\right\},\left\{\gamma _2 = 0,\lambda _2 =
   0\right\},\left\{\gamma _2 = 0,\lambda _3 = 0\right\},\left\{\gamma _2 = 0,\lambda _4 = 0\right\},
\nonumber\\
& & \hskip0.75cm \left\{\gamma _2 = 0,\lambda _6 = 0\right\},\left\{\gamma _2 = 0,\lambda _7 = 0\right\},\left\{\gamma _3 = 0,\gamma _4 = 0\right\},\left\{\gamma _3 = 0,\lambda _1 =
   0\right\},\left\{\gamma _3 = 0,\lambda _2 = 0\right\},
\nonumber\\
& & \hskip0.75cm \left\{\gamma _3 = 0,\lambda _3 = 0\right\},\left\{\gamma _3 = 0,\lambda _4 = 0\right\},\left\{\gamma _3 = 0,\lambda _6 = 0\right\},\left\{\gamma _3 = 0,\lambda _7 = 0\right\},\left\{\gamma _4 = 0,\lambda _1 =
   0\right\},
\nonumber\\
& & \hskip0.75cm \left\{\gamma _4 = 0,\lambda _2 = 0\right\},\left\{\gamma _4 = 0,\lambda _3 = 0\right\},\left\{\gamma _4 = 0,\lambda _4 = 0\right\},\left\{\gamma _4 = 0,\lambda _6 = 0\right\},\left\{\gamma _4 = 0,\lambda _7 = 0\right\},
\nonumber\\
& & \hskip0.75cm \left\{\lambda _1 = 0,\lambda _3 =
   0\right\},\left\{\lambda _1 = 0,\lambda _4 = 0\right\},\left\{\lambda _1 = 0,\lambda _6 = 0\right\},\left\{\lambda _1 = 0,\lambda _7 = 0\right\},\left\{\lambda _2 = 0,\lambda _4 = 0\right\},
\nonumber\\
& & \hskip0.75cm \left\{\lambda _2 = 0,\lambda _6 = 0\right\},\left\{\lambda _2 = 0,\lambda _7 =
   0\right\},\left\{\lambda _3 = 0,\lambda _6 = 0\right\},\left\{\lambda _3 = 0,\lambda _7 = 0\right\},\left\{\lambda _4 = 0,\lambda _6 = 0\right\},
\nonumber\\
& & \hskip0.75cm \left\{\lambda _4 = 0,\lambda _7 = 0\right\},\left\{\lambda _5 = 0,\lambda _9 = 0\right\},\left\{\lambda _5 = 0,\lambda _{10} =
   0\right\},\left\{\lambda _6 = 0,\lambda _7 = 0\right\},\left\{\lambda _8 = 0,\lambda_9 = 0\right\},
\nonumber\\
& & \hskip0.75cm \left\{\lambda _8 = 0,\lambda _{10}= 0\right\}\bigr\}. \nonumber
\eea
\bea
& & \hskip-0.25cm {\cal S}_3 \equiv
\bigl\{\left\{\gamma _1 = 0,\gamma _2 = 0,\lambda _1 = 0\right\},\left\{\gamma _1 = 0,\gamma _2 = 0,\lambda _2 = 0\right\},\left\{\gamma _1 = 0,\gamma _2 = 0,\lambda _6 = 0\right\},
\nonumber\\
& & \hskip0.75cm \left\{\gamma _1 = 0,\gamma _2 = 0,\lambda _7 = 0\right\},\left\{\gamma _1 = 0,\gamma _3 = 0,\lambda
   _1 = 0\right\},\left\{\gamma _1 = 0,\gamma _3 = 0,\lambda _3 = 0\right\},
\nonumber\\
& & \hskip0.75cm \left\{\gamma _1 = 0,\gamma _3 = 0,\lambda _6 = 0\right\},\left\{\gamma _1 = 0,\gamma _3 = 0,\lambda _7 = 0\right\},\left\{\gamma _1 = 0,\gamma _4 = 0,\lambda _1 = 0\right\},
\nonumber\\
& & \hskip0.75cm \left\{\gamma _1 = 0,\gamma _4\to
   0,\lambda _4 = 0\right\},\left\{\gamma _1 = 0,\gamma _4 = 0,\lambda _6 = 0\right\},\left\{\gamma _1 = 0,\gamma _4 = 0,\lambda _7 = 0\right\},
\nonumber\\
& & \hskip0.75cm \left\{\gamma _1 = 0,\lambda _1 = 0,\lambda _3 = 0\right\},\left\{\gamma _1 = 0,\lambda _1 = 0,\lambda _6 = 0\right\},\left\{\gamma _1\to
   0,\lambda _1 = 0,\lambda _7 = 0\right\},
\nonumber\\
& & \hskip0.75cm \left\{\gamma _1 = 0,\lambda _2 = 0,\lambda _6 = 0\right\},\left\{\gamma _1 = 0,\lambda _2 = 0,\lambda _7 = 0\right\},\left\{\gamma _1 = 0,\lambda _3 = 0,\lambda _6 = 0\right\},
\nonumber\\
& & \hskip0.75cm \left\{\gamma _1 = 0,\lambda _3 = 0,\lambda _7\to
   0\right\},\left\{\gamma _1 = 0,\lambda _4 = 0,\lambda _6 = 0\right\},\left\{\gamma _1 = 0,\lambda _4 = 0,\lambda _7 = 0\right\},
   \nonumber\\
& & \hskip0.75cm \left\{\gamma _1 = 0,\lambda _5 = 0,\lambda _9 = 0\right\},\left\{\gamma _1 = 0,\lambda _5 = 0,\lambda _{10} = 0\right\},\left\{\gamma _1 = 0,\lambda
   _6 = 0,\lambda _7 = 0\right\},
\nonumber\\
& & \hskip0.75cm \left\{\gamma _1 = 0,\lambda _8 = 0,\lambda _9 = 0\right\},\left\{\gamma _1 = 0,\lambda _8 = 0,\lambda _{10} = 0\right\},\left\{\gamma _2 = 0,\gamma _3 = 0,\lambda _2 = 0\right\},
\nonumber\\
& & \hskip0.75cm \left\{\gamma _2 = 0,\gamma _3 = 0,\lambda _3 = 0\right\},\left\{\gamma
   _2 = 0,\gamma _3 = 0,\lambda _6 = 0\right\},\left\{\gamma _2 = 0,\gamma _3 = 0,\lambda _7 = 0\right\},
\nonumber\\
& & \hskip0.75cm \left\{\gamma _2 = 0,\gamma _4 = 0,\lambda _2 = 0\right\},\left\{\gamma _2 = 0,\gamma _4 = 0,\lambda _4 = 0\right\},\left\{\gamma _2 = 0,\gamma _4 = 0,\lambda _6\to
   0\right\},
\nonumber\\
& & \hskip0.75cm \left\{\gamma _2 = 0,\gamma _4 = 0,\lambda _7 = 0\right\},\left\{\gamma _2 = 0,\lambda _1 = 0,\lambda _6 = 0\right\},\left\{\gamma _2 = 0,\lambda _1 = 0,\lambda _7 = 0\right\},
\nonumber\\
& & \hskip0.75cm \left\{\gamma _2 = 0,\lambda _2 = 0,\lambda _4 = 0\right\},\left\{\gamma _2 = 0,\lambda _2\to
   0,\lambda _6 = 0\right\},\left\{\gamma _2 = 0,\lambda _2 = 0,\lambda _7 = 0\right\},
\nonumber\\
& & \hskip0.75cm \left\{\gamma _2 = 0,\lambda _3 = 0,\lambda _6 = 0\right\},\left\{\gamma _2 = 0,\lambda _3 = 0,\lambda _7 = 0\right\},\left\{\gamma _2 = 0,\lambda _4 = 0,\lambda _6 = 0\right\},
\nonumber\\
& & \hskip0.75cm \left\{\gamma _2\to
   0,\lambda _4 = 0,\lambda _7 = 0\right\},\left\{\gamma _2 = 0,\lambda _5 = 0,\lambda _9 = 0\right\},\left\{\gamma _2 = 0,\lambda _5 = 0,\lambda _{10} = 0\right\},
\nonumber\\
& & \hskip0.75cm \left\{\gamma _2 = 0,\lambda _6 = 0,\lambda _7 = 0\right\},\left\{\gamma _2 = 0,\lambda _8 = 0,\lambda _9\to
   0\right\},\left\{\gamma _2 = 0,\lambda _8 = 0,\lambda _{10} = 0\right\},
\nonumber\\
& & \hskip0.75cm \left\{\gamma _3 = 0,\gamma _4 = 0,\lambda _3 = 0\right\},\left\{\gamma _3 = 0,\gamma _4 = 0,\lambda _4 = 0\right\},\left\{\gamma _3 = 0,\gamma _4 = 0,\lambda _6 = 0\right\},
\nonumber\\
& & \hskip0.75cm \left\{\gamma _3 = 0,\gamma _4\to
   0,\lambda _7 = 0\right\},\left\{\gamma _3 = 0,\lambda _1 = 0,\lambda _3 = 0\right\},\left\{\gamma _3 = 0,\lambda _1 = 0,\lambda _6 = 0\right\},
\nonumber\\
& & \hskip0.75cm \left\{\gamma _3 = 0,\lambda _1 = 0,\lambda _7 = 0\right\},\left\{\gamma _3 = 0,\lambda _2 = 0,\lambda _6 = 0\right\},\left\{\gamma _3\to
   0,\lambda _2 = 0,\lambda _7 = 0\right\},
\nonumber\\
& & \hskip0.75cm \left\{\gamma _3 = 0,\lambda _3 = 0,\lambda _6 = 0\right\},\left\{\gamma _3 = 0,\lambda _3 = 0,\lambda _7 = 0\right\},\left\{\gamma _3 = 0,\lambda _4 = 0,\lambda _6 = 0\right\},
\nonumber\\
& & \hskip0.75cm \left\{\gamma _3 = 0,\lambda _4 = 0,\lambda _7\to
   0\right\},\left\{\gamma _3 = 0,\lambda _5 = 0,\lambda _9 = 0\right\},\left\{\gamma _3 = 0,\lambda _5 = 0,\lambda _{10} = 0\right\},
\nonumber\\
& & \hskip0.75cm \left\{\gamma _3 = 0,\lambda _6 = 0,\lambda _7 = 0\right\},\left\{\gamma _3 = 0,\lambda _8 = 0,\lambda _9 = 0\right\},\left\{\gamma _3 = 0,\lambda
   _8 = 0,\lambda _{10} = 0\right\},
\nonumber\\
& & \hskip0.75cm \left\{\gamma _4 = 0,\lambda _1 = 0,\lambda _6 = 0\right\},\left\{\gamma _4 = 0,\lambda _1 = 0,\lambda _7 = 0\right\},\left\{\gamma _4 = 0,\lambda _2 = 0,\lambda _4 = 0\right\},
\nonumber\\
& & \hskip0.75cm \left\{\gamma _4 = 0,\lambda _2 = 0,\lambda _6\to
   0\right\},\left\{\gamma _4 = 0,\lambda _2 = 0,\lambda _7 = 0\right\},\left\{\gamma _4 = 0,\lambda _3 = 0,\lambda _6 = 0\right\},
\nonumber\\
& & \hskip0.75cm \left\{\gamma _4 = 0,\lambda _3 = 0,\lambda _7 = 0\right\},\left\{\gamma _4 = 0,\lambda _4 = 0,\lambda _6 = 0\right\},\left\{\gamma _4 = 0,\lambda _4\to
   0,\lambda _7 = 0\right\},
\nonumber\\
& & \hskip0.75cm \left\{\gamma _4 = 0,\lambda _5 = 0,\lambda _9 = 0\right\},\left\{\gamma _4 = 0,\lambda _5 = 0,\lambda _{10} = 0\right\},\left\{\gamma _4 = 0,\lambda _6 = 0,\lambda _7 = 0\right\},
\nonumber\\
& & \hskip0.75cm \left\{\gamma _4 = 0,\lambda _8 = 0,\lambda _9 = 0\right\},\left\{\gamma
   _4 = 0,\lambda _8 = 0,\lambda _{10} = 0\right\},\left\{\lambda _1 = 0,\lambda _5 = 0,\lambda _{10} = 0\right\},
\nonumber\\
& & \hskip0.75cm \left\{\lambda _1 = 0,\lambda _8 = 0,\lambda _9 = 0\right\},\left\{\lambda _1 = 0,\lambda _8 = 0,\lambda _{10} = 0\right\},\left\{\lambda _2 = 0,\lambda _5 = 0,\lambda
   _{10} = 0\right\},
\nonumber\\
& & \hskip0.75cm \left\{\lambda _2 = 0,\lambda _6 = 0,\lambda _7 = 0\right\},\left\{\lambda _2 = 0,\lambda _8 = 0,\lambda _9 = 0\right\},\left\{\lambda _3 = 0,\lambda _5 = 0,\lambda _{10} = 0\right\},
\nonumber\\
& & \hskip0.75cm \left\{\lambda _3 = 0,\lambda _6 = 0,\lambda _7 = 0\right\},\left\{\lambda _3\to
   0,\lambda _8 = 0,\lambda _9 = 0\right\},\left\{\lambda _4 = 0,\lambda _5 = 0,\lambda _9 = 0\right\},
\nonumber\\
& & \hskip0.75cm \left\{\lambda _4 = 0,\lambda _5 = 0,\lambda _{10} = 0\right\},\left\{\lambda _4 = 0,\lambda _8 = 0,\lambda _9 = 0\right\},\left\{\lambda _5 = 0,\lambda _7 = 0,\lambda _9\to
   0\right\},
\nonumber\\
& & \hskip0.75cm \left\{\lambda _5 = 0,\lambda _7 = 0,\lambda _{10} = 0\right\},\left\{\lambda _5 = 0,\lambda _8 = 0,\lambda _9 = 0\right\},\left\{\lambda _5 = 0,\lambda _8 = 0,\lambda _{10} = 0\right\},
\nonumber\\
& & \hskip0.75cm \left\{\lambda _6 = 0,\lambda _8 = 0,\lambda _9 = 0\right\},\left\{\lambda _6\to
   0,\lambda _8 = 0,\lambda _{10} = 0\right\}\bigr\}. \nonumber
\eea
and
\bea
& & \hskip-0.25cm {\cal S}_4 \equiv \bigl\{\left\{\gamma _1 = 0,\gamma _3 = 0,\lambda _2 = 0,\lambda _4 = 0\right\},\left\{\gamma _1 = 0,\lambda _4 = 0,\lambda _5 = 0,\lambda _9 = 0\right\},
\nonumber\\
& & \hskip0.75cm \left\{\gamma _2 = 0,\gamma _4 = 0,\lambda _1 = 0,\lambda _3 = 0\right\},\left\{\gamma _2 = 0,\lambda _2 = 0,\lambda _5\to
   0,\lambda _{10} = 0\right\},
   \nonumber\\
& & \hskip0.75cm \left\{\gamma _2 = 0,\lambda _3 = 0,\lambda _8 = 0,\lambda _9 = 0\right\},\left\{\gamma _3 = 0,\lambda _2 = 0,\lambda _5 = 0,\lambda _{10} = 0\right\},
\nonumber\\
& & \hskip0.75cm \left\{\lambda _5 = 0,\lambda _6 = 0,\lambda _9 = 0,\lambda _{10} = 0\right\}\bigr\}. \nonumber
\eea


\bibliographystyle{JHEP}
\bibliography{reference}

\providecommand{\href}[2]{#2}\begingroup\raggedright\begin{thebibliography}{100}

\bibitem{Aldazabal:2006up}
G.~Aldazabal, P.~G. Camara, A.~Font and L.~Ibanez, \emph{{More dual fluxes and
  moduli fixing}},
  \href{https://doi.org/10.1088/1126-6708/2006/05/070}{\emph{JHEP} {\bfseries
  0605} (2006) 070}, [\href{https://arxiv.org/abs/hep-th/0602089}{{\ttfamily
  hep-th/0602089}}].

\bibitem{deCarlos:2009qm}
B.~de~Carlos, A.~Guarino and J.~M. Moreno, \emph{{Complete classification of
  Minkowski vacua in generalised flux models}},
  \href{https://doi.org/10.1007/JHEP02(2010)076}{\emph{JHEP} {\bfseries 1002}
  (2010) 076}, [\href{https://arxiv.org/abs/0911.2876}{{\ttfamily 0911.2876}}].

\bibitem{Danielsson:2012by}
U.~Danielsson and G.~Dibitetto, \emph{{On the distribution of stable de Sitter
  vacua}}, \href{https://doi.org/10.1007/JHEP03(2013)018}{\emph{JHEP}
  {\bfseries 1303} (2013) 018},
  [\href{https://arxiv.org/abs/1212.4984}{{\ttfamily 1212.4984}}].

\bibitem{Blaback:2013ht}
J.~Blaback, U.~Danielsson and G.~Dibitetto, \emph{{Fully stable dS vacua from
  generalised fluxes}},
  \href{https://doi.org/10.1007/JHEP08(2013)054}{\emph{JHEP} {\bfseries 1308}
  (2013) 054}, [\href{https://arxiv.org/abs/1301.7073}{{\ttfamily 1301.7073}}].

\bibitem{Damian:2013dq}
C.~Damian, L.~R. Diaz-Barron, O.~Loaiza-Brito and M.~Sabido, \emph{{Slow-Roll
  Inflation in Non-geometric Flux Compactification}},
  \href{https://doi.org/10.1007/JHEP06(2013)109}{\emph{JHEP} {\bfseries 1306}
  (2013) 109}, [\href{https://arxiv.org/abs/1302.0529}{{\ttfamily 1302.0529}}].

\bibitem{Damian:2013dwa}
C.~Damian and O.~Loaiza-Brito, \emph{{More stable de Sitter vacua from S-dual
  nongeometric fluxes}},
  \href{https://doi.org/10.1103/PhysRevD.88.046008}{\emph{Phys.Rev.} {\bfseries
  D88} (2013) 046008}, [\href{https://arxiv.org/abs/1304.0792}{{\ttfamily
  1304.0792}}].

\bibitem{Hassler:2014mla}
F.~Hassler, D.~Lust and S.~Massai, \emph{{On Inflation and de Sitter in
  Non‐Geometric String Backgrounds}},
  \href{https://doi.org/10.1002/prop.201700062}{\emph{Fortsch. Phys.}
  {\bfseries 65} (2017) 1700062},
  [\href{https://arxiv.org/abs/1405.2325}{{\ttfamily 1405.2325}}].

\bibitem{Blumenhagen:2015qda}
R.~Blumenhagen, A.~Font, M.~Fuchs, D.~Herschmann and E.~Plauschinn,
  \emph{{Towards Axionic Starobinsky-like Inflation in String Theory}},
  \href{https://doi.org/10.1016/j.physletb.2015.05.001}{\emph{Phys. Lett.}
  {\bfseries B746} (2015) 217--222},
  [\href{https://arxiv.org/abs/1503.01607}{{\ttfamily 1503.01607}}].

\bibitem{Blumenhagen:2015kja}
R.~Blumenhagen, A.~Font, M.~Fuchs, D.~Herschmann, E.~Plauschinn, Y.~Sekiguchi
  et~al., \emph{{A Flux-Scaling Scenario for High-Scale Moduli Stabilization in
  String Theory}},
  \href{https://doi.org/10.1016/j.nuclphysb.2015.06.003}{\emph{Nucl. Phys.}
  {\bfseries B897} (2015) 500--554},
  [\href{https://arxiv.org/abs/1503.07634}{{\ttfamily 1503.07634}}].

\bibitem{Blumenhagen:2015jva}
R.~Blumenhagen, A.~Font, M.~Fuchs, D.~Herschmann and E.~Plauschinn,
  \emph{{Large field inflation and string moduli stabilization}}, {\emph{PoS}
  {\bfseries PLANCK2015} (2015) 021},
  [\href{https://arxiv.org/abs/1510.04059}{{\ttfamily 1510.04059}}].

\bibitem{Blumenhagen:2015xpa}
R.~Blumenhagen, C.~Damian, A.~Font, D.~Herschmann and R.~Sun, \emph{{The
  Flux-Scaling Scenario: De Sitter Uplift and Axion Inflation}},
  \href{https://doi.org/10.1002/prop.201600030}{\emph{Fortsch. Phys.}
  {\bfseries 64} (2016) 536--550},
  [\href{https://arxiv.org/abs/1510.01522}{{\ttfamily 1510.01522}}].

\bibitem{Li:2015taa}
T.~Li, Z.~Li and D.~V. Nanopoulos, \emph{{Helical Phase Inflation via
  Non-Geometric Flux Compactifications: from Natural to Starobinsky-like
  Inflation}}, \href{https://doi.org/10.1007/JHEP10(2015)138}{\emph{JHEP}
  {\bfseries 10} (2015) 138},
  [\href{https://arxiv.org/abs/1507.04687}{{\ttfamily 1507.04687}}].

\bibitem{Plauschinn:2020ram}
E.~Plauschinn, \emph{{Moduli Stabilization with Non-Geometric Fluxes
  \textemdash{} Comments on Tadpole Contributions and de-Sitter Vacua}},
  \href{https://doi.org/10.1002/prop.202100003}{\emph{Fortsch. Phys.}
  {\bfseries 69} (2021) 2100003},
  [\href{https://arxiv.org/abs/2011.08227}{{\ttfamily 2011.08227}}].

\bibitem{Shukla:2022srx}
P.~Shukla, \emph{{On stable type IIA de-Sitter vacua with geometric flux}},
  \href{https://doi.org/10.1140/epjc/s10052-023-11361-w}{\emph{Eur. Phys. J. C}
  {\bfseries 83} (2023) 196},
  [\href{https://arxiv.org/abs/2202.12840}{{\ttfamily 2202.12840}}].

\bibitem{Damian:2023ote}
C.~Damian and O.~Loaiza-Brito, \emph{{Galois groups of uplifted de Sitter
  vacua}}, \href{https://doi.org/10.1016/j.aop.2024.169697}{\emph{Annals Phys.}
  {\bfseries 467} (2024) 169697},
  [\href{https://arxiv.org/abs/2307.08468}{{\ttfamily 2307.08468}}].

\bibitem{Plauschinn:2018wbo}
E.~Plauschinn, \emph{{Non-geometric backgrounds in string theory}},
  \href{https://doi.org/10.1016/j.physrep.2018.12.002}{\emph{Phys. Rept.}
  {\bfseries 798} (2019) 1--122},
  [\href{https://arxiv.org/abs/1811.11203}{{\ttfamily 1811.11203}}].

\bibitem{Shukla:2019wfo}
P.~Shukla, \emph{{Dictionary for the type II nongeometric flux
  compactifications}},
  \href{https://doi.org/10.1103/PhysRevD.103.086009}{\emph{Phys. Rev. D}
  {\bfseries 103} (2021) 086009},
  [\href{https://arxiv.org/abs/1909.07391}{{\ttfamily 1909.07391}}].

\bibitem{Derendinger:2004jn}
J.-P. Derendinger, C.~Kounnas, P.~M. Petropoulos and F.~Zwirner,
  \emph{{Superpotentials in IIA compactifications with general fluxes}},
  \href{https://doi.org/10.1016/j.nuclphysb.2005.02.038}{\emph{Nucl.Phys.}
  {\bfseries B715} (2005) 211--233},
  [\href{https://arxiv.org/abs/hep-th/0411276}{{\ttfamily hep-th/0411276}}].

\bibitem{Shelton:2005cf}
J.~Shelton, W.~Taylor and B.~Wecht, \emph{{Nongeometric flux
  compactifications}},
  \href{https://doi.org/10.1088/1126-6708/2005/10/085}{\emph{JHEP} {\bfseries
  0510} (2005) 085}, [\href{https://arxiv.org/abs/hep-th/0508133}{{\ttfamily
  hep-th/0508133}}].

\bibitem{Grana:2012rr}
M.~Gra\~{n}a and D.~Marques, \emph{{Gauged Double Field Theory}},
  \href{https://doi.org/10.1007/JHEP04(2012)020}{\emph{JHEP} {\bfseries 1204}
  (2012) 020}, [\href{https://arxiv.org/abs/1201.2924}{{\ttfamily 1201.2924}}].

\bibitem{Dibitetto:2012rk}
G.~Dibitetto, J.~Fernandez-Melgarejo, D.~Marques and D.~Roest, \emph{{Duality
  orbits of non-geometric fluxes}},
  \href{https://doi.org/10.1002/prop.201200078}{\emph{Fortsch.Phys.} {\bfseries
  60} (2012) 1123--1149}, [\href{https://arxiv.org/abs/1203.6562}{{\ttfamily
  1203.6562}}].

\bibitem{Ihl:2007ah}
M.~Ihl, D.~Robbins and T.~Wrase, \emph{{Toroidal orientifolds in IIA with
  general NS-NS fluxes}},
  \href{https://doi.org/10.1088/1126-6708/2007/08/043}{\emph{JHEP} {\bfseries
  0708} (2007) 043}, [\href{https://arxiv.org/abs/0705.3410}{{\ttfamily
  0705.3410}}].

\bibitem{Danielsson:2009ff}
U.~H. Danielsson, S.~S. Haque, G.~Shiu and T.~Van~Riet, \emph{{Towards
  Classical de Sitter Solutions in String Theory}},
  \href{https://doi.org/10.1088/1126-6708/2009/09/114}{\emph{JHEP} {\bfseries
  09} (2009) 114}, [\href{https://arxiv.org/abs/0907.2041}{{\ttfamily
  0907.2041}}].

\bibitem{Blaback:2015zra}
J.~Blaback, U.~H. Danielsson, G.~Dibitetto and S.~C. Vargas, \emph{{Universal
  dS vacua in STU-models}},
  \href{https://doi.org/10.1007/JHEP10(2015)069}{\emph{JHEP} {\bfseries 10}
  (2015) 069}, [\href{https://arxiv.org/abs/1505.04283}{{\ttfamily
  1505.04283}}].

\bibitem{Dibitetto:2011qs}
G.~Dibitetto, A.~Guarino and D.~Roest, \emph{{Vacua Analysis in Extended
  Supersymmetry Compactifications}},
  \href{https://doi.org/10.1002/prop.201200004}{\emph{Fortsch. Phys.}
  {\bfseries 60} (2012) 987--990},
  [\href{https://arxiv.org/abs/1112.1306}{{\ttfamily 1112.1306}}].

\bibitem{Kachru:2003aw}
S.~Kachru, R.~Kallosh, A.~D. Linde and S.~P. Trivedi, \emph{{De Sitter vacua in
  string theory}},
  \href{https://doi.org/10.1103/PhysRevD.68.046005}{\emph{Phys. Rev.}
  {\bfseries D68} (2003) 046005},
  [\href{https://arxiv.org/abs/hep-th/0301240}{{\ttfamily hep-th/0301240}}].

\bibitem{Balasubramanian:2005zx}
V.~Balasubramanian, P.~Berglund, J.~P. Conlon and F.~Quevedo,
  \emph{{Systematics of moduli stabilisation in Calabi-Yau flux
  compactifications}},
  \href{https://doi.org/10.1088/1126-6708/2005/03/007}{\emph{JHEP} {\bfseries
  03} (2005) 007}, [\href{https://arxiv.org/abs/hep-th/0502058}{{\ttfamily
  hep-th/0502058}}].

\bibitem{Angelantonj:2002ct}
C.~Angelantonj and A.~Sagnotti, \emph{{Open strings}},
  \href{https://doi.org/10.1016/S0370-1573(02)00273-9}{\emph{Phys. Rept.}
  {\bfseries 371} (2002) 1--150},
  [\href{https://arxiv.org/abs/hep-th/0204089}{{\ttfamily hep-th/0204089}}].

\bibitem{Sagnotti:1987tw}
A.~Sagnotti, \emph{{Open Strings and their Symmetry Groups}},  in \emph{{NATO
  Advanced Summer Institute on Nonperturbative Quantum Field Theory (Cargese
  Summer Institute)}}, 9, 1987,
  \href{https://arxiv.org/abs/hep-th/0208020}{{\ttfamily hep-th/0208020}}.

\bibitem{Grana:2005jc}
M.~Grana, \emph{{Flux compactifications in string theory: A Comprehensive
  review}}, \href{https://doi.org/10.1016/j.physrep.2005.10.008}{\emph{Phys.
  Rept.} {\bfseries 423} (2006) 91--158},
  [\href{https://arxiv.org/abs/hep-th/0509003}{{\ttfamily hep-th/0509003}}].

\bibitem{Blumenhagen:2006ci}
R.~Blumenhagen, B.~Kors, D.~Lust and S.~Stieberger, \emph{{Four-dimensional
  String Compactifications with D-Branes, Orientifolds and Fluxes}},
  \href{https://doi.org/10.1016/j.physrep.2007.04.003}{\emph{Phys. Rept.}
  {\bfseries 445} (2007) 1--193},
  [\href{https://arxiv.org/abs/hep-th/0610327}{{\ttfamily hep-th/0610327}}].

\bibitem{Douglas:2006es}
M.~R. Douglas and S.~Kachru, \emph{{Flux compactification}},
  \href{https://doi.org/10.1103/RevModPhys.79.733}{\emph{Rev. Mod. Phys.}
  {\bfseries 79} (2007) 733--796},
  [\href{https://arxiv.org/abs/hep-th/0610102}{{\ttfamily hep-th/0610102}}].

\bibitem{Denef:2005mm}
F.~Denef, M.~R. Douglas, B.~Florea, A.~Grassi and S.~Kachru, \emph{{Fixing all
  moduli in a simple f-theory compactification}},
  \href{https://doi.org/10.4310/ATMP.2005.v9.n6.a1}{\emph{Adv. Theor. Math.
  Phys.} {\bfseries 9} (2005) 861--929},
  [\href{https://arxiv.org/abs/hep-th/0503124}{{\ttfamily hep-th/0503124}}].

\bibitem{Blumenhagen:2007sm}
R.~Blumenhagen, S.~Moster and E.~Plauschinn, \emph{{Moduli Stabilisation versus
  Chirality for MSSM like Type IIB Orientifolds}},
  \href{https://doi.org/10.1088/1126-6708/2008/01/058}{\emph{JHEP} {\bfseries
  01} (2008) 058}, [\href{https://arxiv.org/abs/0711.3389}{{\ttfamily
  0711.3389}}].

\bibitem{Blumenhagen:2013hva}
R.~Blumenhagen, X.~Gao, D.~Herschmann and P.~Shukla, \emph{{Dimensional
  Oxidation of Non-geometric Fluxes in Type II Orientifolds}},
  \href{https://doi.org/10.1007/JHEP10(2013)201}{\emph{JHEP} {\bfseries 1310}
  (2013) 201}, [\href{https://arxiv.org/abs/1306.2761}{{\ttfamily 1306.2761}}].

\bibitem{Villadoro:2005cu}
G.~Villadoro and F.~Zwirner, \emph{{N=1 effective potential from dual type-IIA
  D6/O6 orientifolds with general fluxes}},
  \href{https://doi.org/10.1088/1126-6708/2005/06/047}{\emph{JHEP} {\bfseries
  0506} (2005) 047}, [\href{https://arxiv.org/abs/hep-th/0503169}{{\ttfamily
  hep-th/0503169}}].

\bibitem{Robbins:2007yv}
D.~Robbins and T.~Wrase, \emph{{D-terms from generalized NS-NS fluxes in type
  II}}, \href{https://doi.org/10.1088/1126-6708/2007/12/058}{\emph{JHEP}
  {\bfseries 0712} (2007) 058},
  [\href{https://arxiv.org/abs/0709.2186}{{\ttfamily 0709.2186}}].

\bibitem{Gao:2015nra}
X.~Gao and P.~Shukla, \emph{{Dimensional oxidation and modular completion of
  non-geometric type IIB action}},
  \href{https://doi.org/10.1007/JHEP05(2015)018}{\emph{JHEP} {\bfseries 1505}
  (2015) 018}, [\href{https://arxiv.org/abs/1501.07248}{{\ttfamily
  1501.07248}}].

\bibitem{Shukla:2015rua}
P.~Shukla, \emph{{On modular completion of generalized flux orbits}},
  \href{https://doi.org/10.1007/JHEP11(2015)075}{\emph{JHEP} {\bfseries 11}
  (2015) 075}, [\href{https://arxiv.org/abs/1505.00544}{{\ttfamily
  1505.00544}}].

\bibitem{Shukla:2015bca}
P.~Shukla, \emph{{Implementing odd-axions in dimensional oxidation of 4D
  non-geometric type IIB scalar potential}},
  \href{https://doi.org/10.1016/j.nuclphysb.2015.11.020}{\emph{Nucl. Phys.}
  {\bfseries B902} (2016) 458--482},
  [\href{https://arxiv.org/abs/1507.01612}{{\ttfamily 1507.01612}}].

\bibitem{Gao:2017gxk}
X.~Gao, P.~Shukla and R.~Sun, \emph{{Symplectic formulation of the type IIA
  nongeometric scalar potential}},
  \href{https://doi.org/10.1103/PhysRevD.98.046009}{\emph{Phys. Rev.}
  {\bfseries D98} (2018) 046009},
  [\href{https://arxiv.org/abs/1712.07310}{{\ttfamily 1712.07310}}].

\bibitem{Leontaris:2023lfc}
G.~K. Leontaris and P.~Shukla, \emph{{Taxonomy of scalar potential with U-dual
  fluxes}}, \href{https://doi.org/10.1103/PhysRevD.108.126020}{\emph{Phys. Rev.
  D} {\bfseries 108} (2023) 126020},
  [\href{https://arxiv.org/abs/2308.15529}{{\ttfamily 2308.15529}}].

\bibitem{Leontaris:2023mmm}
G.~K. Leontaris and P.~Shukla, \emph{{Symplectic formulation of the type IIB
  scalar potential with U-dual fluxes}},
  \href{https://doi.org/10.1103/PhysRevD.109.066018}{\emph{Phys. Rev. D}
  {\bfseries 109} (2024) 066018},
  [\href{https://arxiv.org/abs/2309.08664}{{\ttfamily 2309.08664}}].

\bibitem{Ceresole:1995ca}
A.~Ceresole, R.~D'Auria and S.~Ferrara, \emph{{The Symplectic structure of N=2
  supergravity and its central extension}},
  \href{https://doi.org/10.1016/0920-5632(96)00008-4}{\emph{Nucl.Phys.Proc.Suppl.}
  {\bfseries 46} (1996) 67--74},
  [\href{https://arxiv.org/abs/hep-th/9509160}{{\ttfamily hep-th/9509160}}].

\bibitem{Taylor:1999ii}
T.~R. Taylor and C.~Vafa, \emph{{R R flux on Calabi-Yau and partial
  supersymmetry breaking}},
  \href{https://doi.org/10.1016/S0370-2693(00)00005-8}{\emph{Phys.Lett.}
  {\bfseries B474} (2000) 130--137},
  [\href{https://arxiv.org/abs/hep-th/9912152}{{\ttfamily hep-th/9912152}}].

\bibitem{D'Auria:2007ay}
R.~D'Auria, S.~Ferrara and M.~Trigiante, \emph{{On the supergravity formulation
  of mirror symmetry in generalized Calabi-Yau manifolds}},
  \href{https://doi.org/10.1016/j.nuclphysb.2007.04.009}{\emph{Nucl. Phys.}
  {\bfseries B780} (2007) 28--39},
  [\href{https://arxiv.org/abs/hep-th/0701247}{{\ttfamily hep-th/0701247}}].

\bibitem{Shukla:2015hpa}
P.~Shukla, \emph{{A symplectic rearrangement of the four dimensional
  non-geometric scalar potential}},
  \href{https://doi.org/10.1007/JHEP11(2015)162}{\emph{JHEP} {\bfseries 11}
  (2015) 162}, [\href{https://arxiv.org/abs/1508.01197}{{\ttfamily
  1508.01197}}].

\bibitem{Blumenhagen:2015lta}
R.~Blumenhagen, A.~Font and E.~Plauschinn, \emph{{Relating double field theory
  to the scalar potential of N = 2 gauged supergravity}},
  \href{https://doi.org/10.1007/JHEP12(2015)122}{\emph{JHEP} {\bfseries 12}
  (2015) 122}, [\href{https://arxiv.org/abs/1507.08059}{{\ttfamily
  1507.08059}}].

\bibitem{Shukla:2016hyy}
P.~Shukla, \emph{{Reading off the nongeometric scalar potentials via the
  topological data of the compactifying Calabi-Yau manifolds}},
  \href{https://doi.org/10.1103/PhysRevD.94.086003}{\emph{Phys. Rev.}
  {\bfseries D94} (2016) 086003},
  [\href{https://arxiv.org/abs/1603.01290}{{\ttfamily 1603.01290}}].

\bibitem{Shukla:2019dqd}
P.~Shukla, \emph{{$T$-dualizing de Sitter no-go scenarios}},
  \href{https://doi.org/10.1103/PhysRevD.102.026014}{\emph{Phys. Rev. D}
  {\bfseries 102} (2020) 026014},
  [\href{https://arxiv.org/abs/1909.08630}{{\ttfamily 1909.08630}}].

\bibitem{Shukla:2019akv}
P.~Shukla, \emph{{Rigid nongeometric orientifolds and the swampland}},
  \href{https://doi.org/10.1103/PhysRevD.103.086010}{\emph{Phys. Rev. D}
  {\bfseries 103} (2021) 086010},
  [\href{https://arxiv.org/abs/1909.10993}{{\ttfamily 1909.10993}}].

\bibitem{Biswas:2024ewk}
S.~Biswas, G.~K. Leontaris and P.~Shukla, \emph{{Reading-off the non-geometric
  scalar potentials with U-dual fluxes}},
  \href{https://doi.org/10.1007/JHEPxx(2025)xxx}{\emph{JHEP} {\bfseries xx}
  (2025) xxx}, [\href{https://arxiv.org/abs/2407.15822}{{\ttfamily
  2407.15822}}].

\bibitem{Marchesano:2020uqz}
F.~Marchesano, D.~Prieto, J.~Quirant and P.~Shukla, \emph{{Systematics of Type
  IIA moduli stabilisation}},
  \href{https://doi.org/10.1007/JHEP11(2020)113}{\emph{JHEP} {\bfseries 11}
  (2020) 113}, [\href{https://arxiv.org/abs/2007.00672}{{\ttfamily
  2007.00672}}].

\bibitem{Prieto:2024shz}
D.~Prieto, J.~Quirant and P.~Shukla, \emph{{On the limitations of non-geometric
  fluxes to realize dS vacua}},
  \href{https://doi.org/10.1007/JHEP05(2024)008}{\emph{JHEP} {\bfseries 05}
  (2024) 008}, [\href{https://arxiv.org/abs/2402.13899}{{\ttfamily
  2402.13899}}].

\bibitem{Marchesano:2021gyv}
F.~Marchesano, D.~Prieto and M.~Wiesner, \emph{{F-theory flux vacua at large
  complex structure}},
  \href{https://doi.org/10.1007/JHEP08(2021)077}{\emph{JHEP} {\bfseries 08}
  (2021) 077}, [\href{https://arxiv.org/abs/2105.09326}{{\ttfamily
  2105.09326}}].

\bibitem{Gukov:1999ya}
S.~Gukov, C.~Vafa and E.~Witten, \emph{{CFT's from Calabi-Yau four folds}},
  \href{https://doi.org/10.1016/S0550-3213(01)00289-9,
  10.1016/S0550-3213(00)00373-4}{\emph{Nucl. Phys.} {\bfseries B584} (2000)
  69--108}, [\href{https://arxiv.org/abs/hep-th/9906070}{{\ttfamily
  hep-th/9906070}}].

\bibitem{Dasgupta:1999ss}
K.~Dasgupta, G.~Rajesh and S.~Sethi, \emph{{M theory, orientifolds and G -
  flux}}, \href{https://doi.org/10.1088/1126-6708/1999/08/023}{\emph{JHEP}
  {\bfseries 08} (1999) 023},
  [\href{https://arxiv.org/abs/hep-th/9908088}{{\ttfamily hep-th/9908088}}].

\bibitem{Blumenhagen:2003vr}
R.~Blumenhagen, D.~Lust and T.~R. Taylor, \emph{{Moduli stabilization in chiral
  type IIB orientifold models with fluxes}},
  \href{https://doi.org/10.1016/S0550-3213(03)00392-4}{\emph{Nucl.Phys.}
  {\bfseries B663} (2003) 319--342},
  [\href{https://arxiv.org/abs/hep-th/0303016}{{\ttfamily hep-th/0303016}}].

\bibitem{Aldazabal:2008zza}
G.~Aldazabal, P.~G. Camara and J.~Rosabal, \emph{{Flux algebra, Bianchi
  identities and Freed-Witten anomalies in F-theory compactifications}},
  \href{https://doi.org/10.1016/j.nuclphysb.2009.01.006}{\emph{Nucl.Phys.}
  {\bfseries B814} (2009) 21--52},
  [\href{https://arxiv.org/abs/0811.2900}{{\ttfamily 0811.2900}}].

\bibitem{Font:2008vd}
A.~Font, A.~Guarino and J.~M. Moreno, \emph{{Algebras and non-geometric flux
  vacua}}, \href{https://doi.org/10.1088/1126-6708/2008/12/050}{\emph{JHEP}
  {\bfseries 0812} (2008) 050},
  [\href{https://arxiv.org/abs/0809.3748}{{\ttfamily 0809.3748}}].

\bibitem{Guarino:2008ik}
A.~Guarino and G.~J. Weatherill, \emph{{Non-geometric flux vacua, S-duality and
  algebraic geometry}},
  \href{https://doi.org/10.1088/1126-6708/2009/02/042}{\emph{JHEP} {\bfseries
  0902} (2009) 042}, [\href{https://arxiv.org/abs/0811.2190}{{\ttfamily
  0811.2190}}].

\bibitem{Hull:2004in}
C.~Hull, \emph{{A Geometry for non-geometric string backgrounds}},
  \href{https://doi.org/10.1088/1126-6708/2005/10/065}{\emph{JHEP} {\bfseries
  0510} (2005) 065}, [\href{https://arxiv.org/abs/hep-th/0406102}{{\ttfamily
  hep-th/0406102}}].

\bibitem{Kumar:1996zx}
A.~Kumar and C.~Vafa, \emph{{U manifolds}},
  \href{https://doi.org/10.1016/S0370-2693(97)00108-1}{\emph{Phys.Lett.}
  {\bfseries B396} (1997) 85--90},
  [\href{https://arxiv.org/abs/hep-th/9611007}{{\ttfamily hep-th/9611007}}].

\bibitem{Hull:2003kr}
C.~M. Hull and A.~Catal-Ozer, \emph{{Compactifications with S duality twists}},
  \href{https://doi.org/10.1088/1126-6708/2003/10/034}{\emph{JHEP} {\bfseries
  0310} (2003) 034}, [\href{https://arxiv.org/abs/hep-th/0308133}{{\ttfamily
  hep-th/0308133}}].

\bibitem{Aldazabal:2010ef}
G.~Aldazabal, E.~Andres, P.~G. Camara and M.~Grana, \emph{{U-dual fluxes and
  Generalized Geometry}},
  \href{https://doi.org/10.1007/JHEP11(2010)083}{\emph{JHEP} {\bfseries 11}
  (2010) 083}, [\href{https://arxiv.org/abs/1007.5509}{{\ttfamily 1007.5509}}].

\bibitem{Lombardo:2016swq}
D.~M. Lombardo, F.~Riccioni and S.~Risoli, \emph{{$P$ fluxes and exotic
  branes}}, \href{https://doi.org/10.1007/JHEP12(2016)114}{\emph{JHEP}
  {\bfseries 12} (2016) 114},
  [\href{https://arxiv.org/abs/1610.07975}{{\ttfamily 1610.07975}}].

\bibitem{Lombardo:2017yme}
D.~M. Lombardo, F.~Riccioni and S.~Risoli, \emph{{Non-geometric fluxes \&
  tadpole conditions for exotic branes}},
  \href{https://doi.org/10.1007/JHEP10(2017)134}{\emph{JHEP} {\bfseries 10}
  (2017) 134}, [\href{https://arxiv.org/abs/1704.08566}{{\ttfamily
  1704.08566}}].

\bibitem{Shukla:2016xdy}
P.~Shukla, \emph{{Revisiting the two formulations of Bianchi identities and
  their implications on moduli stabilization}},
  \href{https://doi.org/10.1007/JHEP08(2016)146}{\emph{JHEP} {\bfseries 08}
  (2016) 146}, [\href{https://arxiv.org/abs/1603.08545}{{\ttfamily
  1603.08545}}].

\bibitem{Plauschinn:2021hkp}
E.~Plauschinn, \emph{{The tadpole conjecture at large complex-structure}},
  \href{https://doi.org/10.1007/JHEP02(2022)206}{\emph{JHEP} {\bfseries 02}
  (2022) 206}, [\href{https://arxiv.org/abs/2109.00029}{{\ttfamily
  2109.00029}}].

\bibitem{Maldacena:2000mw}
J.~M. Maldacena and C.~Nunez, \emph{{Supergravity description of field theories
  on curved manifolds and a no go theorem}},
  \href{https://doi.org/10.1142/S0217751X01003935,
  10.1142/S0217751X01003937}{\emph{Int. J. Mod. Phys.} {\bfseries A16} (2001)
  822--855}, [\href{https://arxiv.org/abs/hep-th/0007018}{{\ttfamily
  hep-th/0007018}}].

\bibitem{Hertzberg:2007wc}
M.~P. Hertzberg, S.~Kachru, W.~Taylor and M.~Tegmark, \emph{{Inflationary
  Constraints on Type IIA String Theory}},
  \href{https://doi.org/10.1088/1126-6708/2007/12/095}{\emph{JHEP} {\bfseries
  12} (2007) 095}, [\href{https://arxiv.org/abs/0711.2512}{{\ttfamily
  0711.2512}}].

\bibitem{Hertzberg:2007ke}
M.~P. Hertzberg, M.~Tegmark, S.~Kachru, J.~Shelton and O.~Ozcan,
  \emph{{Searching for Inflation in Simple String Theory Models: An
  Astrophysical Perspective}},
  \href{https://doi.org/10.1103/PhysRevD.76.103521}{\emph{Phys. Rev.}
  {\bfseries D76} (2007) 103521},
  [\href{https://arxiv.org/abs/0709.0002}{{\ttfamily 0709.0002}}].

\bibitem{Haque:2008jz}
S.~S. Haque, G.~Shiu, B.~Underwood and T.~Van~Riet, \emph{{Minimal simple de
  Sitter solutions}},
  \href{https://doi.org/10.1103/PhysRevD.79.086005}{\emph{Phys. Rev.}
  {\bfseries D79} (2009) 086005},
  [\href{https://arxiv.org/abs/0810.5328}{{\ttfamily 0810.5328}}].

\bibitem{Flauger:2008ad}
R.~Flauger, S.~Paban, D.~Robbins and T.~Wrase, \emph{{Searching for slow-roll
  moduli inflation in massive type IIA supergravity with metric fluxes}},
  \href{https://doi.org/10.1103/PhysRevD.79.086011}{\emph{Phys. Rev.}
  {\bfseries D79} (2009) 086011},
  [\href{https://arxiv.org/abs/0812.3886}{{\ttfamily 0812.3886}}].

\bibitem{Caviezel:2008tf}
C.~Caviezel, P.~Koerber, S.~Kors, D.~Lust, T.~Wrase and M.~Zagermann, \emph{{On
  the Cosmology of Type IIA Compactifications on SU(3)-structure Manifolds}},
  \href{https://doi.org/10.1088/1126-6708/2009/04/010}{\emph{JHEP} {\bfseries
  04} (2009) 010}, [\href{https://arxiv.org/abs/0812.3551}{{\ttfamily
  0812.3551}}].

\bibitem{Covi:2008ea}
L.~Covi, M.~Gomez-Reino, C.~Gross, J.~Louis, G.~A. Palma and C.~A. Scrucca,
  \emph{{de Sitter vacua in no-scale supergravities and Calabi-Yau string
  models}}, \href{https://doi.org/10.1088/1126-6708/2008/06/057}{\emph{JHEP}
  {\bfseries 06} (2008) 057},
  [\href{https://arxiv.org/abs/0804.1073}{{\ttfamily 0804.1073}}].

\bibitem{deCarlos:2009fq}
B.~de~Carlos, A.~Guarino and J.~M. Moreno, \emph{{Flux moduli stabilisation,
  Supergravity algebras and no-go theorems}},
  \href{https://doi.org/10.1007/JHEP01(2010)012}{\emph{JHEP} {\bfseries 01}
  (2010) 012}, [\href{https://arxiv.org/abs/0907.5580}{{\ttfamily 0907.5580}}].

\bibitem{Caviezel:2009tu}
C.~Caviezel, T.~Wrase and M.~Zagermann, \emph{{Moduli Stabilization and
  Cosmology of Type IIB on SU(2)-Structure Orientifolds}},
  \href{https://doi.org/10.1007/JHEP04(2010)011}{\emph{JHEP} {\bfseries 04}
  (2010) 011}, [\href{https://arxiv.org/abs/0912.3287}{{\ttfamily 0912.3287}}].

\bibitem{Danielsson:2010bc}
U.~H. Danielsson, P.~Koerber and T.~Van~Riet, \emph{{Universal de Sitter
  solutions at tree-level}},
  \href{https://doi.org/10.1007/JHEP05(2010)090}{\emph{JHEP} {\bfseries 05}
  (2010) 090}, [\href{https://arxiv.org/abs/1003.3590}{{\ttfamily 1003.3590}}].

\bibitem{Wrase:2010ew}
T.~Wrase and M.~Zagermann, \emph{{On Classical de Sitter Vacua in String
  Theory}}, \href{https://doi.org/10.1002/prop.201000053}{\emph{Fortsch. Phys.}
  {\bfseries 58} (2010) 906--910},
  [\href{https://arxiv.org/abs/1003.0029}{{\ttfamily 1003.0029}}].

\bibitem{Shiu:2011zt}
G.~Shiu and Y.~Sumitomo, \emph{{Stability Constraints on Classical de Sitter
  Vacua}}, \href{https://doi.org/10.1007/JHEP09(2011)052}{\emph{JHEP}
  {\bfseries 09} (2011) 052},
  [\href{https://arxiv.org/abs/1107.2925}{{\ttfamily 1107.2925}}].

\bibitem{McOrist:2012yc}
J.~McOrist and S.~Sethi, \emph{{M-theory and Type IIA Flux Compactifications}},
  \href{https://doi.org/10.1007/JHEP12(2012)122}{\emph{JHEP} {\bfseries 12}
  (2012) 122}, [\href{https://arxiv.org/abs/1208.0261}{{\ttfamily 1208.0261}}].

\bibitem{Dasgupta:2014pma}
K.~Dasgupta, R.~Gwyn, E.~McDonough, M.~Mia and R.~Tatar, \emph{{de Sitter Vacua
  in Type IIB String Theory: Classical Solutions and Quantum Corrections}},
  \href{https://doi.org/10.1007/JHEP07(2014)054}{\emph{JHEP} {\bfseries 07}
  (2014) 054}, [\href{https://arxiv.org/abs/1402.5112}{{\ttfamily 1402.5112}}].

\bibitem{Gautason:2015tig}
F.~F. Gautason, M.~Schillo, T.~Van~Riet and M.~Williams, \emph{{Remarks on
  scale separation in flux vacua}},
  \href{https://doi.org/10.1007/JHEP03(2016)061}{\emph{JHEP} {\bfseries 03}
  (2016) 061}, [\href{https://arxiv.org/abs/1512.00457}{{\ttfamily
  1512.00457}}].

\bibitem{Junghans:2016uvg}
D.~Junghans, \emph{{Tachyons in Classical de Sitter Vacua}},
  \href{https://doi.org/10.1007/JHEP06(2016)132}{\emph{JHEP} {\bfseries 06}
  (2016) 132}, [\href{https://arxiv.org/abs/1603.08939}{{\ttfamily
  1603.08939}}].

\bibitem{Andriot:2016xvq}
D.~Andriot and J.~Blaback, \emph{{Refining the boundaries of the classical de
  Sitter landscape}}, \href{https://doi.org/10.1007/JHEP03(2017)102,
  10.1007/JHEP03(2018)083}{\emph{JHEP} {\bfseries 03} (2017) 102},
  [\href{https://arxiv.org/abs/1609.00385}{{\ttfamily 1609.00385}}].

\bibitem{Andriot:2017jhf}
D.~Andriot, \emph{{On classical de Sitter and Minkowski solutions with
  intersecting branes}},
  \href{https://doi.org/10.1007/JHEP03(2018)054}{\emph{JHEP} {\bfseries 03}
  (2018) 054}, [\href{https://arxiv.org/abs/1710.08886}{{\ttfamily
  1710.08886}}].

\bibitem{Ooguri:2006in}
H.~Ooguri and C.~Vafa, \emph{{On the Geometry of the String Landscape and the
  Swampland}},
  \href{https://doi.org/10.1016/j.nuclphysb.2006.10.033}{\emph{Nucl. Phys.}
  {\bfseries B766} (2007) 21--33},
  [\href{https://arxiv.org/abs/hep-th/0605264}{{\ttfamily hep-th/0605264}}].

\bibitem{Obied:2018sgi}
G.~Obied, H.~Ooguri, L.~Spodyneiko and C.~Vafa, \emph{{De Sitter Space and the
  Swampland}},  \href{https://arxiv.org/abs/1806.08362}{{\ttfamily
  1806.08362}}.

\bibitem{Danielsson:2018ztv}
U.~H. Danielsson and T.~Van~Riet, \emph{{What if string theory has no de Sitter
  vacua?}}, \href{https://doi.org/10.1142/S0218271818300070}{\emph{Int. J. Mod.
  Phys.} {\bfseries D27} (2018) 1830007},
  [\href{https://arxiv.org/abs/1804.01120}{{\ttfamily 1804.01120}}].

\bibitem{Burgess:2003ic}
C.~P. Burgess, R.~Kallosh and F.~Quevedo, \emph{{De Sitter string vacua from
  supersymmetric D terms}},
  \href{https://doi.org/10.1088/1126-6708/2003/10/056}{\emph{JHEP} {\bfseries
  10} (2003) 056}, [\href{https://arxiv.org/abs/hep-th/0309187}{{\ttfamily
  hep-th/0309187}}].

\bibitem{Achucarro:2006zf}
A.~Achucarro, B.~de~Carlos, J.~A. Casas and L.~Doplicher, \emph{{De Sitter
  vacua from uplifting D-terms in effective supergravities from realistic
  strings}}, \href{https://doi.org/10.1088/1126-6708/2006/06/014}{\emph{JHEP}
  {\bfseries 06} (2006) 014},
  [\href{https://arxiv.org/abs/hep-th/0601190}{{\ttfamily hep-th/0601190}}].

\bibitem{Westphal:2006tn}
A.~Westphal, \emph{{de Sitter string vacua from Kahler uplifting}},
  \href{https://doi.org/10.1088/1126-6708/2007/03/102}{\emph{JHEP} {\bfseries
  03} (2007) 102}, [\href{https://arxiv.org/abs/hep-th/0611332}{{\ttfamily
  hep-th/0611332}}].

\bibitem{Silverstein:2007ac}
E.~Silverstein, \emph{{Simple de Sitter Solutions}},
  \href{https://doi.org/10.1103/PhysRevD.77.106006}{\emph{Phys. Rev.}
  {\bfseries D77} (2008) 106006},
  [\href{https://arxiv.org/abs/0712.1196}{{\ttfamily 0712.1196}}].

\bibitem{Rummel:2011cd}
M.~Rummel and A.~Westphal, \emph{{A sufficient condition for de Sitter vacua in
  type IIB string theory}},
  \href{https://doi.org/10.1007/JHEP01(2012)020}{\emph{JHEP} {\bfseries 01}
  (2012) 020}, [\href{https://arxiv.org/abs/1107.2115}{{\ttfamily 1107.2115}}].

\bibitem{Cicoli:2012fh}
M.~Cicoli, A.~Maharana, F.~Quevedo and C.~P. Burgess, \emph{{De Sitter String
  Vacua from Dilaton-dependent Non-perturbative Effects}},
  \href{https://doi.org/10.1007/JHEP06(2012)011}{\emph{JHEP} {\bfseries 06}
  (2012) 011}, [\href{https://arxiv.org/abs/1203.1750}{{\ttfamily 1203.1750}}].

\bibitem{Louis:2012nb}
J.~Louis, M.~Rummel, R.~Valandro and A.~Westphal, \emph{{Building an explicit
  de Sitter}}, \href{https://doi.org/10.1007/JHEP10(2012)163}{\emph{JHEP}
  {\bfseries 10} (2012) 163},
  [\href{https://arxiv.org/abs/1208.3208}{{\ttfamily 1208.3208}}].

\bibitem{Cicoli:2013cha}
M.~Cicoli, D.~Klevers, S.~Krippendorf, C.~Mayrhofer, F.~Quevedo and
  R.~Valandro, \emph{{Explicit de Sitter Flux Vacua for Global String Models
  with Chiral Matter}},
  \href{https://doi.org/10.1007/JHEP05(2014)001}{\emph{JHEP} {\bfseries 05}
  (2014) 001}, [\href{https://arxiv.org/abs/1312.0014}{{\ttfamily 1312.0014}}].

\bibitem{Cicoli:2015ylx}
M.~Cicoli, F.~Quevedo and R.~Valandro, \emph{{De Sitter from T-branes}},
  \href{https://doi.org/10.1007/JHEP03(2016)141}{\emph{JHEP} {\bfseries 03}
  (2016) 141}, [\href{https://arxiv.org/abs/1512.04558}{{\ttfamily
  1512.04558}}].

\bibitem{Cicoli:2017shd}
M.~Cicoli, I.~Garcìa-Etxebarria, C.~Mayrhofer, F.~Quevedo, P.~Shukla and
  R.~Valandro, \emph{{Global Orientifolded Quivers with Inflation}},
  \href{https://doi.org/10.1007/JHEP11(2017)134}{\emph{JHEP} {\bfseries 11}
  (2017) 134}, [\href{https://arxiv.org/abs/1706.06128}{{\ttfamily
  1706.06128}}].

\bibitem{Akrami:2018ylq}
Y.~Akrami, R.~Kallosh, A.~Linde and V.~Vardanyan, \emph{{The Landscape, the
  Swampland and the Era of Precision Cosmology}},
  \href{https://doi.org/10.1002/prop.201800075}{\emph{Fortsch. Phys.}
  {\bfseries 67} (2019) 1800075},
  [\href{https://arxiv.org/abs/1808.09440}{{\ttfamily 1808.09440}}].

\bibitem{Antoniadis:2018hqy}
I.~Antoniadis, Y.~Chen and G.~K. Leontaris, \emph{{Perturbative moduli
  stabilisation in type IIB/F-theory framework}},
  \href{https://doi.org/10.1140/epjc/s10052-018-6248-4}{\emph{Eur. Phys. J.}
  {\bfseries C78} (2018) 766},
  [\href{https://arxiv.org/abs/1803.08941}{{\ttfamily 1803.08941}}].

\bibitem{Antoniadis:2018ngr}
I.~Antoniadis, Y.~Chen and G.~K. Leontaris, \emph{{Inflation from the internal
  volume in type IIB/F-theory compactification}},
  \href{https://doi.org/10.1142/S0217751X19500428}{\emph{Int. J. Mod. Phys.}
  {\bfseries A34} (2019) 1950042},
  [\href{https://arxiv.org/abs/1810.05060}{{\ttfamily 1810.05060}}].

\bibitem{Antoniadis:2019rkh}
I.~Antoniadis, Y.~Chen and G.~K. Leontaris, \emph{{Logarithmic loop
  corrections, moduli stabilisation and de Sitter vacua in string theory}},
  \href{https://arxiv.org/abs/1909.10525}{{\ttfamily 1909.10525}}.

\bibitem{Basiouris:2020jgp}
V.~Basiouris and G.~K. Leontaris, \emph{{Note on de Sitter vacua from
  perturbative and non-perturbative dynamics in type IIB/F-theory
  compactifications}},
  \href{https://doi.org/10.1016/j.physletb.2020.135809}{\emph{Phys. Lett. B}
  {\bfseries 810} (2020) 135809},
  [\href{https://arxiv.org/abs/2007.15423}{{\ttfamily 2007.15423}}].

\bibitem{Antoniadis:2020stf}
I.~Antoniadis, O.~Lacombe and G.~K. Leontaris, \emph{{Inflation near a
  metastable de Sitter vacuum from moduli stabilisation}},
  \href{https://doi.org/10.1140/epjc/s10052-020-08581-9}{\emph{Eur. Phys. J. C}
  {\bfseries 80} (2020) 1014},
  [\href{https://arxiv.org/abs/2007.10362}{{\ttfamily 2007.10362}}].

\bibitem{Cicoli:2018kdo}
M.~Cicoli, S.~De~Alwis, A.~Maharana, F.~Muia and F.~Quevedo, \emph{{De Sitter
  vs Quintessence in String Theory}},
  \href{https://doi.org/10.1002/prop.201800079}{\emph{Fortsch. Phys.}
  {\bfseries 67} (2019) 1800079},
  [\href{https://arxiv.org/abs/1808.08967}{{\ttfamily 1808.08967}}].

\bibitem{Crino:2020qwk}
C.~Crin\`o, F.~Quevedo and R.~Valandro, \emph{{On de Sitter String Vacua from
  Anti-D3-Branes in the Large Volume Scenario}},
  \href{https://doi.org/10.1007/JHEP03(2021)258}{\emph{JHEP} {\bfseries 03}
  (2021) 258}, [\href{https://arxiv.org/abs/2010.15903}{{\ttfamily
  2010.15903}}].

\bibitem{Cicoli:2021dhg}
M.~Cicoli, I.~n.~G. Etxebarria, F.~Quevedo, A.~Schachner, P.~Shukla and
  R.~Valandro, \emph{{The Standard Model quiver in de Sitter string
  compactifications}},
  \href{https://doi.org/10.1007/JHEP08(2021)109}{\emph{JHEP} {\bfseries 08}
  (2021) 109}, [\href{https://arxiv.org/abs/2106.11964}{{\ttfamily
  2106.11964}}].

\bibitem{Andriot:2022way}
D.~Andriot, L.~Horer and P.~Marconnet, \emph{{Charting the landscape of (anti-)
  de Sitter and Minkowski solutions of 10d supergravities}},
  \href{https://arxiv.org/abs/2201.04152}{{\ttfamily 2201.04152}}.

\bibitem{Heckman:2019dsj}
J.~J. Heckman, C.~Lawrie, L.~Lin, J.~Sakstein and G.~Zoccarato,
  \emph{{Pixelated Dark Energy}},
  \href{https://arxiv.org/abs/1901.10489}{{\ttfamily 1901.10489}}.

\bibitem{Heckman:2018mxl}
J.~J. Heckman, C.~Lawrie, L.~Lin and G.~Zoccarato, \emph{{F-theory and Dark
  Energy}},  \href{https://arxiv.org/abs/1811.01959}{{\ttfamily 1811.01959}}.

\bibitem{Danielsson:2011au}
U.~H. Danielsson, S.~S. Haque, P.~Koerber, G.~Shiu, T.~Van~Riet and T.~Wrase,
  \emph{{De Sitter hunting in a classical landscape}},
  \href{https://doi.org/10.1002/prop.201100047}{\emph{Fortsch. Phys.}
  {\bfseries 59} (2011) 897--933},
  [\href{https://arxiv.org/abs/1103.4858}{{\ttfamily 1103.4858}}].

\bibitem{Chen:2011ac}
X.~Chen, G.~Shiu, Y.~Sumitomo and S.~H.~H. Tye, \emph{{A Global View on The
  Search for de-Sitter Vacua in (type IIA) String Theory}},
  \href{https://doi.org/10.1007/JHEP04(2012)026}{\emph{JHEP} {\bfseries 04}
  (2012) 026}, [\href{https://arxiv.org/abs/1112.3338}{{\ttfamily 1112.3338}}].

\bibitem{Danielsson:2012et}
U.~H. Danielsson, G.~Shiu, T.~Van~Riet and T.~Wrase, \emph{{A note on obstinate
  tachyons in classical dS solutions}},
  \href{https://doi.org/10.1007/JHEP03(2013)138}{\emph{JHEP} {\bfseries 03}
  (2013) 138}, [\href{https://arxiv.org/abs/1212.5178}{{\ttfamily 1212.5178}}].

\bibitem{Garg:2018reu}
S.~K. Garg and C.~Krishnan, \emph{{Bounds on Slow Roll and the de Sitter
  Swampland}},  \href{https://arxiv.org/abs/1807.05193}{{\ttfamily
  1807.05193}}.

\bibitem{Agrawal:2018own}
P.~Agrawal, G.~Obied, P.~J. Steinhardt and C.~Vafa, \emph{{On the Cosmological
  Implications of the String Swampland}},
  \href{https://doi.org/10.1016/j.physletb.2018.07.040}{\emph{Phys. Lett.}
  {\bfseries B784} (2018) 271--276},
  [\href{https://arxiv.org/abs/1806.09718}{{\ttfamily 1806.09718}}].

\bibitem{Andriot:2018wzk}
D.~Andriot, \emph{{On the de Sitter swampland criterion}},
  \href{https://doi.org/10.1016/j.physletb.2018.09.022}{\emph{Phys. Lett.}
  {\bfseries B785} (2018) 570--573},
  [\href{https://arxiv.org/abs/1806.10999}{{\ttfamily 1806.10999}}].

\bibitem{Andriot:2018ept}
D.~Andriot, \emph{{New constraints on classical de Sitter: flirting with the
  swampland}}, \href{https://doi.org/10.1002/prop.201800103}{\emph{Fortsch.
  Phys.} {\bfseries 67} (2019) 1800103},
  [\href{https://arxiv.org/abs/1807.09698}{{\ttfamily 1807.09698}}].

\bibitem{Denef:2018etk}
F.~Denef, A.~Hebecker and T.~Wrase, \emph{{de Sitter swampland conjecture and
  the Higgs potential}},
  \href{https://doi.org/10.1103/PhysRevD.98.086004}{\emph{Phys. Rev.}
  {\bfseries D98} (2018) 086004},
  [\href{https://arxiv.org/abs/1807.06581}{{\ttfamily 1807.06581}}].

\bibitem{Conlon:2018eyr}
J.~P. Conlon, \emph{{The de Sitter swampland conjecture and supersymmetric AdS
  vacua}}, \href{https://doi.org/10.1142/S0217751X18501786}{\emph{Int. J. Mod.
  Phys.} {\bfseries A33} (2018) 1850178},
  [\href{https://arxiv.org/abs/1808.05040}{{\ttfamily 1808.05040}}].

\bibitem{Roupec:2018mbn}
C.~Roupec and T.~Wrase, \emph{{de Sitter Extrema and the Swampland}},
  \href{https://doi.org/10.1002/prop.201800082}{\emph{Fortsch. Phys.}
  {\bfseries 67} (2019) 1800082},
  [\href{https://arxiv.org/abs/1807.09538}{{\ttfamily 1807.09538}}].

\bibitem{Murayama:2018lie}
H.~Murayama, M.~Yamazaki and T.~T. Yanagida, \emph{{Do We Live in the
  Swampland?}}, \href{https://doi.org/10.1007/JHEP12(2018)032}{\emph{JHEP}
  {\bfseries 12} (2018) 032},
  [\href{https://arxiv.org/abs/1809.00478}{{\ttfamily 1809.00478}}].

\bibitem{Choi:2018rze}
K.~Choi, D.~Chway and C.~S. Shin, \emph{{The dS swampland conjecture with the
  electroweak symmetry and QCD chiral symmetry breaking}},
  \href{https://doi.org/10.1007/JHEP11(2018)142}{\emph{JHEP} {\bfseries 11}
  (2018) 142}, [\href{https://arxiv.org/abs/1809.01475}{{\ttfamily
  1809.01475}}].

\bibitem{Hamaguchi:2018vtv}
K.~Hamaguchi, M.~Ibe and T.~Moroi, \emph{{The swampland conjecture and the
  Higgs expectation value}},
  \href{https://doi.org/10.1007/JHEP12(2018)023}{\emph{JHEP} {\bfseries 12}
  (2018) 023}, [\href{https://arxiv.org/abs/1810.02095}{{\ttfamily
  1810.02095}}].

\bibitem{Olguin-Tejo:2018pfq}
Y.~Olguin-Trejo, S.~L. Parameswaran, G.~Tasinato and I.~Zavala, \emph{{Runaway
  Quintessence, Out of the Swampland}},
  \href{https://doi.org/10.1088/1475-7516/2019/01/031}{\emph{JCAP} {\bfseries
  1901} (2019) 031}, [\href{https://arxiv.org/abs/1810.08634}{{\ttfamily
  1810.08634}}].

\bibitem{Blanco-Pillado:2018xyn}
J.~J. Blanco-Pillado, M.~A. Urkiola and J.~M. Wachter, \emph{{Racetrack
  Potentials and the de Sitter Swampland Conjectures}},
  \href{https://doi.org/10.1007/JHEP01(2019)187}{\emph{JHEP} {\bfseries 01}
  (2019) 187}, [\href{https://arxiv.org/abs/1811.05463}{{\ttfamily
  1811.05463}}].

\bibitem{Lin:2018kjm}
C.-M. Lin, K.-W. Ng and K.~Cheung, \emph{{Chaotic inflation on the brane and
  the Swampland Criteria}},
  \href{https://doi.org/10.1103/PhysRevD.100.023545}{\emph{Phys. Rev.}
  {\bfseries D100} (2019) 023545},
  [\href{https://arxiv.org/abs/1810.01644}{{\ttfamily 1810.01644}}].

\bibitem{Han:2018yrk}
C.~Han, S.~Pi and M.~Sasaki, \emph{{Quintessence Saves Higgs Instability}},
  \href{https://doi.org/10.1016/j.physletb.2019.02.037}{\emph{Phys. Lett.}
  {\bfseries B791} (2019) 314--318},
  [\href{https://arxiv.org/abs/1809.05507}{{\ttfamily 1809.05507}}].

\bibitem{Raveri:2018ddi}
M.~Raveri, W.~Hu and S.~Sethi, \emph{{Swampland Conjectures and Late-Time
  Cosmology}}, \href{https://doi.org/10.1103/PhysRevD.99.083518}{\emph{Phys.
  Rev.} {\bfseries D99} (2019) 083518},
  [\href{https://arxiv.org/abs/1812.10448}{{\ttfamily 1812.10448}}].

\bibitem{Dasgupta:2018rtp}
K.~Dasgupta, M.~Emelin, E.~McDonough and R.~Tatar, \emph{{Quantum Corrections
  and the de Sitter Swampland Conjecture}},
  \href{https://doi.org/10.1007/JHEP01(2019)145}{\emph{JHEP} {\bfseries 01}
  (2019) 145}, [\href{https://arxiv.org/abs/1808.07498}{{\ttfamily
  1808.07498}}].

\bibitem{Danielsson:2018qpa}
U.~Danielsson, \emph{{The quantum swampland}},
  \href{https://doi.org/10.1007/JHEP04(2019)095}{\emph{JHEP} {\bfseries 04}
  (2019) 095}, [\href{https://arxiv.org/abs/1809.04512}{{\ttfamily
  1809.04512}}].

\bibitem{Andriolo:2018yrz}
S.~Andriolo, G.~Shiu, H.~Triendl, T.~Van~Riet, G.~Venken and G.~Zoccarato,
  \emph{{Compact G2 holonomy spaces from SU(3) structures}},
  \href{https://doi.org/10.1007/JHEP03(2019)059}{\emph{JHEP} {\bfseries 03}
  (2019) 059}, [\href{https://arxiv.org/abs/1811.00063}{{\ttfamily
  1811.00063}}].

\bibitem{Dasgupta:2019gcd}
K.~Dasgupta, M.~Emelin, M.~M. Faruk and R.~Tatar, \emph{{de Sitter Vacua in the
  String Landscape}},  \href{https://arxiv.org/abs/1908.05288}{{\ttfamily
  1908.05288}}.

\bibitem{Andriot:2019wrs}
D.~Andriot, \emph{{Open problems on classical de Sitter solutions}},
  \href{https://doi.org/10.1002/prop.201900026}{\emph{Fortsch. Phys.}
  {\bfseries 67} (2019) 1900026},
  [\href{https://arxiv.org/abs/1902.10093}{{\ttfamily 1902.10093}}].

\bibitem{Palti:2019pca}
E.~Palti, \emph{{The Swampland: Introduction and Review}},
  \href{https://doi.org/10.1002/prop.201900037}{\emph{Fortsch. Phys.}
  {\bfseries 67} (2019) 1900037},
  [\href{https://arxiv.org/abs/1903.06239}{{\ttfamily 1903.06239}}].

\bibitem{Blumenhagen:2017cxt}
R.~Blumenhagen, I.~Valenzuela and F.~Wolf, \emph{{The Swampland Conjecture and
  F-term Axion Monodromy Inflation}},
  \href{https://doi.org/10.1007/JHEP07(2017)145}{\emph{JHEP} {\bfseries 07}
  (2017) 145}, [\href{https://arxiv.org/abs/1703.05776}{{\ttfamily
  1703.05776}}].

\bibitem{Blumenhagen:2018nts}
R.~Blumenhagen, D.~Kläwer, L.~Schlechter and F.~Wolf, \emph{{The Refined
  Swampland Distance Conjecture in Calabi-Yau Moduli Spaces}},
  \href{https://doi.org/10.1007/JHEP06(2018)052}{\emph{JHEP} {\bfseries 06}
  (2018) 052}, [\href{https://arxiv.org/abs/1803.04989}{{\ttfamily
  1803.04989}}].

\bibitem{Blumenhagen:2018hsh}
R.~Blumenhagen, \emph{{Large Field Inflation/Quintessence and the Refined
  Swampland Distance Conjecture}},
  \href{https://doi.org/10.22323/1.318.0175}{\emph{PoS} {\bfseries CORFU2017}
  (2018) 175}, [\href{https://arxiv.org/abs/1804.10504}{{\ttfamily
  1804.10504}}].

\bibitem{Palti:2017elp}
E.~Palti, \emph{{The Weak Gravity Conjecture and Scalar Fields}},
  \href{https://doi.org/10.1007/JHEP08(2017)034}{\emph{JHEP} {\bfseries 08}
  (2017) 034}, [\href{https://arxiv.org/abs/1705.04328}{{\ttfamily
  1705.04328}}].

\bibitem{Conlon:2016aea}
J.~P. Conlon and S.~Krippendorf, \emph{{Axion decay constants away from the
  lamppost}}, \href{https://doi.org/10.1007/JHEP04(2016)085}{\emph{JHEP}
  {\bfseries 04} (2016) 085},
  [\href{https://arxiv.org/abs/1601.00647}{{\ttfamily 1601.00647}}].

\bibitem{Hebecker:2017lxm}
A.~Hebecker, P.~Henkenjohann and L.~T. Witkowski, \emph{{Flat Monodromies and a
  Moduli Space Size Conjecture}},
  \href{https://doi.org/10.1007/JHEP12(2017)033}{\emph{JHEP} {\bfseries 12}
  (2017) 033}, [\href{https://arxiv.org/abs/1708.06761}{{\ttfamily
  1708.06761}}].

\bibitem{Klaewer:2016kiy}
D.~Klaewer and E.~Palti, \emph{{Super-Planckian Spatial Field Variations and
  Quantum Gravity}}, \href{https://doi.org/10.1007/JHEP01(2017)088}{\emph{JHEP}
  {\bfseries 01} (2017) 088},
  [\href{https://arxiv.org/abs/1610.00010}{{\ttfamily 1610.00010}}].

\bibitem{Baume:2016psm}
F.~Baume and E.~Palti, \emph{{Backreacted Axion Field Ranges in String
  Theory}}, \href{https://doi.org/10.1007/JHEP08(2016)043}{\emph{JHEP}
  {\bfseries 08} (2016) 043},
  [\href{https://arxiv.org/abs/1602.06517}{{\ttfamily 1602.06517}}].

\bibitem{Landete:2018kqf}
A.~Landete and G.~Shiu, \emph{{Mass Hierarchies and Dynamical Field Range}},
  \href{https://doi.org/10.1103/PhysRevD.98.066012}{\emph{Phys. Rev.}
  {\bfseries D98} (2018) 066012},
  [\href{https://arxiv.org/abs/1806.01874}{{\ttfamily 1806.01874}}].

\bibitem{Cicoli:2018tcq}
M.~Cicoli, D.~Ciupke, C.~Mayrhofer and P.~Shukla, \emph{{A Geometrical Upper
  Bound on the Inflaton Range}},
  \href{https://doi.org/10.1007/JHEP05(2018)001}{\emph{JHEP} {\bfseries 05}
  (2018) 001}, [\href{https://arxiv.org/abs/1801.05434}{{\ttfamily
  1801.05434}}].

\bibitem{Font:2019cxq}
A.~Font, A.~Herráez and L.~E. Ibáñez, \emph{{The Swampland Distance
  Conjecture and Towers of Tensionless Branes}},
  \href{https://doi.org/10.1007/JHEP08(2019)044}{\emph{JHEP} {\bfseries 08}
  (2019) 044}, [\href{https://arxiv.org/abs/1904.05379}{{\ttfamily
  1904.05379}}].

\bibitem{Grimm:2018cpv}
T.~W. Grimm, C.~Li and E.~Palti, \emph{{Infinite Distance Networks in Field
  Space and Charge Orbits}},
  \href{https://doi.org/10.1007/JHEP03(2019)016}{\emph{JHEP} {\bfseries 03}
  (2019) 016}, [\href{https://arxiv.org/abs/1811.02571}{{\ttfamily
  1811.02571}}].

\bibitem{Hebecker:2018fln}
A.~Hebecker, D.~Junghans and A.~Schachner, \emph{{Large Field Ranges from
  Aligned and Misaligned Winding}},
  \href{https://doi.org/10.1007/JHEP03(2019)192}{\emph{JHEP} {\bfseries 03}
  (2019) 192}, [\href{https://arxiv.org/abs/1812.05626}{{\ttfamily
  1812.05626}}].

\bibitem{Banlaki:2018ayh}
A.~Banlaki, A.~Chowdhury, C.~Roupec and T.~Wrase, \emph{{Scaling limits of dS
  vacua and the swampland}},
  \href{https://doi.org/10.1007/JHEP03(2019)065}{\emph{JHEP} {\bfseries 03}
  (2019) 065}, [\href{https://arxiv.org/abs/1811.07880}{{\ttfamily
  1811.07880}}].

\bibitem{Junghans:2018gdb}
D.~Junghans, \emph{{Weakly Coupled de Sitter Vacua with Fluxes and the
  Swampland}}, \href{https://doi.org/10.1007/JHEP03(2019)150}{\emph{JHEP}
  {\bfseries 03} (2019) 150},
  [\href{https://arxiv.org/abs/1811.06990}{{\ttfamily 1811.06990}}].

\bibitem{Junghans:2020acz}
D.~Junghans, \emph{{O-Plane Backreaction and Scale Separation in Type IIA Flux
  Vacua}}, \href{https://doi.org/10.1002/prop.202000040}{\emph{Fortsch. Phys.}
  {\bfseries 68} (2020) 2000040},
  [\href{https://arxiv.org/abs/2003.06274}{{\ttfamily 2003.06274}}].

\bibitem{Apers:2022zjx}
F.~Apers, M.~Montero, T.~Van~Riet and T.~Wrase, \emph{{Comments on classical
  AdS flux vacua with scale separation}},
  \href{https://arxiv.org/abs/2202.00682}{{\ttfamily 2202.00682}}.

\bibitem{Bena:2020xrh}
I.~Bena, J.~Blaback, M.~Gra\~na and S.~L\"ust, \emph{{The tadpole problem}},
  \href{https://doi.org/10.1007/JHEP11(2021)223}{\emph{JHEP} {\bfseries 11}
  (2021) 223}, [\href{https://arxiv.org/abs/2010.10519}{{\ttfamily
  2010.10519}}].

\bibitem{Cicoli:2007xp}
M.~Cicoli, J.~P. Conlon and F.~Quevedo, \emph{{Systematics of String Loop
  Corrections in Type IIB Calabi-Yau Flux Compactifications}},
  \href{https://doi.org/10.1088/1126-6708/2008/01/052}{\emph{JHEP} {\bfseries
  01} (2008) 052}, [\href{https://arxiv.org/abs/0708.1873}{{\ttfamily
  0708.1873}}].

\bibitem{AbdusSalam:2020ywo}
S.~AbdusSalam, S.~Abel, M.~Cicoli, F.~Quevedo and P.~Shukla, \emph{{A
  systematic approach to K\"ahler moduli stabilisation}},
  \href{https://doi.org/10.1007/JHEP08(2020)047}{\emph{JHEP} {\bfseries 08}
  (2020) 047}, [\href{https://arxiv.org/abs/2005.11329}{{\ttfamily
  2005.11329}}].

\bibitem{Cicoli:2021tzt}
M.~Cicoli, A.~Schachner and P.~Shukla, \emph{{Systematics of type IIB moduli
  stabilisation with odd axions}},
  \href{https://doi.org/10.1007/JHEP04(2022)003}{\emph{JHEP} {\bfseries 04}
  (2022) 003}, [\href{https://arxiv.org/abs/2109.14624}{{\ttfamily
  2109.14624}}].

\bibitem{Leontaris:2022rzj}
G.~K. Leontaris and P.~Shukla, \emph{{Stabilising all K\"ahler moduli in
  perturbative LVS}},
  \href{https://doi.org/10.1007/JHEP07(2022)047}{\emph{JHEP} {\bfseries 07}
  (2022) 047}, [\href{https://arxiv.org/abs/2203.03362}{{\ttfamily
  2203.03362}}].

\bibitem{Ihl:2006pp}
M.~Ihl and T.~Wrase, \emph{{Towards a Realistic Type IIA T**6/Z(4) Orientifold
  Model with Background Fluxes. Part 1. Moduli Stabilization}},
  \href{https://doi.org/10.1088/1126-6708/2006/07/027}{\emph{JHEP} {\bfseries
  07} (2006) 027}, [\href{https://arxiv.org/abs/hep-th/0604087}{{\ttfamily
  hep-th/0604087}}].

\bibitem{Gao:2018ayp}
X.~Gao, P.~Shukla and R.~Sun, \emph{{On Missing Bianchi Identities in
  Cohomology Formulation}},
  \href{https://doi.org/10.1140/epjc/s10052-019-7291-5}{\emph{Eur. Phys. J.}
  {\bfseries C79} (2019) 781},
  [\href{https://arxiv.org/abs/1805.05748}{{\ttfamily 1805.05748}}].

\bibitem{Damian:2018tlf}
C.~Damian and O.~Loaiza-Brito, \emph{{Two‐Field Axion Inflation and the
  Swampland Constraint in the Flux‐Scaling Scenario}},
  \href{https://doi.org/10.1002/prop.201800072}{\emph{Fortsch. Phys.}
  {\bfseries 67} (2019) 1800072},
  [\href{https://arxiv.org/abs/1808.03397}{{\ttfamily 1808.03397}}].

\end{thebibliography}\endgroup


\end{document}